\documentclass[prx,aps,twocolumn,english,superscriptaddress,longbibliography,nobibnotes]{revtex4-2}
\usepackage[T1]{fontenc}
\usepackage[latin9]{inputenc}
\usepackage{graphicx}
\usepackage{amsmath}  
\usepackage{xfrac}
\usepackage{txfonts}
\usepackage[mathscr]{euscript}

\makeatletter
\usepackage{babel}
\usepackage{color}
\usepackage[colorlinks = true,
            linkcolor = blue,
            urlcolor  = blue,
            citecolor = blue,
            anchorcolor = blue]{hyperref}
\makeatother

\begin{document}

%%TC:ignore

%%%%%%%%%%%%%%%%%%%%%%%%%%%%%% Mandatory DOE disclaimer
%\setcounter{page}{0}
%
%Notice: This manuscript has been authored by UT-Battelle, LLC, under contract DE-AC05-00OR22725 with the US Department of Energy (DOE). The US government retains and the publisher, by accepting the article for publication, acknowledges that the US government retains a nonexclusive, paid-up, irrevocable, worldwide license to publish or reproduce the published form of this manuscript, or allow others to do so, for US government purposes. DOE will provide public access to these results of federally sponsored research in accordance with the DOE Public Access Plan (http://energy.gov/downloads/doe-public-access-plan).
%
%\newpage
%\quad
%\pagebreak

%%%%%%%%%%%%%%%%%%%%%%%%%%%%%% Title and authors
%\title{Witnessing Entanglement in Quantum Magnets using Neutron Scattering}
\title{Witnessing entanglement in quantum magnets using neutron scattering}
	
%	\title{Entanglement witnesses from neutron scattering in a spin-\texorpdfstring{$1/2$}{1/2} Heisenberg chain KCuF\texorpdfstring{$_3$}{3}}

	%\title{Entanglement witnesses from neutron scattering on a spin-\texorpdfstring{$1/2$}{1/2} Heisenberg chain}
	%https://latex.ornl.gov/project/5e99a105c1e1240096ed69cf
	\author{A. Scheie}
	\email{scheieao@ornl.gov}
	\thanks{These authors contributed equally to this work.}
	
	\affiliation{Neutron Scattering Division, Oak Ridge National Laboratory, Oak Ridge, TN 37831, USA}
	
	\author{Pontus Laurell}
	\email{laurellp@ornl.gov}
	\thanks{These authors contributed equally to this work.}
	
	\affiliation{Center for Nanophase Materials Sciences, Oak Ridge National Laboratory, Oak Ridge, TN 37831, USA}
	\affiliation{Computational Science and Engineering Division, Oak Ridge National Laboratory, Oak Ridge, TN 37831, USA}

	\author{A. M. Samarakoon}
	\affiliation{Neutron Scattering Division, Oak Ridge National Laboratory, Oak Ridge, TN 37831, USA}
	
	\author{B. Lake}
	\affiliation{Helmholtz-Zentrum Berlin f{\"u}r Materialien und Energie GmbH, Hahn-Meitner Platz 1, D-14109 Berlin, Germany}
	\affiliation{Institut f{\"u}r Festk{\"o}rperphysik, Technische Universit\"at Berlin, Hardenbergstra{\ss}e 36, D-10623 Berlin, Germany}
	
	\author{S. E. Nagler}
	\affiliation{Neutron Scattering Division, Oak Ridge National Laboratory, Oak Ridge, TN 37831, USA}
	\affiliation{Quantum Science Center, Oak Ridge National Laboratory, Tennessee 37831, USA}

	\author{G. E. Granroth}
	\affiliation{Neutron Scattering Division, Oak Ridge National Laboratory, Oak Ridge, TN 37831, USA}

    %\author{J-S. Caux}
    %\address{Institute of Physics, University of Amsterdam, Postbus 94485, 1090 GL Amsterdam, The Netherlands} 

	\author{S. Okamoto}
	\affiliation{Materials Science and Technology Division, Oak Ridge National Laboratory, Oak Ridge, TN 37831, USA.}
	\affiliation{Quantum Science Center, Oak Ridge National Laboratory, Tennessee 37831, USA}
	
	\author{G. Alvarez}
	\affiliation{Center for Nanophase Materials Sciences, Oak Ridge National Laboratory, Oak Ridge, TN 37831, USA}
	\affiliation{Computational Science and Engineering Division, Oak Ridge National Laboratory, Oak Ridge, TN 37831, USA}

	\author{D. A. Tennant}
    \email{tennantda@ornl.gov}
	\affiliation{Neutron Scattering Division, Oak Ridge National Laboratory, Oak Ridge, TN 37831, USA}
	%\affiliation{Materials Science and Technology Division, Oak Ridge National Laboratory, Oak Ridge, TN 37831, USA.}
	\affiliation{Quantum Science Center, Oak Ridge National Laboratory, Tennessee 37831, USA}
	\affiliation{Shull Wollan Center - A Joint Institute for Neutron Sciences, Oak Ridge National Laboratory, TN 37831. USA}
	
	\date{\today}

	\begin{abstract}
	    We demonstrate how quantum entanglement can be directly witnessed in the quasi-1D Heisenberg antiferromagnet KCuF$_3$. We apply three entanglement witnesses --- one-tangle, two-tangle, and quantum Fisher information --- to its inelastic neutron spectrum, and compare with spectra simulated by finite-temperature density matrix renormalization group (DMRG) and classical Monte Carlo methods. We find that each witness provides direct access to entanglement. Of these, quantum Fisher information is the most robust experimentally, and indicates the presence of at least bipartite entanglement up to at least $50$ K, corresponding to around 10\% of the spinon zone-boundary energy. We apply quantum Fisher information to higher spin-$S$ Heisenberg chains, and show theoretically that the witnessable entanglement gets suppressed to lower temperatures as the quantum number increases. Finally, we outline how these results can be applied to higher dimensional quantum materials to witness and quantify entanglement.
	\end{abstract}
	
	\maketitle
	
%%TC:endignore	
	
%%%%%%%%%%%%%%%%%%%%%%%%%%%%%% Main Text
	
\section{Introduction}
Quantum entanglement (QE) is intrinsically linked to measurements of correlations between observables. %and measured correlations between observables are intrinsically linked. 
Celebrated examples of this relationship are the Bell \cite{PhysicsPhysiqueFizika.1.195} and Clauser-Horne-Shimony-Holt (CHSH) \cite{PhysRevLett.23.880} inequalities involving correlations of {\it e.g.} photon polarization, used to demonstrate entanglement in few-particle systems \cite{Aspect_1999,RevModPhys.71.S288,RevModPhys.81.865}. 
Recently, such experiments have been successfully extended to systems of many particles \cite{Schmied16,PhysRevLett.118.140401}. Indeed, entanglement in many-body systems is attracting great interest \cite{RevModPhys.80.517, Vedral2008, RevModPhys.82.277, Laflorencie2016, Chiara2018, Friis2019} for potential technological application as well as a route to new insight into novel states of matter --- particularly ones with interesting emergent \cite{PhysRevLett.90.227902} and topological \cite{RevModPhys.89.041004} states and dynamics \cite{RevModPhys.91.021001}. Experimentally detecting and quantifying entanglement in macroscopic systems, though, is challenging \cite{Guehne2009,Ghosh2003,Islam2015,Kaufman2016, Friis2019} especially in the solid state. In cases where a quantum system can be quantitatively modeled, insight into quantum behavior can be gained by measuring correlation functions and carefully comparing experiment and theory %Techniques to measure correlation functions, such as inelastic neutron scattering (INS), can give significant insights into quantum behavior in materials through careful comparison between experiment and theory
\cite{Tennant1993, PhysRevA.73.012110, Christensen2007, mourigal2013fractional,  Piazza2015}. 
However, model-independent approaches to verifying and quantifying entanglement would give a more direct route to quantum properties of materials and their enhancement.

The most commonly studied form of QE within condensed matter physics is bipartite entanglement, defined as follows. If one considers a quantum system described by a density matrix $\rho$, in a Hilbert space $\mathcal{H}$, one can (bi)-partition $\mathcal{H}$ into two parts, $A$ and $B$. The trace of $\rho$ over the degrees of freedom in $B$ yields the reduced density matrix, $\rho_A=\mathrm{Tr}_B \left[ \rho\right]$. From $\rho_A$ we can obtain e.g. (i) the von Neumann entanglement entropy $S_\mathrm{vN}=-\mathrm{Tr} \left[ \rho_A\, \mathrm{ln} \rho_A \right]$, which provides a natural quantitative measure of the entanglement between regions $A$ and $B$, and (ii) the related entanglement spectrum \cite{PhysRevLett.101.010504} given by the eigenvalues of $\rho_A$. This formalism has allowed significant progress, including a deep understanding of critical systems \cite{PhysRevLett.90.227902}, topological states \cite{RevModPhys.89.041004}, and development of advanced numerical methods \cite{RevModPhys.82.277}. In particular, both entanglement entropy \cite{PhysRevLett.96.110404, PhysRevLett.96.110405, Jiang2012} and spectrum \cite{PhysRevLett.101.010504, PhysRevLett.105.116805, PhysRevLett.113.060501, PhysRevLett.113.256404, PhysRevB.94.081112} can be used to identify topological ground states and low-energy field theories of quantum systems. 
Despite being a nonlocal measure, the entanglement entropy has been directly probed in cold atom \cite{Islam2015,Kaufman2016} and photonic \cite{Pitsios2017} systems. %Unfortunately, this kind of non-local measure does not readily lend itself to experimental detection in solid systems.
%Fortunately however, other entanglement measures exist. 
These measures do not readily lend themselves to experimental detection in solid systems, where the number of particles is very large.  Fortunately, however, other entanglement measures exist. 
Researchers studying quantum information have defined many types of entanglement, and have introduced measures to detect, quantify and witness them. Since quantum information is mainly concerned with systems of qubits, which are mathematically equivalent to spin-$1/2$ systems, we can directly apply many of its insights to quantum materials.

Entanglement witnesses (EWs) \cite{RevModPhys.80.517, Vedral2008, Guehne2009} are functionals of the density matrix $\rho$ used to identify specific sets of entangled states and distinguish them from separable (unentangled) states. Every entangled state can, in principle, be detected by some EW \cite{Guehne2009}. However, the construction of an EW capable of detecting all possible entangled states would be equivalent to a solution of the so-called separability problem, which is $NP$-hard, meaning that the runtime of any algorithm solving the problem is believed to grow exponentially with the size of the Hilbert space. Hence, for practical purposes, no single EW can detect all entangled states, much like a single order parameter cannot describe all phase transitions. To be practically useful in experiment an EW (like an order parameter) should correspond to a quantity that can be directly measured or calculated from measurable quantities. While many EWs have been proposed, in the present work we choose to focus on three EWs suited for magnetic systems, where the QE is reflected in spin-spin correlations. These are (i) the one-tangle \cite{PhysRevA.61.052306, PhysRevA.69.022304, Roscilde_2004}, (ii) concurrence \cite{PhysRevA.61.052306, Roscilde_2004, Baroni_2007, PhysRevA.73.012110, Amico_2006} and the related two-tangle \cite{PhysRevA.61.052306, PhysRevA.69.022304, Roscilde_2004}, and (iii)  quantum Fisher information (QFI) \cite{Hauke2016,PhysRevA.85.022321}. These probe different types of entanglement, reflecting the rich mathematical structure of many-body states. %As we will demonstrate, the chosen EWs are practical to apply in the analysis of neutron scattering data, allowing for a protocol of entanglement identification in spin systems. 
As the results on the transverse-field XXZ spin chain material Cs$_2$CoCl$_4$ \cite{Laurell2020} demonstrate, the chosen EWs are practical to apply in the analysis of neutron scattering data, allowing for a protocol of entanglement identification in spin systems. By utilizing multiple witnesses, this approach goes beyond previous neutron scattering measurements of concurrence as applied to dimerized alternating chains  \cite{PhysRevA.73.012110,StonePhysRevLett.99.087204} and molecular magnet systems \cite{Garlatti2017}. 
By obtaining the scattering intensity in absolute units we also go beyond a recent study \cite{PhysRevResearch.2.043329} of temperature scaling of QFI in an isotropic spin chain, allowing for a quantitative determination of the entanglement present. 
Our method may further be combined with independent measurements of EWs based on e.g. static susceptibility \cite{Wiesniak_2005} or magnetic specific heat \cite{PhysRevB.78.064108}, which have been applied to a number of spin chain materials \cite{PhysRevA.73.012110, PhysRevB.75.054422, Das_2013, Singh2013, Sahling2015}. We thus believe our approach is widely applicable to quantum spin systems, and can allow rapid identification of materials hosting highly entangled states, such as quantum spin liquids \cite{broholm2020quantum,knolle2019field,Savary_2016} %, some of which may be useful for quantum computing \cite{kitaev2006anyons} or other quantum information applications. 
with stringent but feasible measurements.

\begin{figure}
	\centering\includegraphics[width=0.26\textwidth]{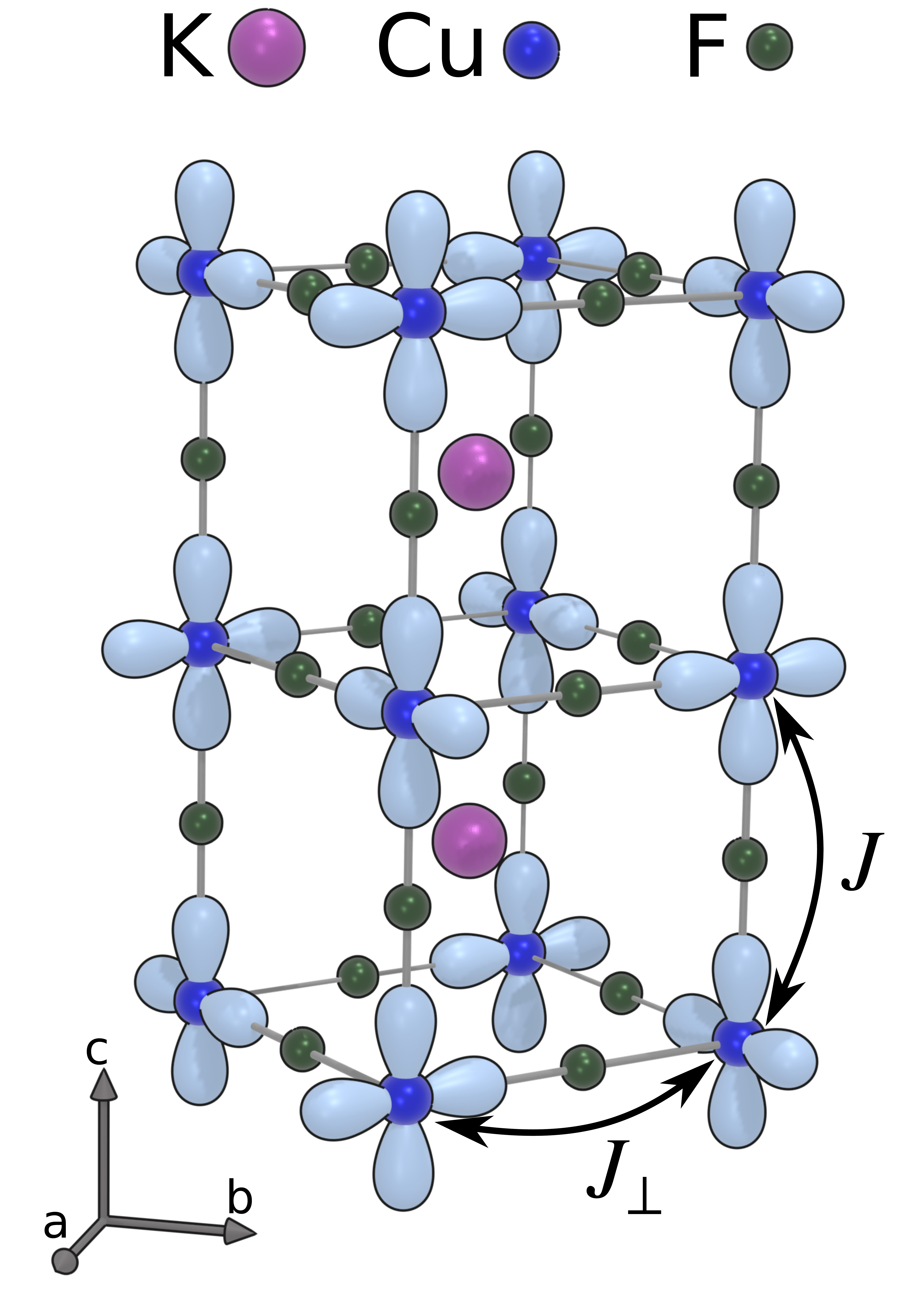}
	
	\caption{Crystal structure of KCuF$_3$. The Cu orbital order makes interchain exchange, $J_{\perp}$, much weaker than intrachain exchange, $J$.}
	\label{flo:KCuF3}
\end{figure}

In this study we apply our protocol to high-quality INS data on the one-dimensional $S=1/2$ Heisenberg antiferromagnetic (HAF) chain material KCuF$_3$. Previous literature has established that KCuF$_3$ provides an excellent realization of the isotropic HAF chain model, 
\begin{equation}
    \mathcal{H}=J\sum_{i=1}^{N}{\Vec{S}_i}\cdot{\Vec{S}_{i+1}},    \label{eq:Heisenberg}
\end{equation}
both qualitatively and quantitatively \cite{Lake2013}. 
KCuF$_3$ consists of chains of interacting Cu$^{2+}$ ions extending along the $c$-axis [Fig. \ref{flo:KCuF3}], with intra-chain coupling $J=33.5$ meV. Weak inter-chain coupling ($J_\perp=-1.6$~meV) causes the system to order magnetically at $39$~K, nevertheless at low temperatures the spectrum is dominated by a continuum of scattering above $~\sim 15$~meV \cite{Lake2005_PRB}, and is essentially 1D above $30$~meV \cite{Lake2013}. The form of the scattering continuum is a signature of fractionalized excitations (spinons) and long-range entanglement in 1D \cite{Tennant1993,Lake2013}. 

Importantly, the one-dimensional setting allows accurate simulation of the system at various temperatures using the numerically exact Density Matrix Renormalization Group (DMRG) technique, and analytical Bethe ansatz calculations at low temperature. Thus we can compare the entanglement quantified from data to accurate theoretical values.
Since both excitation spectra and entanglement properties of the Heisenberg model are relatively well-understood, this system is an ideal platform for testing the use of many-body EWs.

We find that, when experimental conditions are taken into account, the entanglement inferred from the data agrees with theoretical predictions of entanglement witnesses. However, each EW has strengths and weaknesses. The one-tangle is straightforward to calculate but suffers from strictly being applicable only at zero temperature. The concurrence and two-tangle, on the other hand, measure finite-temperature two-spin entanglement--- %Their functionals involve inequalities of combinations of real space correlation functions. 
but the precision required in the low-energy correlations make their quantification difficult for KCuF$_3$, so for gapless or very small-gap systems they may be of limited utility. In contrast, the quantum Fisher information is found to be a more practical measure of entanglement. It involves an integral that can be determined from inelastic neutron scattering data and gives a measure of multi-partite entanglement making it well suited for strongly fluctuating quantum magnets. %Further, the temperature and spin length dependence give insight into the emergence of semi-classical behavior at large $S$ and $T$. %In higher dimensional materials, carefully designed experiments involving full polarization analysis and high resolution will be required to conduct entanglement studies.

We provide here a theoretical and experimental examination of applying entanglement witnesses to a model $S=1/2$ HAF chain. In Sec. \ref{sec:EWs} we briefly review the notions of separability and entanglement, and describe the studied entanglement witnesses. We present our methods in Sec \ref{sec:methods} and results in Sec. \ref{sec:results}. Section \ref{sec:discussion} discusses the results and guidelines for future experiments probing entanglement properties in quantum matter. We end with conclusions in Sec. \ref{sec:conclusion}, and provide appendices with further technical details.

\section{Entanglement witnesses}\label{sec:EWs}

What does it mean for a state to be entangled? A state is entangled if its density matrix is not separable. An arbitrary state can be described with the density matrix $\rho=\sum_i p_i |\phi_i\rangle\langle \phi_i|$, where $p_i$ are probabilities of individual pure states $|\phi_i\rangle$. In the case of bipartite entanglement we say that $\rho$ is a separable state if it can be expressed $\rho=\sum_i p_i \rho^A_i\otimes \rho^B_i$, where $\rho^A_i$ ($\rho^B_i$) is constructed from the states in region $A$ ($B$) in $\mathcal{H}$. Product states are a special class of separable states with $\rho=\rho^A \otimes \rho^B$. The state $\rho$ is then called entangled if it is not separable \cite{Guehne2009}. Note that some separable states have genuine quantum correlations (quantified by e.g. quantum discord) despite not being entangled \cite{Chiara2018, PhysRevLett.88.017901, PhysRevA.81.052318}, and may be of use in quantum information applications \cite{PhysRevLett.101.200501}. Identifying whether a given state $\rho$ is separable or not has been shown to be $NP$-hard \cite{Guehne2009}. This is known as the separability problem.

The definitions above can be generalized to the multipartite case \cite{Guehne2009, PhysRevA.85.022321}. We say that a state is fully separable if it can be expressed $\rho=\sum_i p_i \rho_i^{(1)}\otimes \dots \otimes \rho_i^{(N)}$, where $N$ is the number of regions in the Hilbert space $\mathcal{H}$, e.g. the number of lattice sites or particles in the spin system described by Eq.~\eqref{eq:Heisenberg}. If a state cannot be expressed this way, it possesses some entanglement. However, this does not require that all $N$ particles are entangled. Indeed, we generally only expect full entanglement in specially engineered states, and not in typical condensed matter systems. To quantify how many particles are entangled, we first need two more definitions. We say that a pure state is $m$-separable if it can be written $|\phi_{m-\mathrm{sep}}\rangle=\otimes_{l=1}^M |\phi_l\rangle$, where $| \phi_l\rangle$ is a state of $N_l\leq m$ particles and $\sum_{l=1}^M N_l = N$. The pure state has $m$-partite entanglement if it is $m$-separable but not $(m-1)$-separable. A mixed state has $m$-partite entanglement if it can be written as a mixture of $(m_l\leq m)$-separable pure states, i.e. $\rho_{m-\mathrm{sep}} = \sum_l p_l |\phi_{m-\mathrm{sep}}\rangle\langle \phi_{m-\mathrm{sep}}|$, where  $|\phi_{m-\mathrm{sep}}\rangle=\otimes_{l=1}^M |\phi_l\rangle$.

The above may seem rather formal, but provides background necessary to appreciate entanglement witnesses \cite{RevModPhys.80.517, Vedral2008, Guehne2009}. As mentioned earlier, these are functionals of the density matrix, $\rho$, that identify some set of (bi- or multi-partite) entangled states without having to solve the separability problem in general. If the EW corresponds to an observable $\mathcal{O}$ it can be used to identify entangled states without full knowledge of $\rho$, since any measurement gives $\langle \mathcal{O}\rangle = \mathrm{Tr}\left[ \rho \mathcal{O} \right]$. EWs thus provide a way to experimentally detect entanglement in materials. The choice of witness (or witnesses) will depend on the system or state of interest, and the type of entanglement to be probed.

In this study we focus on three entanglement witnesses expressible as spin-spin correlation functions measurable by neutron scattering.

%\paragraph{One-tangle:}
\subsection{One-tangle}
The one-tangle, $\tau_1$, which quantifies entanglement of a single spin with the rest of the system \cite{Wootters_1998, PhysRevA.61.052306, PhysRevA.69.022304} gives a measure of {\it total} entanglement. For translation-invariant $S=1/2$ systems it can be expressed in terms of the ordered moment $M^\alpha= \langle S^\alpha \rangle,$     $\alpha\in\{x,y,z\}$ as 
\begin{align}
\tau_1=1-4\sum_\alpha (M^\alpha)^2.
\label{eq:onetangle}
\end{align}
It vanishes for a classical magnetic order and reaches its maximum in the absence of order due to quantum fluctuations. However, it is only defined at $T=0$, restricting its experimental use to the lowest temperatures. We are not aware of a finite-temperature generalization.

%\paragraph{Two-tangle:}
\subsection{Two-tangle}
The two-tangle, $\tau_2$, quantifies the total entanglement stored in pairwise correlations \cite{Roscilde_2004,Amico_2006} and satisfies $\tau_2<\tau_1$ \cite{PhysRevA.61.052306,PhysRevLett.96.220503}. It is defined as $\tau_2=2\sum_{r\neq 0}C_{r}^2,$ where $C_r$ is the concurrence \cite{PhysRevA.61.052306, Roscilde_2004, Baroni_2007, Amico_2006} for a pair of spins separated by a distance $r$. The concurrence is itself an entanglement witness that quantifies the pairwise entanglement of two spins and is closely related to Bell's type inequalities. For the isotropic $S=1/2$ HAF chain in the absence of order it simplifies to,
\begin{align}
    C_{r}  &=  2\,\mathrm{max}\left\{\, 0,\, 2|g_{r}^{zz}| -\left|\frac{1}{4}+g_{r}^{zz}\right|\,\right\},
    \label{eq:twotangle}
\end{align}
where $g_{r}^{zz} = \langle S_i^z S_{i+r}^z\rangle$. In general, concurrence for $S=1/2$ systems is a function of real-space spin correlations and magnetization components \cite{PhysRevA.69.022304}.
The concurrence remains short-ranged and $\tau_2$ is non-infinite even at quantum critical points where correlations become long-ranged; a consequence of quantum monogamy  (the tradeoff in bipartite entanglement between multiple spins) \cite{PhysRevA.61.052306,Roscilde_2004,PhysRevLett.96.220503}, which is itself linked to frustration effects in spin-spin correlations \cite{PhysRevLett.96.220503}. 

One can see the limitations inherent in pairwise EWs by considering resonating valence bond type states in higher dimensional lattices. Monogamy will mean the correlations between pairs will be reduced due to sharing of singlets in the ground state. For such a state, although clearly quantum entangled, the strict condition of exceeding the classical correlation value of 1/4 may not be met and this can be expected to be a problem for most quantum magnets beyond explicitly dimerized systems and low-dimensional geometries. For example, the concurrence vanishes for the highly entangled Kitaev spin liquid \cite{PhysRevLett.98.247201}, reinforcing the point that a single EW cannot detect all non-separable states. There are generalizations of concurrence to $S>1/2$ \cite{Li2008, Osterloh2015, Bahmani2020}, but to our knowledge there are so far no simple expressions for spin models of interest. Thus concurrence and two-tangle are currently only useful for $S=1/2$ systems.
%\tau_2 includes a factor of 2 in the \sum_r form of the expression of Amico et al. (2006), to account for the double counting in the \sum_{j\neq i} formulation used by Roscilde et al. (2004)

%\paragraph{QFI:}
\subsection{Quantum Fisher information}
Finally, quantum Fisher information (QFI) originates from quantum metrology in analogy with classical Fisher information. It puts precision bounds on parameter estimation through the quantum Cram{\'e}r-Rao bound \cite{PhysRevLett.72.3439, Toth_2014}, and has been shown to act as a witness of multi-partite entanglement \cite{PhysRevA.85.022321, PhysRevA.85.022322}. In non-integrable systems, QFI could also be used to test the eigenstate thermalization hypothesis \cite{PhysRevLett.124.040605}. %There are close relations between quantum measurement and information. 
%For a system of $N$ spin-1/2's in a separable state the QFI is limited to $f_\mathcal{Q}[\rho, S_\mathrm{tot}^{\alpha}]=F_\mathcal{Q}[\rho, S_\mathrm{tot}^{\alpha}]/N\leq 1$ (where $S_\mathrm{tot}^{\alpha}=\sum_{i=1}^N S_{i}^{\alpha}$ and  $\mathcal{Q}$ in the subscript denotes ``Quantum'')---whereas the maximum is $f_{\mathcal{Q}}[\rho, S_\mathrm{tot}^{\alpha}])\leq N$ for a completely entangled quantum state.
For a system of $N$ spin-1/2's in a separable state the QFI $F_\mathcal{Q}$ is limited to $F_\mathcal{Q}[\rho; S_\mathrm{tot}^{\alpha}]\leq N$ (where $S_\mathrm{tot}^{\alpha}=\sum_{i=1}^N S_{i}^{\alpha}$ and  $\mathcal{Q}$ in the subscript denotes ``Quantum'')---whereas the maximum is $F_{\mathcal{Q}}[\rho; S_\mathrm{tot}^{\alpha}]\leq N^2$ for a completely entangled quantum state. It is convenient to define the QFI density $f_\mathcal{Q}[\rho; S_\mathrm{tot}^{\alpha}]=F_\mathcal{Q}[\rho; S_\mathrm{tot}^{\alpha}]/N$.

The QFI is rigorously related to the dynamical susceptibility of the observable $\mathcal{O}$ \cite{Hauke2016}. For spin systems, where the dynamical susceptibility is accessible to INS experiments, we have the QFI density: 
\begin{equation}
    { f_\mathcal{Q}({T})} = \frac{4\hbar}{\pi} \int_0^{\infty} \mathrm{d} \left( \hbar\omega\right) \tanh \left( \frac{\hbar \omega}{2 k_B T} \right) \chi\prime\prime (\hbar \omega, T), \label{eq:QFI}
\end{equation}
%for spins of length $S$
where the dynamical susceptibility, $\chi\prime\prime$, is measured at a specific point in reciprocal space. For a $S=1/2$ antiferromagnetic chain, QFI is evaluated at the nearest neighbor correlation $k=\pi$ (which would be the ordering wavevector of an equivalent classical system). If the QFI density satisfies the bound
\begin{align}
    f_\mathcal{Q}   &> m \left( h_\mathrm{max}-h_\mathrm{min} \right)^2,    \label{eq:QFIbound}
\end{align}
where $h_\mathrm{max}$ and $h_\mathrm{min}$ are the maximum and minimum eigenvalues of the observable $\mathcal{O}$, and $m$ is an integer, then the system must be at least $(m+1)$-partite entangled \cite{PhysRevA.85.022321}. (Strictly speaking this holds only if $m$ is a divisor of $N$. We assume that $N$ in experiment is large enough and indeterminate, such that $N$ is divisible by all $m\ll N$. Note also that, unlike Ref. \cite{Hauke2016}, we treat $\chi''$ as an intensive quantity, i.e. it includes a factor $1/N$, as is conventional in the study of magnetism.) To determine if this bound is met, it is thus necessary to obtain the inelastic scattering in absolute units.

%Here, the QFI is normalized to the difference of the maximum and minimum values of the observable $\mathcal{O}$ squared. For spin components $S^{\alpha}$ where $\alpha\in\{x,y,z\}$ this gives $(h_{max}-h_{min})^2/4=S^2$ in the denominator. A $f_{\mathcal{Q}}$ larger than some integer $m$ strictly requires at least $m+1$ components in the system are entangled per site. (Strictly speaking this holds only if $m$ is a divisor of $N$. We assume that $N$ in experiment is large enough and indeterminate, such that $N$ is divisible by all $m\ll N$. Note also that, unlike Ref. \cite{Hauke2016}, we treat $\chi''$ as an intensive quantity, i.e. it includes a factor $1/N$, as is conventional in the study of magnetism.) %Because of the relation between entanglement and variance QFI finds immediate application in {\it e.g.} squeezed states of cold atoms and photons. 

Here, the fluctuation-dissipation theorem,  $\chi\prime\prime\left(k,\omega\right)=\frac{1}{\hbar}\tanh\left(\sfrac{\hbar\omega}{2k_BT}\right)S\left(k,\omega\right)$,
links $\chi\prime\prime$ to the dynamical spin structure factor $S(k,\omega)$
measured by neutron scattering. Sum rules for total scattering, {\it e.g.}
\begin{equation}
\sum_{\alpha\in\{x,y,z\}}\int_{-\infty}^{\infty}\int_{0}^{2\pi}
\mathrm{d}\omega \mathrm{d}k \, S^{\alpha\alpha} \left(k, \omega\right) =S(S+1) \label{eq:sumrule}
\end{equation}
in the isotropic case, constrain the dynamical response. It is evident then that Eq. \eqref{eq:QFI} relates QFI to a quantum enhancement in the linear response of a system, and thus provides a potentially useful discriminator for quantum materials. For neutron scattering studies of spin-$S$ systems satisfying Eq.~\eqref{eq:sumrule}, the bound \eqref{eq:QFIbound} becomes \cite{Laurell2020}
% \begin{align}
%     f_\mathcal{Q}   &< 12m S^2.
% \end{align}
\begin{align}
    {\rm nQFI} = \frac{f_\mathcal{Q}}{12 S^2}   &> m.    \label{eq:QFIbound2}
\end{align}
 This is the relevant bound for systems of arbitrary spin.
Throughout this work we will call the left hand side $(\frac{f_\mathcal{Q}}{12 S^2})$ ``normalized QFI'' (nQFI). 
Unlike the other EWs we discuss, QFI is generally applicable to physical systems over all physical conditions ({\it e.g.} temperature) reinforcing its usefulness.

\section{Data analysis and numerical methods}\label{sec:methods}
\subsection{Analysis of INS data}

\begin{figure*}
	\centering\includegraphics[width=0.99\textwidth]{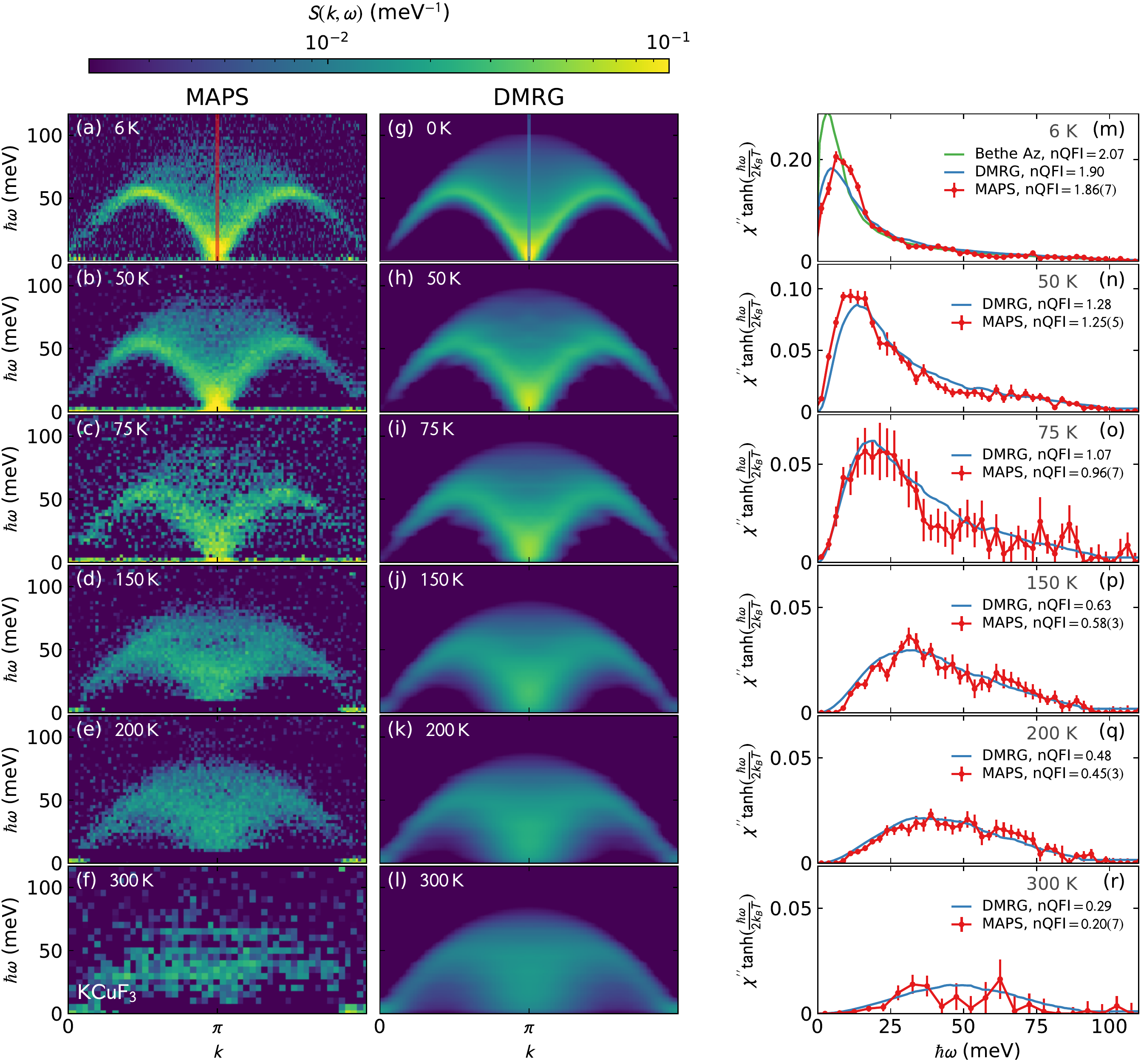}
	
	\caption{Spectra and quantum Fisher information. (a)-(f): Neutron scattering of KCuF$_3$ measured on MAPS at different temperatures. (g)-(l): DMRG simulated scattering from a 1D 
	HAF with experimental resolution broadening  applied. (m)-(r): %\textcolor{red}{Semiclassical Monte Carlo simulation of a 1D HAF.} 
	%Classical Monte Carlo simulation of a 1D HAF, showing the classical contribution. 
	%(s)-(x):
	QFI integrand at $k=\pi$, shown with normalized Quantum Fisher Information ($\frac{f_\mathcal{Q}}{12 \cdot S^2}$) calculated at that point.  At $6$ K the data is also compared with the algebraic Bethe ansatz result (m). %\textcolor{red}{The dashed purple line shows the semiclassical approximation, which gives a finite integrand. When the classical limit is taken we recover $\mathrm{nQFI}=0.$}
	}
	\label{flo:QFIa}
\end{figure*}

We use inelastic neutron scattering data on KCuF$_3$ from Refs. \cite{Lake2005,Lake2013} to evaluate experimental entanglement witnesses.
The spectra were measured on the MAPS time-of-flight spectrometer at the ISIS pulsed neutron source and cover the full frequency and wave-vector response of the material over temperatures (6, 50, 75, 150, 200, and 300~K) up to of order the Curie-Weiss temperature $\Theta_{CW}=JS(S+1)=274$~K. Data were corrected for anisotropic Cu$^{2+}$ form factor and converted to absolute units to obtain $\mathcal{S}(k,\omega)$, see Appendix~\ref{app:DataProcessing} for details. At high temperatures the low energy scattering is dominated by phonons and the estimated phonon contributions were subtracted from the data at all temperatures prior to form-factor correction. To ensure this was done accurately the phonon spectrum was remeasured carefully using the ARCS spectrometer at Oak RIdge National Laboratory; see Appendix~\ref{app:Phonons} for details. Phonon-subtracted and form-factor corrected $\mathcal{S}(k,\omega)$ %and model %at $k=\pi$, transformed to dynamic susceptibility, 
are shown in Fig. \ref{flo:QFIa}.

%Data and model at $k=\pi$, transformed to dynamic susceptibility, are shown in Fig. \ref{flo:QFIa}. Theoretically, the $k=\pi$ intensity diverges as $T \rightarrow 0$ for a pure $S=1/2$ HAF chain, and thus QFI diverges as well \cite{Hauke2016}. However, a system at finite temperature measured with a finite energy and momentum resolution will have a finite QFI at low temperatures. 

\begin{figure}
	\centering\includegraphics[width=0.48\textwidth]{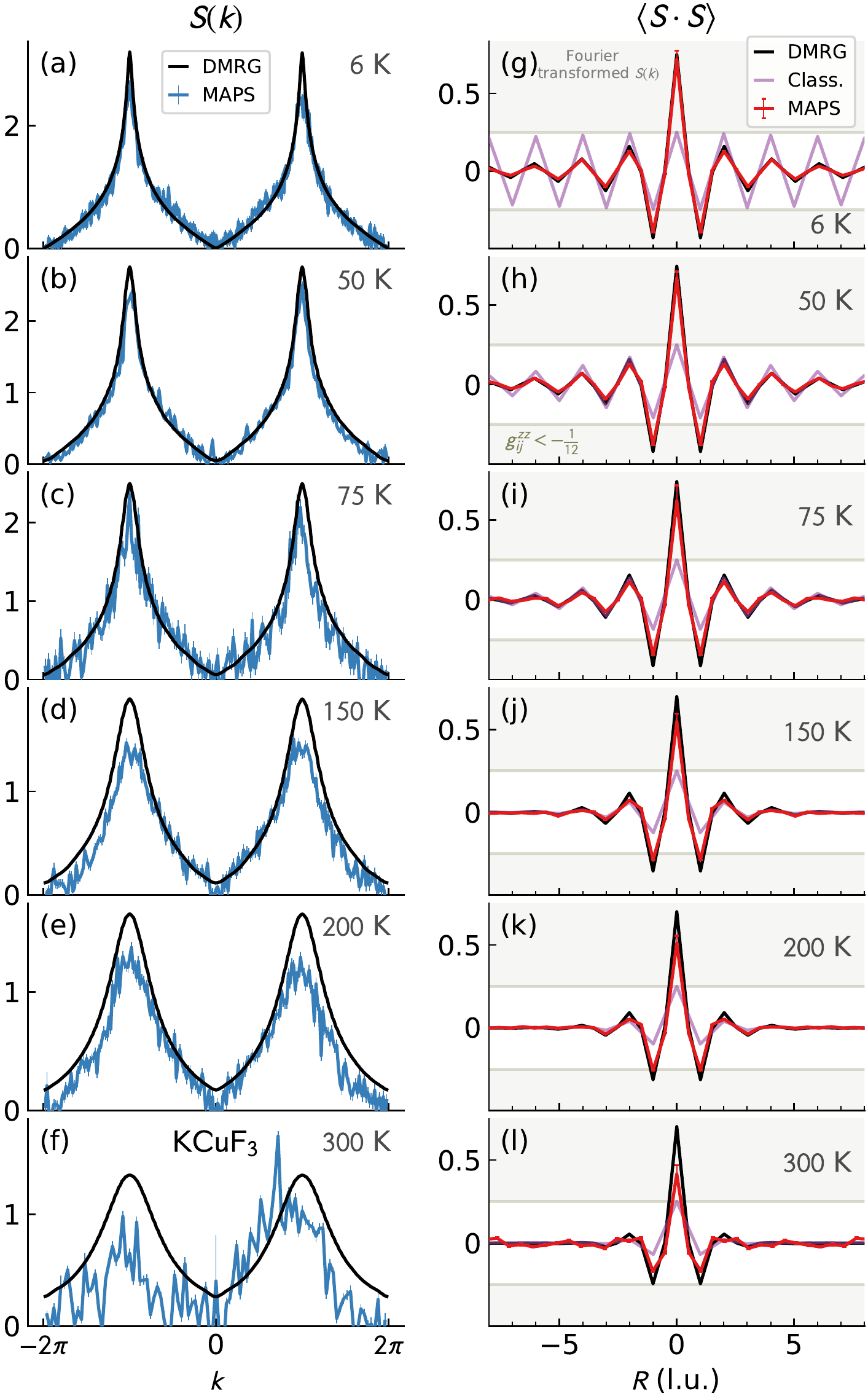}
	
	\caption{Spin-spin correlations. The left column (a)-(f) shows the energy-integrated scattering $S(k)$ (blue) from the MAPS KCuF$_3$ data compared to DMRG $S(k)$ (black). The right column (g)-(l) shows the spin-spin correlation calculated as the Fourier transform of $S(k)$. The $x$ axis in this case gives neighbor distance along the 1D chain. The horizontal grey bars show the threshold of quantum correlations. Any value within the shaded regions indicates quantum entanglement. The comparison of classical Monte Carlo, DMRG and experimental results clearly shows the quantum behavior of the system, with enhanced on-site correlations and decay of correlation functions at low temperatures driven by quantum fluctuations.}
	\label{flo:SdotSMaps}
\end{figure}

The concurrence and two-tangle require the distance dependent equal-time correlations $g_r^{\alpha\alpha}$. These are extracted from $\mathcal{S}^{\alpha\alpha}(k)=\int_{-\infty}^{\infty}d\omega \mathcal{S}^{\alpha\alpha}(k,\omega)$ by an inverse Fourier transform [see Fig. \ref{flo:SdotSMaps}(a-f)]. The elastic line has been masked due to the incoherent scattering from other sources. Negative-energy transfers were not measured in the experiments so they are calculated from the positive energy scattering using detailed balance---see Appendix \ref{app:DataProcessing} for details. From this spin-spin correlation we calculate the concurrence and two-tangle, using Eq. \eqref{eq:twotangle}.  %Inverse Fourier transformation gives the spin-spin correlation function $g_r^{zz}$ used to calculate the concurrence and two-tangle, Eq. \eqref{eq:twotangle}.

\subsection{Simulations}

To compare these calculated quantities with the theoretical behavior of a pure $S=1/2$ HAF chain, neutron spectra were simulated with DMRG \cite{PhysRevLett.69.2863, PhysRevB.48.10345, Alvarez2009}. We used the Krylov-space correction vector approach \cite{PhysRevB.60.335, PhysRevE.94.053308} to calculate $S(k,\omega)$, allowing accurate results at all $\omega$. Due to finite-size limitations the spectra were calculated with a Lorentzian energy broadening with half-width at half-maximum (HWHM) $\eta=0.1J$. To simulate experimental conditions, the DMRG spectra were convolved with a resolution function using the ms\_simulate package of the MSLICE program (see Appendix \ref{app:DataProcessing} for details). The simulated spectra are shown in Fig. \ref{flo:QFIa}(g-l).

The DMRG calculations were carried out with the DMRG++ code \cite{Alvarez2009}, keeping a minimum of $100$ and up to $1000$ states in the calculation, while targeting a truncation error below $10^{-8}$. In practice, the actual truncation error in obtaining wave functions was $\lesssim 10^{-10}$. The $6$ K result was approximated with a $T=0$ DMRG calculation on a chain with $100$ sites and open boundary conditions (OBC). For $T>0$ calculations we used the ancilla (or purification) method \cite{PhysRevB.72.220401,PhysRevB.81.075108,PhysRevB.93.045137} with a system consisting of $50$ physical and $50$ ancilla sites, also with OBC. Details on how to reproduce our results are given in Appendix \ref{app:DMRG} and the Supplemental Material \cite{SuppMat}. Based on finite-size scaling between 50, 100, and 120 site DMRG, we estimate an uncertainty of 0.4\% in the overall intensity of the DMRG simulations.

To highlight the quantum nature of entanglement, we also consider a fully classical system where entanglement is strictly absent. For this purpose, we simulated a HAF 
%To compare our results to a fully classical system (in which entanglement cannot be present), we simulated a HAF 
chain using Landau-Lifshitz dynamics (LLD) followed by Metropolis annealing \cite{samarakoon2017comprehensive}. A spin chain of 2000 classical vector spins of length $S=1/2$ was solved for LLD starting from a thermalized spin configuration at a given temperature. A standard Metropolis sampling algorithm has been used to thermalize the spin system starting from a long-ranged anti-ferromagnetic configuration. 
Correlation functions were calculated by averaging over 192 independent simulation runs.

\section{Results}\label{sec:results}

\subsection{One-tangle}

The low-temperature ($T \ll T_N$) ordered moment for KCuF$_3$ is $\mu=0.49(7)$~$\mu_B$ \cite{Hutchings_1969} ($\langle S^z \rangle = 0.24(3)$). This gives a one-tangle value, Eq. \eqref{eq:onetangle}, of $\tau_1 = 0.76 \pm 0.14$.
Theoretically, the $S=1/2$ HAF chain does not order (giving $\tau_1 = 1$), but the ordering in KCuF$_3$ %and the ordering
is due to inter-chain coupling \cite{Lake2005_PRB}. %Quantum field theories provide a quantitative account of the ordering as well as a longitudinal Higgs mode associated with the order. 
Although $\tau_1$ is reduced due to long-range order it still indicates substantial entanglement.

\subsection{Two-tangle}

\begin{figure}
	\centering\includegraphics[width=0.48\textwidth]{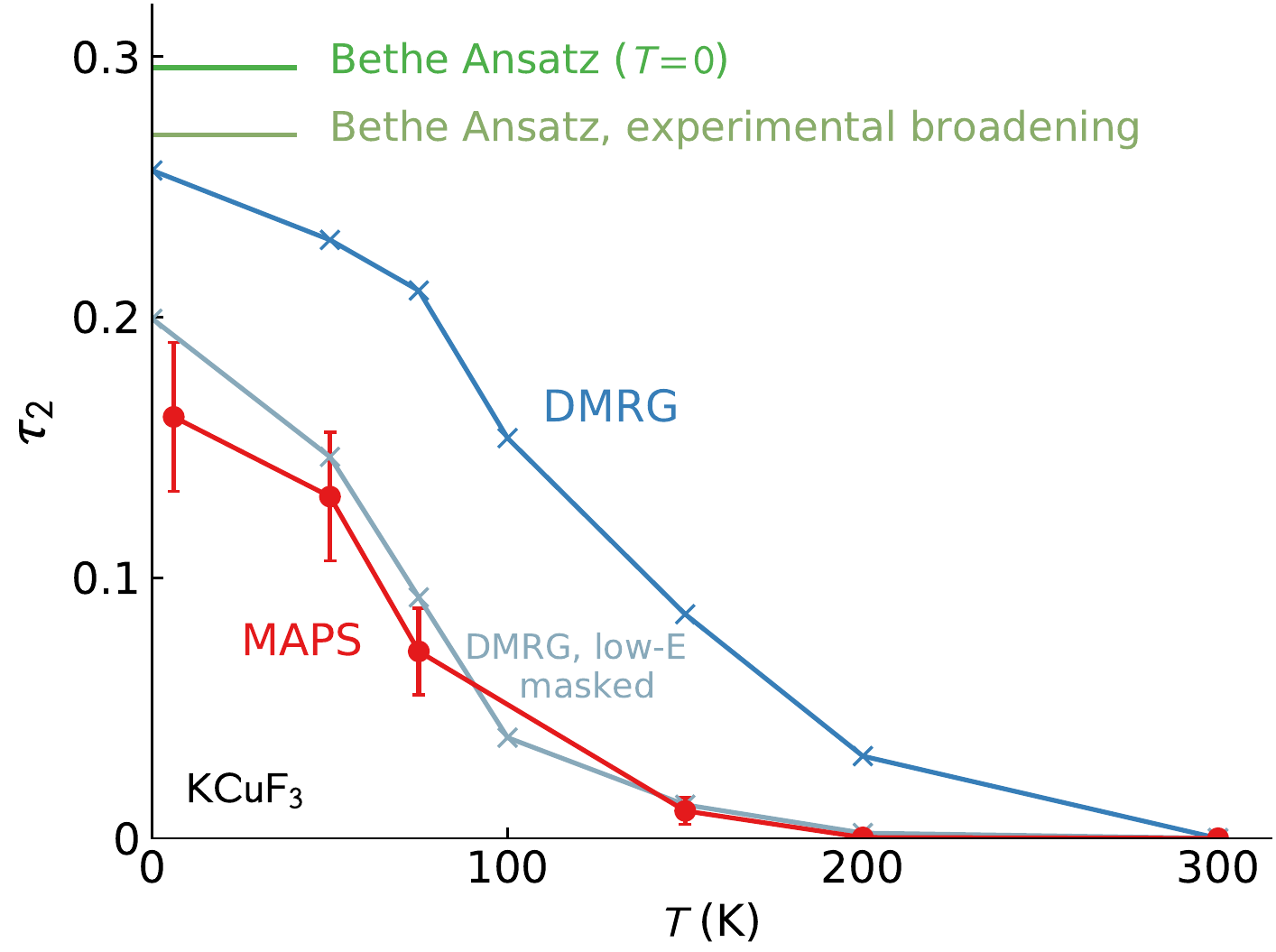}
	
	\caption{Two-tangle. Two-tangle $\tau_2$ for KCuF$_3$ calculated from the spin-spin correlation functions in Fig. \ref{flo:SdotSMaps}. DMRG two-tangle is $\approx 0.1$ higher than the experimental values, but the correspondence is very close if the low-energy scattering is excluded (light blue data), showing the low-energy features are key to accurate two-tangle calculations. Classical MC two-tangle is zero at all temperatures.}
	\label{flo:Two-tangle}
\end{figure}

The calculated two-tangle $\tau_2$ values as a function of temperature are plotted in Fig. \ref{flo:Two-tangle}. In this case, $\tau_2$ extracted directly from DMRG simulations is noticeably higher than the experimental values over the whole temperature range. This discrepancy is surprising: the {\it prima facie} agreement for $S(k)$ in Fig. \ref{flo:SdotSMaps} between theory and experiment appears excellent while the Bethe ansatz calculations (shown by the green bars in Fig. \ref{flo:Two-tangle}) show resolution effects are small.

The origin of the discrepancy can be deduced from a close examination of the data in Fig. \ref{flo:SdotSMaps}(a)-(f). DMRG $S(k)$---which was calculated with resolution broadening---is much sharper than experiment at $k=\pi$ at low temperatures. This is because the elastic line was masked below $\hbar \omega = 4$~meV in the MAPS data to eliminate substantial background contamination from unavoidable incoherent elastic scattering. The most intense magnetic scattering at $k=\pi$ is then masked, resulting in $S(k)$ not being as sharp as theory, nearest-neighbor $\langle S \cdot S \rangle$ being slightly reduced, and the calculated $\tau_2$ is suppressed.

Another thing to note about the data in Fig. \ref{flo:SdotSMaps} is that $\langle S \cdot S \rangle$ at $R=0$ falls off in the experimental data as temperature increases. This is not true for the DMRG---it remains constant for all temperatures. This shows that there is missing magnetic spectral weight for the MAPS data at elevated temperatures. ($\langle S \cdot S \rangle$ at $R=0$ corresponds to the zero-moment sum rule, which for $S=1/2$ should be $S(S+1)=3/4$.) Both MAPS and DMRG satisfy the sum rule at low temperatures, but at high temperatures only the DMRG does. The reason for this can be seen in Fig. \ref{flo:QFIa}, where the high-temperature MAPS data is oversubtracted at low energies due to intense phonon scattering (see Appendix \ref{app:Phonons}). 
To simulate this missing intensity, we masked the low-energy DMRG intensity (details are given in Appendix \ref{app:DataProcessing}), and recalculated two-tangle. As shown by Fig. \ref{flo:Two-tangle}, the DMRG-masked two-tangle closely matches the experimental calculations below 100~K.
This shows that the low-energy scattering has a strong influence on two-tangle calculations, in contrast to QFI.

As a final note, the classical MC simulations, shown in purple in Fig. \ref{flo:SdotSMaps}, have zero concurrence, and thus zero two-tangle at all temperatures. This is as expected for a classical system.

\subsection{Quantum Fisher Information}

\begin{figure}
	\centering\includegraphics[width=0.48\textwidth]{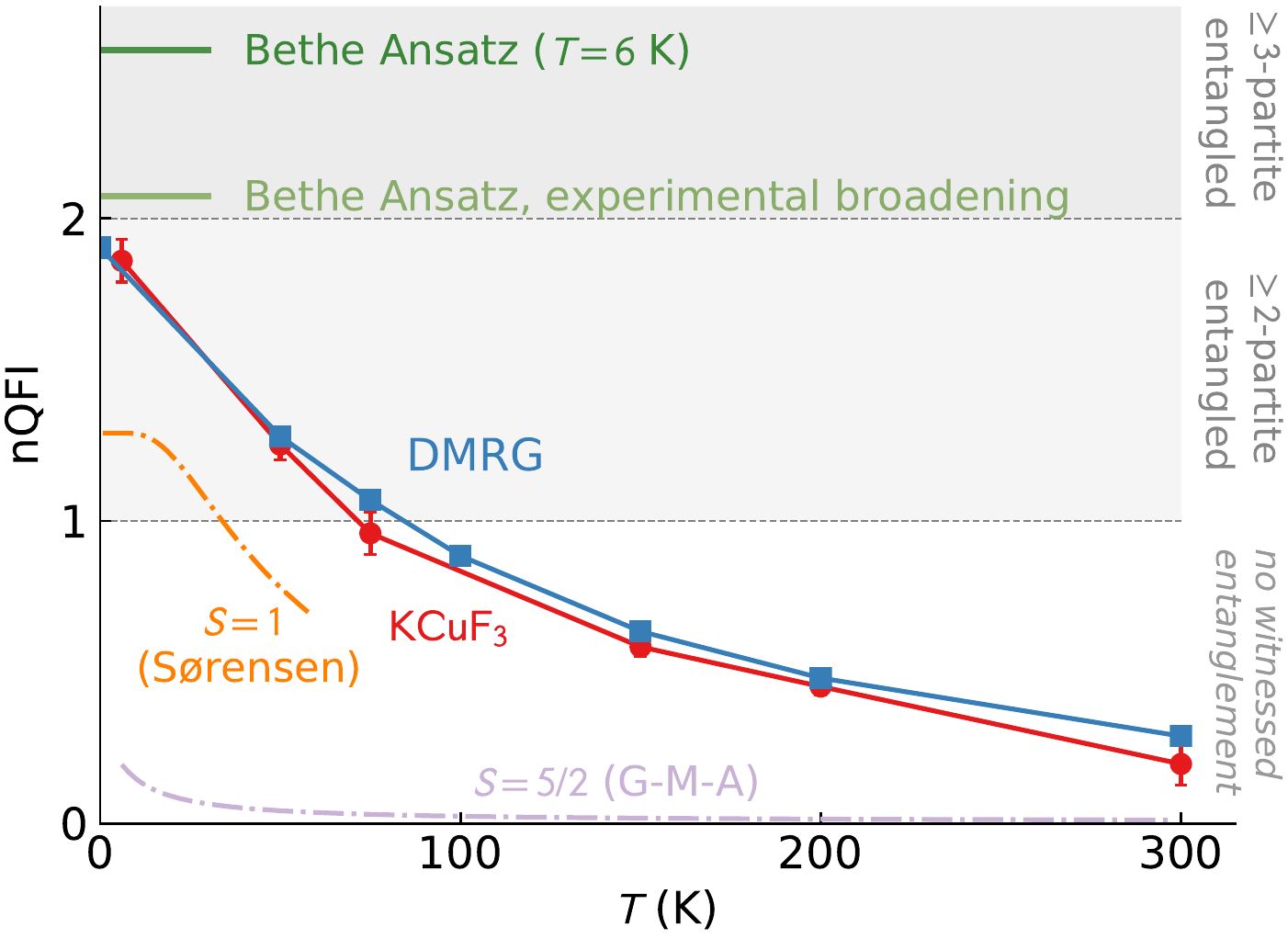}
	
	\caption{Normalized Quantum Fisher information [Eq. \eqref{eq:QFIbound2}] as a function of temperature. The $y$-axis units directly indicate the degree of multipartite entanglement present. When ${\rm nQFI}=f_\mathcal{Q}/\left(12S^2\right)>m,$ where $m>0$ is an integer, the system is in a state with $\geq \left( m+1\right)$-partite entanglement. 
	%indicate $\geq m+1$-partite entangled states. 
	We show nQFI densities calculated according to the formula in Eq. \eqref{eq:QFI} from DMRG simulations and MAPS data. We also show theoretical nQFI calculated for a $S=1$ chain ("S\o{}rensen") \cite{Lambert_2019} and for $S=5/2$ ("G-M-A") \cite{Gozel_2019}. Classical MC nQFI is zero at all temperatures.}
	\label{flo:QFIb}
\end{figure}

The experimental and DMRG-simulated QFI values agree remarkably well with each other, as shown in Fig. \ref{flo:QFIb}.
Such correspondence between theory and experiment is possible because the $\tanh$ function in the finite-temperature QFI integral suppresses the low-energy scattering, which is where the effects of inter-chain coupling and background subtraction are most manifest \cite{Lake2005}. Thus, the theoretical integral is quite close to the experiment as shown in Fig. \ref{flo:QFIa}(m-r).
To show the effects of finite resolution, we include a Bethe ansatz calculation of the $T=0$ HAF $S=1/2$ chain with and without resolution broadening \cite{Lake2013}, shown by the green bars in Fig. \ref{flo:QFIb}. (Bethe ansatz is exact and not subject to finite-size broadening like DMRG is.) To avoid the zero-temperature divergence, we calculated the QFI at $T=6$~K. This shows that resolution effects decreases the normalized QFI by about 1. This effect is noticeable, but it by no means suppresses the normalized QFI. However, there is a qualitative difference in that the Bethe ansatz result indicates the presence of at least tripartite entanglement, while experiment and DMRG witness at least bipartite entanglement.
%Thus QFI values remain meaningful even with finite resolution.

We now consider QFI in higher-spin HAF chains. Partial suppression of  $\mathrm{nQFI}=f_\mathcal{Q}/\left(12S^2\right)$ occurs for the isotropic $S=1$ spin-chain QMC calculation in Ref. \cite{Lambert_2019}, which is plotted in Fig. \ref{flo:QFIb}. We also simulated higher-spin systems with exchange strength $J$ scaled to keep the exchange energy scale (approximated by the Curie-Weiss temperature $\Theta_{CW} \propto J S(S+1)$) constant across all spin values. Appendix~\ref{app:DMRG} shows $T=0$ DMRG spectra for $S=3/2$ and $S=5/2$ HAF chains, which are quantum critical systems with extensive entanglement. We find that $T=0$ nQFI is approximately the same for these $S$ values. 
Meanwhile, finite-temperature $S(k,\omega)$ and nQFI for higher spin chains can be calculated with the semiclassical Gozel-Mila-Affleck approximation in Ref. \cite{Gozel_2019}, and the $S=5/2$ QFI is plotted in Fig. \ref{flo:QFIb}. nQFI
is noticeably suppressed with the larger spin at nonzero temperature. Increasing the spin quantum number further shows a power-law decrease in nQFI as $S \rightarrow \infty$, as shown in Appendix \ref{app:Semiclassical}. Thus, although higher spin chains are highly entangled at $T=0$, the larger spin quantum number suppresses the measurable entanglement to lower and lower temperatures. In the classical limit, which we consider in Appendix~ \ref{app:ClassicalLimit}, the ability to witness entanglement is completely suppressed.

\section{Discussion}\label{sec:discussion}

We have shown three different EWs which quantitatively demonstrate entanglement in KCuF$_3$.
These results highlight the requirements and limitations of measuring the  one-tangle, two-tangle, and QFI.

The one-tangle is the easiest EW to measure and provides an immediately useful number directly related to the entanglement of a spin with the rest of the system \cite{Wootters_1998}. For a translationally invariant Hamiltonian at zero temperature the one-tangle can be extracted from the ordered moment, which is readily measurable with neutron scattering (through magnetic elastic intensity). However, some care is needed in interpreting the results, since $\tau_1$ may be overestimated if the moments are not fully characterized, see Ref.~\cite{Laurell2020}. A major problem though is that $\tau_1$ is derived for a pure eigenstate, and thus restricted to zero temperature. Generalizing this result would be very useful in the experimental quantification of quantum effects in materials. 
Despite the lack of rigorous derivation beyond zero temperature, it is reasonable to expect that the non-entangled contribution will be within $k_{B}T$ of the elastic line at low temperature and instead of the Bragg peak intensity it will be given by the total long-time correlations beyond $t \sim 1/k_BT$, {\it i.e.}
\begin{equation}
\tau_1 \sim 1 - 4\sum_{\alpha,\,\beta = x,y,z} \int_{-k_{B}T}^{k_{B}T} d\omega \int_{B.Z.} d{\bf q}\mathcal{S}^{\alpha\beta}({\bf {q}},\omega).
\end{equation}
Similarly, a useful expression for the one-tangle in disordered systems would provide a way of determining whether an experimental system is of interest as say a quantum spin liquid versus a glassy or thermally disordered state. 

%Speculate that maybe it is ${\int dQ S({\bf Q},{\omega=0})}$.    
Two-tangle is less susceptible to experimental broadening effects than QFI, but it is more susceptible to other experimental artifacts and perturbations away from quantum-criticality. Of all the EWs we considered, two-tangle required the most careful isolation and treatment of magnetic scattering---DMRG had to be masked in accord with missing experimental spectral weight---which may prove a serious limitation to studying less ideal systems than KCuF$_3$. On the other hand, the two-tangle is easy to compute theoretically, and is almost immune to finite-size effects. However, since concurrence is typically short-ranged, $\tau_2$ is less powerful than the QFI in demonstrating true long-range entanglement in topologically ordered states.

Finally, QFI is a powerful measure of finite-temperature entanglement for low-dimensional systems, showing $\geq 2$-partite entanglement in KCuF$_3$. At finite-temperature, QFI remains robust against weak perturbations away from quantum criticality, as shown by the correspondence to DMRG at the quantum critical point. This correspondence also shows the robustness against experimental artifacts in the neutron scattering data.
Nevertheless, there are two limitations to QFI as an EW. First, resolution broadening somewhat suppresses the calculated QFI, and thus good energy resolution is key to a successful calculation.
Second, the $T \rightarrow 0$ QFI for a real experiment will never diverge. This is because (i) resolution effects are always present which smooth over divergent intensity, and (ii) real condensed matter systems are generally not ideal. KCuF$_3$ for example has interchain coupling which brings the system away from criticality, causing the low-energy scales---which determine the low-temperature multipartite entanglement---to deviate from the theoretical $T \rightarrow 0$ behavior.
For other systems and methods, a recent reformulation of QFI \cite{Almeida2020} or the related quantum variance EW \cite{PhysRevB.94.075121, Frerot2019} may prove useful.

Higher spin simulations show that normalized QFI also discriminates between systems of different spin size.
A spin-1/2 chain shows extreme quantum behavior --- a quantum critical ground state and pairs of fractional $S=1/2$ spinons as quasiparticles. The QFI shows an immediate difference between $S=1/2$ and $S=1$ chains, with a low-temperature plateau in $S=1$ due to the Haldane gap \cite{Lambert_2019,Gabbrielli2018}. The difference in the behavior is due to a topological term in the quantum field theory describing the systems. The strength of this term is $J S^2\exp (-\pi S)$ which defines an energy scale above which the dynamics behaves akin to spin-waves. Correlations on temperature and energy scales below this exponentially suppressed crossover will still show divergent QFI in the case of half-odd-integer spins. This agrees with the conformal field theory predictions for the von Neumann entanglement entropy and QFI \cite{PhysRevLett.90.227902,PhysRevD.96.126007}. However, this energy and temperature scale suppression implies that the regime of diverging multipartite entanglement will be extremely hard to access in high-spin systems. DMRG at zero temperature on $S=3/2$ and $5/2$ 1D chains are plotted in Appendix \ref{app:DMRG}, and we expect similar rapid cross over to semiclassical behavior with spin length in other quantum magnets. QFI may then prove to be a very useful experimental indicator for when a fully quantum theoretical approach is required, and when a semi-classical approach may suffice instead.

% We have also shown how the real-space Van Hove function can discriminate between quantum and semiclassical behavior in 1D spin chains.
% These real-space correlators show distinct floquet-like physics induced by spinon quasiparticles. This view provides a sharp distinction from semiclassical magnon propogation, which includes only a perturbation on top of a long range ordered state. This view also provides new insight into the propogation of spinon quasiparticles, showing dynamic antiferromagnetic correlations between every other site. 
% As a model-independent technique, real-space correlations may provide valuable insight into the consequences of long-range entanglement for quantum dynamics.

We can expect, on the basis of the results here, that combining quantum entanglement witnesses could prove useful in a wide range of other magnetic systems. For short-range entangled systems, such as dimerized and molecular magnets the concurrence and two-tangle alone provide a useful measure of the entanglement \cite{Tennant2003,Wernsdorfer2002}. Meanwhile, the combination of two-tangle and one-tangle are able to provide new potential insights into both entanglement and quantum phase transitions by identifying changes in entanglement and quantum wave-functions. A prominent example of this is the proposed entanglement and QPTs in the XXZ model in transverse field, which is explored in Ref. \cite{Laurell2020}.  %(Eq. \eqref{eq:XXZ} for $\epsilon>1$ and with a additional Zeeman term $\sum_i BS^x_i$) \cite{Roscilde_2004, Amico_2006}. Demonstration of entanglement measures in such a system would be highly desirable in developing their application. 
However, the addition of the Quantum Fisher Information provides a powerful, % approach. Its definition is 
system agnostic %, and provides a robust 
indication of the impact of entanglement on the response of the materials. %For the HAF chain with larger spin, the indication of  multi-partite entanglement at low temperatures is a nontrivial insight and the observation of significant multi-partite entanglement in systems where it is not expected could lead to discovery of new quantum states where theories have not yet been developed. 
Further, the observation of significant multi-partite entanglement in systems where it is not expected could lead to discovery of new quantum states where theories have not yet been developed. 

Of particular interest are quantum spin liquids and their discrimination from the effects of other forms of disorder.
As mentioned earlier, quantum monogamy is likely to make the concurrence and two-tangle go to zero between all sites in higher dimension spin-liquids \cite{PhysRevLett.98.247201}. %Indeed in the Kitaev model the concurrence and two tangle are zero . 
Instead, non-zero two-tangle in a higher-dimensional system might be a signature of a random singlet phase \cite{Uematsu_2018,Uematsu_2019,Liu_2018}, and so possibly discriminates spin-liquid-like random singlet phases from true quantum spin liquids.
The QFI on the other hand may well show multipartite entanglement in higher dimensions. Although topological quantum spin liquids such as the Kitaev model have long-range quantum entanglement that cannot be fully quantified by multi-partite entanglement, a combination of (i) substantial $\tau_1$, (ii) $\tau_2=0$, and (iii) finite QFI would strongly indicate long-range entanglement. This would be a useful way of selecting systems on which to undertake experiments to probe topological quantum states (like quantum interference measurements).

As a final note, these results show that neutron scattering is well suited to witnessing entanglement in solid state systems. 
The demands of entanglement witnesses will require high-resolution techniques and carefully designed scattering experiments. For systems more complex than the $S=1/2$ HAF chain, polarized scattering may be required to isolate the magnetic signal. Also, for anisotropic systems, quantifying entanglement witnesses will require measuring the full polarization tensor of the spin-spin correlation functions \cite{Laurell2020}. 
These EW measurements could be aided by self-entangled neutron beams as recently demonstrated for CHSH states \cite{Shen20}. These measure spin-spin correlations like un-entangled beams, but they can be conditioned to simultaneously measure combinations of correlations  of distance, time, and polarization, measuring Fourier components directly. In this respect, these techniques could be used to develop more direct measurements of EWs in materials.

Although our results have focused on neutron scattering, many other experimental techniques can measure EWs (QFI in particular), for example inelastic x-ray scattering and THz spectroscopy. Furthermore, there is rich information content in the correlation functions not used in the present entanglement witnesses, so other insightful neutron scattering witness measures could powerfully elucidate many-body quantum states. Given the potential utility of identifying and quantifying entanglement in the response behavior of quantum materials, experimental and theoretical approaches should be explored further. In theoretical condensed matter physics we are often used to thinking about entanglement exclusively in terms of bipartite entanglement --- e.g. in the form of entanglement entropies and spectra. It is time to broaden this perspective, and more seriously consider entanglement measures that are experimentally accessible.

\section{Conclusion}\label{sec:conclusion}

We have demonstrated several model-independent means of  quantifying entanglement using the neutron spectrum of the 1D Heisenberg antiferromagnet KCuF$_3$. One-tangle, two-tangle, and QFI all show non-zero entanglement. We find each has specific advantages and disadvantages: One-tangle is simple to calculate, but limited to the zero-temperature limit. Two-tangle provides direct insight to the bipartite entanglement, but is easily disrupted by experimental artifacts. QFI we find to be the most robust, giving quantitative agreement with DMRG calculations across the entire temperature range. Further, QFI directly and unambiguously shows that KCuF$_3$ has at least bipartite entanglement, up to at least $50$ K.

These results serve as a proof of principle that meaningful information about quantum entanglement can be extracted from experimentally measured spin-spin correlations. 
Our results call for the development of additional EWs accessible through spin correlation functions. More generally, EWs formulated in terms of accessible observables present a promising direction forward.
Armed with such tools, the study of exotic quantum materials can progress in new ways.

%\emph{Note added:} Upon completion of this work, a related preprint appeared \cite{PhysRevResearch.2.043329}.

\subsection*{Acknowledgments}
We gratefully acknowledge Jean-S\'ebastien Caux for performing the Bethe ansatz calculations in Ref.~\cite{Lake2013}. We thank Matthew Stone for a critical reading of the manuscript. 
D.A.T. acknowledges stimulating and useful discussions with Cristian Batista, Gabor Hal\'asz, Pavel Lougovski, Gerardo Ortiz, and Roger Pynn. 
The research by P.L., S.O., and G.A. was supported by the Scientific Discovery through Advanced Computing (SciDAC) program funded by the US Department of Energy, Office of Science, Advanced Scientific Computing Research and Basic Energy Sciences, Division of Materials Sciences and Engineering. GA was in part supported by the ExaTN ORNL LDRD. This research used resources at the Spallation Neutron Source, a DOE Office of Science User Facility operated by the Oak Ridge National Laboratory. The work by DAT and SEN is supported by the Quantum Science Center (QSC), a National Quantum Information Science Research Center of the U.S. Department of Energy (DOE).  Software development has been partially supported by the Center for Nanophase Materials Sciences, which is a DOE Office of Science User Facility.
%AS was supported by the DOE Office of Science, Basic Energy Sciences, Scientific User Facilities Division. (not needed with the SNS acknowledgement above included)

% \subsection*{Author contributions}

% [insert text]

% \subsection*{Competing financial interests}
% The authors declare no competing financial interests.

%%%%%%%%%%%%%%%%%%%%%%%%%%%%%% Appendices
\appendix

\section{Data processing}\label{app:DataProcessing}

The MAPS KCuF$_3$ neutron scattering data were corrected for the anisotropic $d_{x^2-y^2}$ Cu$^{2+}$ form factor in order to account for the orbital order in Fig. \ref{flo:KCuF3}:
\begin{equation}
\begin{aligned}
    f(\mathbf{k}) = & \langle j_0 \rangle +
    \frac{5}{7}(3 \cos^2 \beta - 1) \langle j_2 \rangle +  
    \frac{3}{56}(35 \cos^4 \beta - \\ & 30 \cos^2 \beta + 35 \sin^4 \beta \cos 4 \alpha + 3)\langle j_4 \rangle,
\end{aligned}
\end{equation}
where $\beta$ is the angle between $\bf k$ and the $d_{x^2-y^2}$ orbital $z$ axis, and $\alpha$ is the angle from the $x$ axis in the $xy$-plane \cite{boothroyd2020principles}. Cu$^{2+}$ $\langle j_n \rangle$ constants were from from Ref. \cite{BrownFF}. To isolate the magnetic scattering, a phonon background was subtracted (described in Ref. \cite{Lake2013}).  This background intensifies as temperature increases (see Appendix \ref{app:Phonons}), so the low-energy scattering at the highest temperatures has a large uncertainty---but the higher energy scattering is clear. We normalized the data to absolute units by setting the inelastic zero moment total sum rule of the 6~K data to 0.75.

In order to compare the DMRG calculations directly with the experimental data, we simulated the dataset that would be collected on the MAPS instrument at ISIS, for a sample with the dynamic structure factor (DSF) of the DMRG, by using the ms\_simulate package of the MSLICE program. Before this was done however a number of corrections were applied to the theoretical DSF. First, in order to model the instrumental resolution, the DSF was convolved numerically by a Gaussian whose width was the energy-dependent resolution obtained from the MCHOP program. Second, to take account of the mosaic spread of the sample, a Gaussian angular broadening was introduced which resulted in an effective wavevector broadening. The resulting simulated datasets were identical in form to the experiment datasets and all manipulations (such as binning) performed on the real data were also performed on the virtual data. Direct comparison between theory and experiment was achieved by using the MSLICE program to perform the same cuts and slices on the virtual and real datasets.

\begin{figure}
	\centering\includegraphics[width=0.48\textwidth]{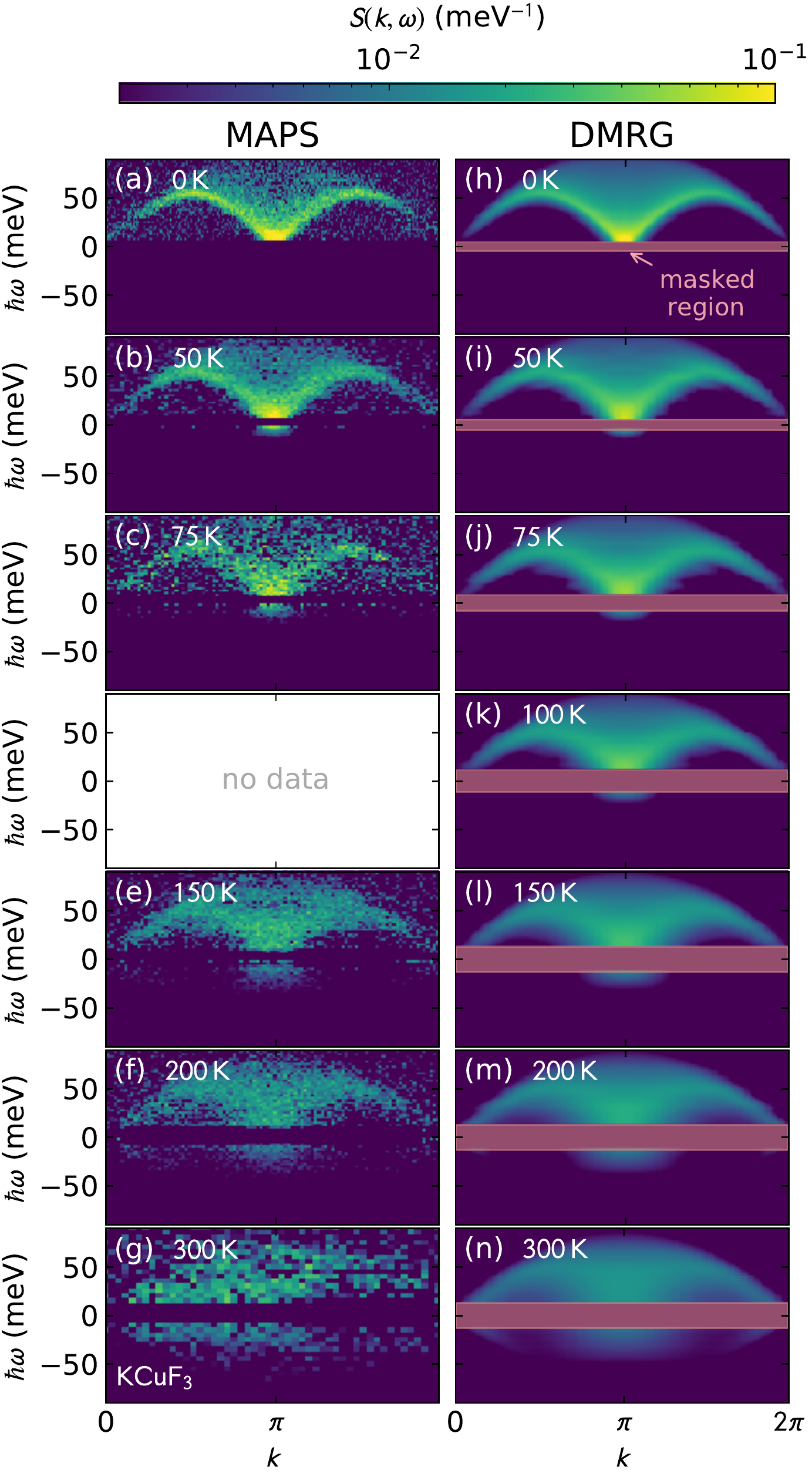}
	
	\caption{Detailed balance applied to MAPS KCuF$_3$ data and DMRG simulations. The red shaded regions in the right column indicate the masked regions used to calculate two-tangle in Fig. \ref{flo:Two-tangle}}
	\label{flo:MaskedRegions}
\end{figure}

To simulate the effects of experimental artifacts and background subtraction, we masked the low-energy DMRG simulated intensity as shown in Fig. \ref{flo:MaskedRegions}.
Because the region of missing intensity grows with temperature [see Fig. \ref{flo:MaskedRegions}(a-g)], we varied the region masked with the phenomenological function
$$
masked < 4\>{\rm meV} + \frac{8.5\>{\rm meV}}{1+ \exp{(-\frac{T-80\>{\rm K}}{15\>{\rm K}})}}.
$$
This function is not meant to be exact, but to approximate the missing intensity in the MAPS data. Below 100~K, it matches the spectrum visually, and matches two-tangle quantitatively.

\section{\texorpdfstring{KCuF$_3$}{KCuF3} phonon spectrum.}\label{app:Phonons}
\begin{figure*}
	\centering\includegraphics[width=0.98\textwidth]{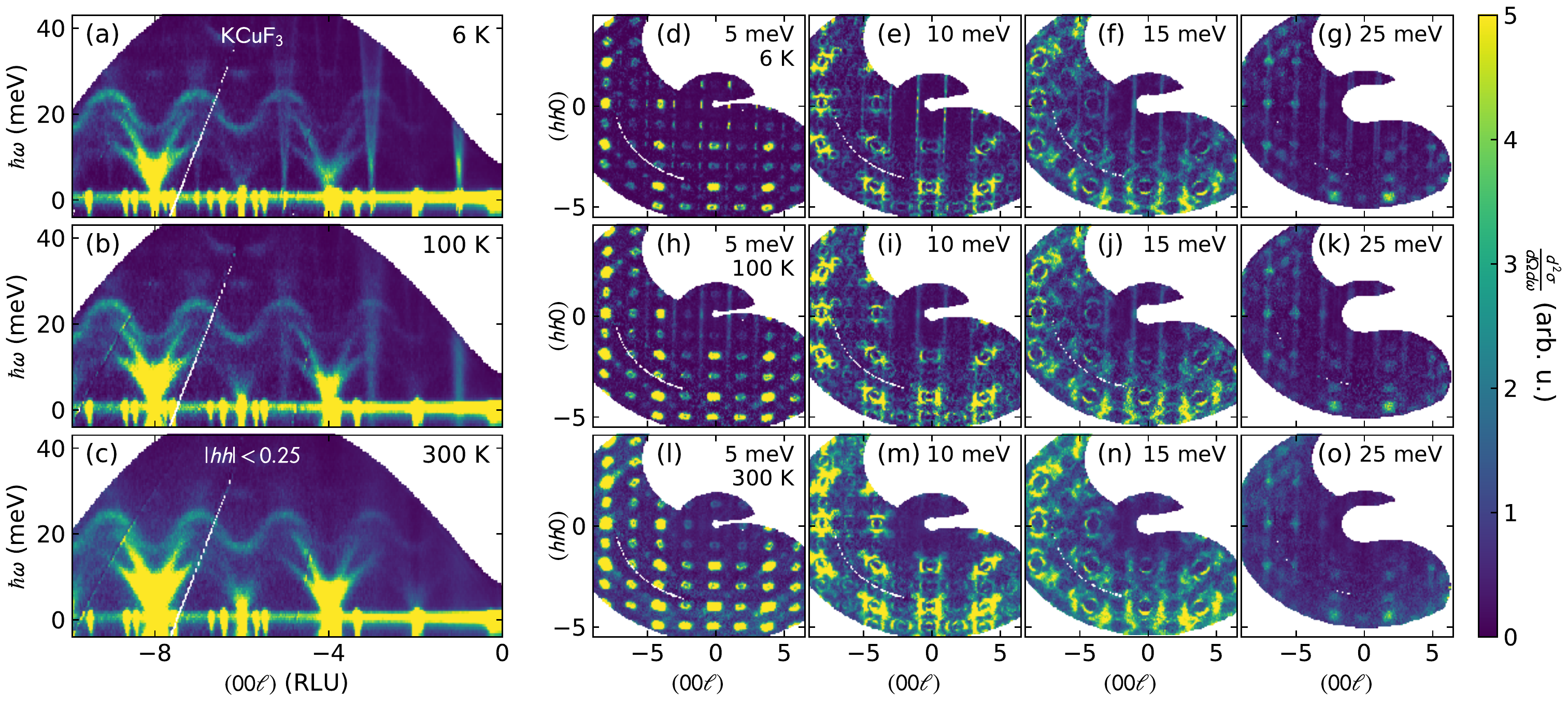}
	
	\caption{Phonon spectrum for KCuF$_3$. (a)-(c) show the phonon spectrum along $(00\ell)$, the direction of magnetic chains, at 6~K, 100~K, and 300~K respectively. Magnetic scattering is visible as the strongly dispersive modes coming out of $\ell=$1, 3, 5, etc. (d)-(o) show constant energy slices ($\hbar \omega \pm 1$~meV) at the three temperatures. The phonon spectrum is mostly below 30 meV, and increases in intensity as temperature increases. Note that above 25 meV the phonons are mostly gone, leaving only magnetic intensity.}
	\label{flo:Phonons}
\end{figure*}

The phonon spectrum of KCuF$_3$ was measured at the ARCS spectrometer \cite{abernathy2012design} at the ORNL SNS in the $(hh\ell)$ scattering plane with $E_i = 50$~meV neutrons ($T_0$ chopper at 90 Hz, Fermi 1 chopper at 120 Hz, Fermi 2 chopper at 420 Hz, slits 40 mm wide and 18 mm tall). The large  scattering vector $k$ coverage of ARCS allows for a much clearer picture of the phonons than MAPS, which are stronger at large $k$. Data were analyzed and plotted using Mantid \cite{arnold2014mantid}. The data at 6~K, 100~K, and 300~K are shown in Fig. \ref{flo:Phonons}. The phonon dispersions are primarily below 30 meV, and grow more intense as temperature increases, confirming the phonon subtraction scheme used for the MAPS data. At high temperatures, the complicated spectrum makes the phonon subtraction difficult: as shown in Fig. \ref{flo:Phonons}, the 300~K low-energy magnetic scattering is much weaker than 6~K, whilst the phonon scattering is much stronger at 300~K than 6~K. This explains why the high temperature KCuF$_3$ data from MAPS that was used to extract the two tangle witness is noisy at low energies.

\section{DMRG calculations}\label{app:DMRG}
In this appendix we provide additional DMRG results. Detailed instructions on how to reproduce the DMRG results are given in the Supplemental Material \cite{SuppMat}.

\subsection{Finite size effects}

\begin{figure*}
	\centering\includegraphics[width=0.88\textwidth]{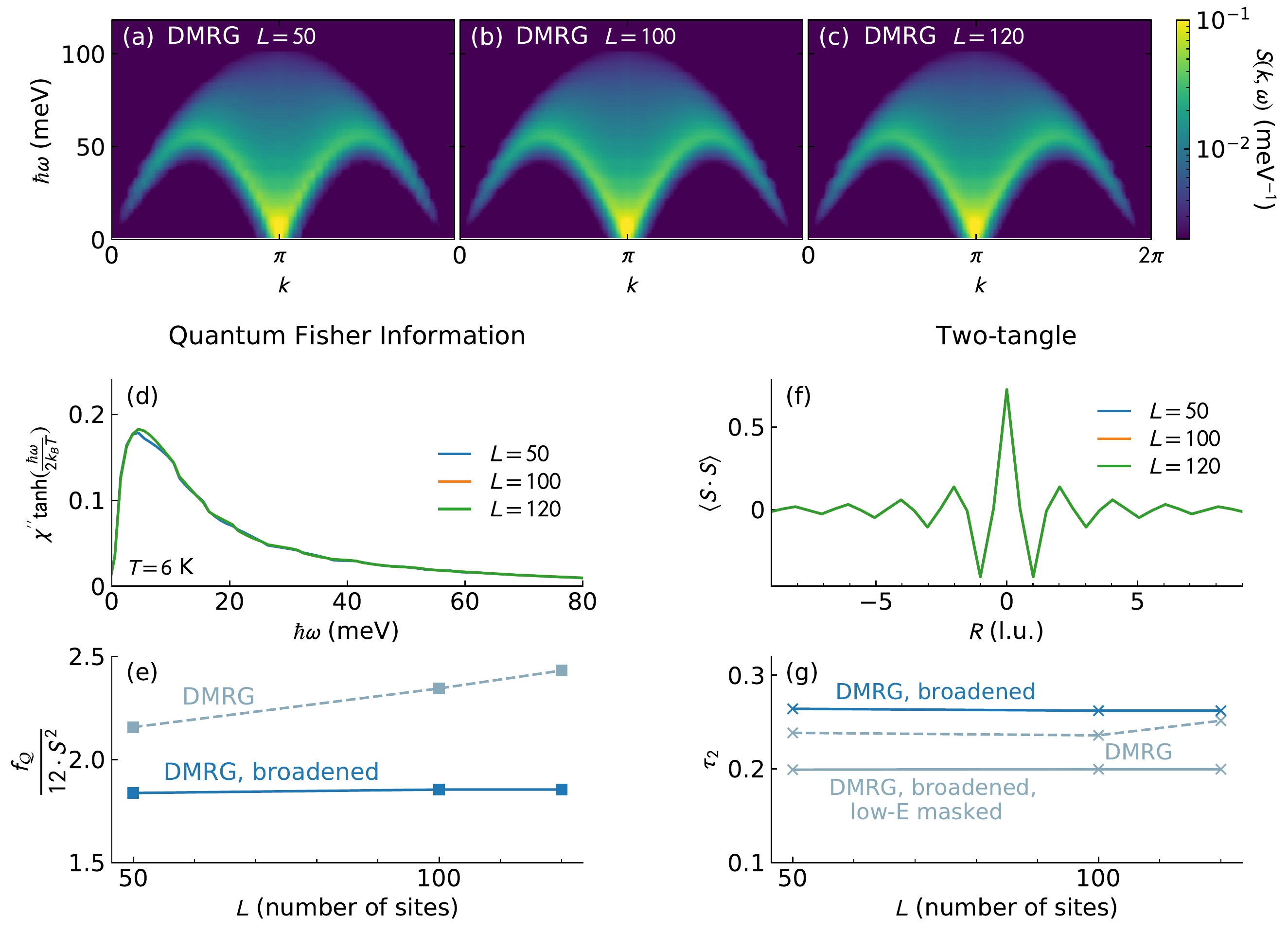}
	
	\caption{Effects of system size scaling in DMRG. (a)-(c) show the $T=0$ DMRG simulated spectra. (d)-(e) show the effect on QFI (assuming $T=6$~K), and (f)-(g) show the effect on two-tangle. QFI is dependent on system size, but two-tangle is nearly independent.}
	\label{flo:FiniteSize}
\end{figure*}

We investigated finite-size scaling effects by running $T=0$ DMRG simulations with 50, 100, and 120 sites, shown in Fig. \ref{flo:FiniteSize}. The Lorentzian broadening was set to $\eta=0.1J$ for $N=100$, and scaled as $\eta\sim 1/N$ for other system sizes following Ref.~\cite{PhysRevB.66.045114}. These results show QFI increases with system size. This is because QFI is strongly dependent upon the low-energy intensity at $k=\pi$, which gets sharper as system size increases. However, experimental broadening suppresses QFI and removes this size-dependence.
Meanwhile, the two-tangle is nearly independent of system size, both with and without experimental effects. This is because $\tau_2$ is dominated by the nearest-neighbor concurrence, which is determined by  nearest-neighbor correlations that are less influenced by the overall size of the simulated system.

\begin{figure*}
	\centering\includegraphics[width=0.90\textwidth]{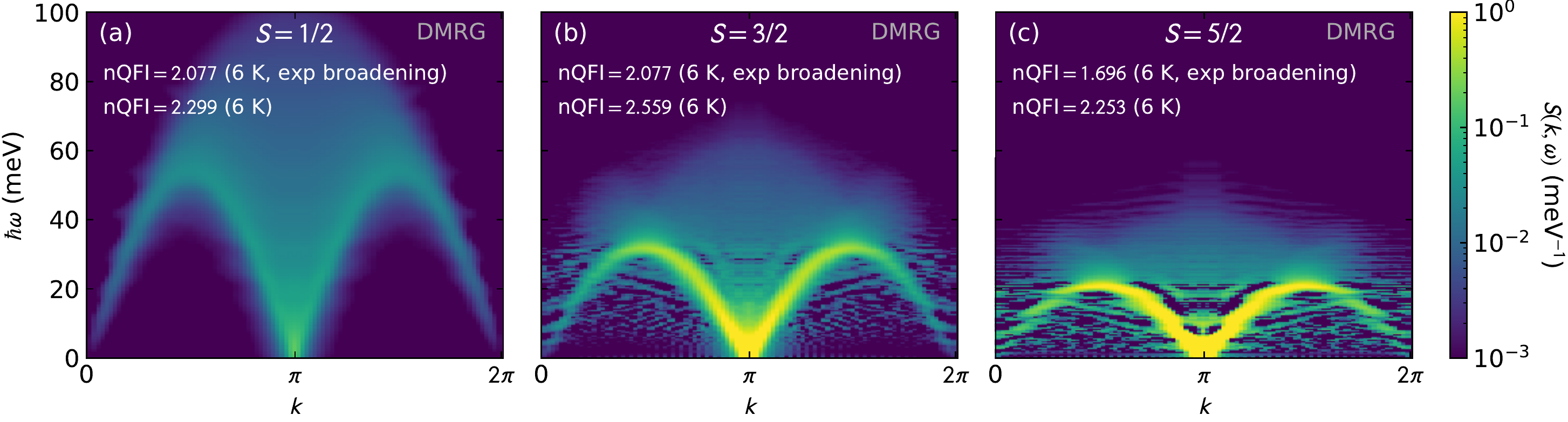}
	
	\caption{DMRG calculated $T=0$ spectrum for (a) $S=1/2$, (b) $S=3/2$, (c) and $S=5/2$, with normalized QFI ($\frac{f_\mathcal{Q}}{12 \cdot S^2}$) calculated at $k=\pi$ displayed on the figures.}
	\label{flo:DMRG_highspin}
\end{figure*}

\subsection{Higher half-integer spin Heisenberg antiferromagnets}\label{app:DMRG:higherS}

We also calculated the DMRG spectrum at $T=0$ for $S=3/2$ and $S=5/2$ HAF spin chains, as shown in Fig. \ref{flo:DMRG_highspin}. Similarly to the main $T=0$ $S=1/2$ calculation, these results were obtained with $J=1$, $\eta=0.1J$ and $N=100$ sites. In order to reduce computational and memory cost, a ground state in the $S^z=0$ sector was targeted. To avoid an unphysical artifact in the spectrum (a line of moderately intense scattering at $k=\pi$ extending to the highest frequencies, due to a combination of diverging intensity as $\omega\rightarrow0$ and the Lorentzian energy broadening) we removed a Lorentzian with height $S(k,0)$ and $\eta=0.1J$ at each $k$-point. This was necessary to avoid unphysical contributions to the QFI values. Following the DMRG computation, $J$ was scaled to keep the Curie-Weiss temperature $\Theta_{CW} \propto J S(S+1)$ constant across all spin values.

\section{Semiclassical approximation}\label{app:Semiclassical}
As the spin size increases, numerically computing the dynamical correlations with e.g. DMRG or Quantum Monte Carlo and therefore calculating the QFI becomes increasingly demanding --- especially at finite temperature. The spin-1 case has, however, been calculated by Lambert and Sorensen \cite{Lambert_2019} in a single-mode approximation. Their results are shown in Fig. \ref{flo:QFIb}, normalized to match the bound given by Eq. \eqref{eq:QFIbound2}. To understand the case of higher $S$ we turn to a recent semiclassical theory work.

Gozel, Mila, and Affleck (GMA) \cite{Gozel_2019} have considered the mapping of the large-spin Heisenberg chain to an $O(3)$ nonlinear $\sigma$-model, and constructed a perturbative spin-wave theory in $1/S$. Exploiting asymptotic freedom and rotational invariance, they obtain analytic expressions for the dynamical spin structure factor valid for distances shorter than $S^{-1}e^{\pi S}$ and energies greater than $JS^2e^{-\pi S}$. These distance and energy scales rapidly lengthen and decrease, respectively, with spin size, and GMA find that their theory is useful mainly to describe $S\geq 5/2$ HAF chains. The scales involved also mean that the semiclassical correlations will rapidly exhaust the experimentally relevant scales. Such spin-wave type excitations/correlations are consistent with inelastic neutron scattering studies of Heisenberg chains with $S=3/2$ \cite{ITOH1995} and $S=5/2$ \cite{Hutchings_1972}.

\begin{figure*}
	\centering\includegraphics[width=0.99\textwidth]{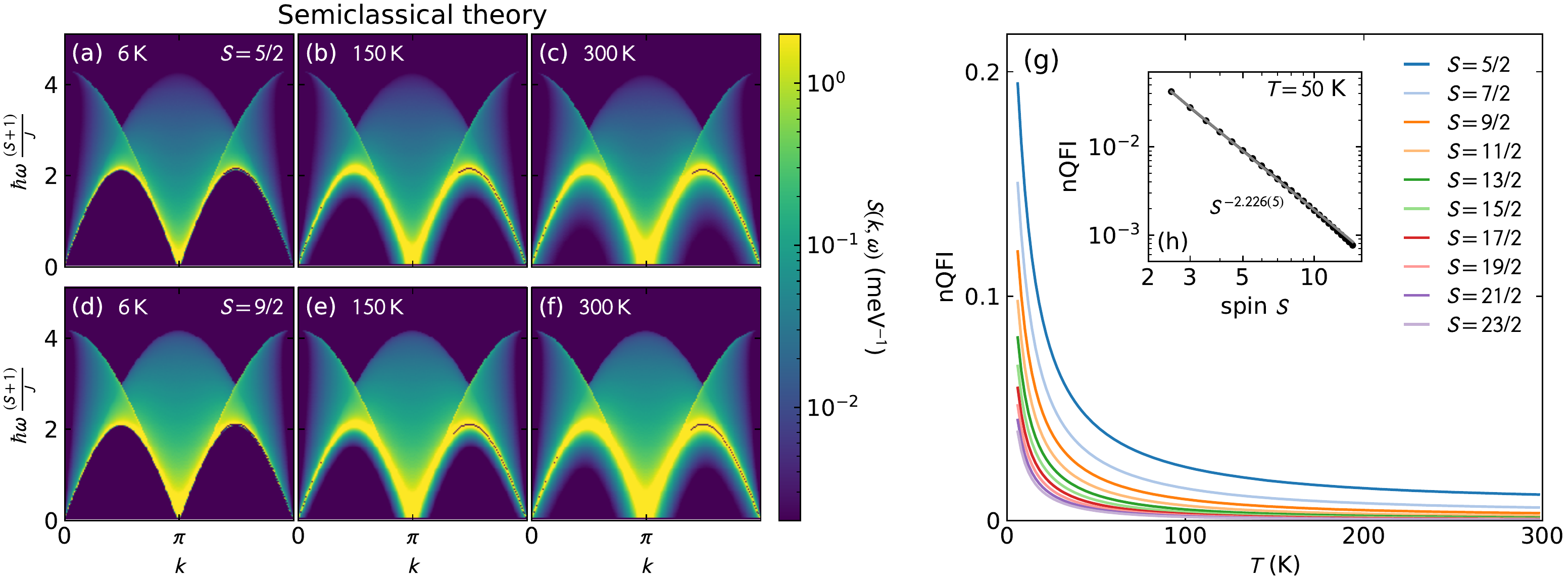}
	
	\caption{Semiclassical inelastic spectrum computed following ref. \cite{Gozel_2019}. (a)-(f) show scattering from $S=5/2$ and $S=9/2$ chains at several temperatures. (g) shows the temperature-dependent normalized QFI ($\frac{f_\mathcal{Q}}{12S^2}$) calculated at $k=\pi$ for $S \geq 5/2$. The inset (h) shows normalized QFI vs $S$ at 50~K, revealing a power-law decay.}
	\label{flo:Semiclassical}
\end{figure*}

We have calculated the QFI of large half-integer spin-$S$ HAF chains using the GMA theory \cite{Gozel_2019} to simulate the neutron spectrum. Several sample spectra are shown in Fig. \ref{flo:Semiclassical}. Since this approach is valid at energies above $\Lambda = J S^2 e^{-\pi S}$, we used $\Lambda$ as a cutoff to define the lower bound of the QFI integral. Figure \ref{flo:Semiclassical}(g) shows the decay of calculated $f_\mathcal{Q}/\left(12S^2\right)$ for $S \geq 5/2$. As spin increases, this quantity falls off with a power-law [Fig. \ref{flo:Semiclassical}(h)].

\section{Classical limit}\label{app:ClassicalLimit}

\begin{figure*}
	\centering\includegraphics[width=\textwidth]{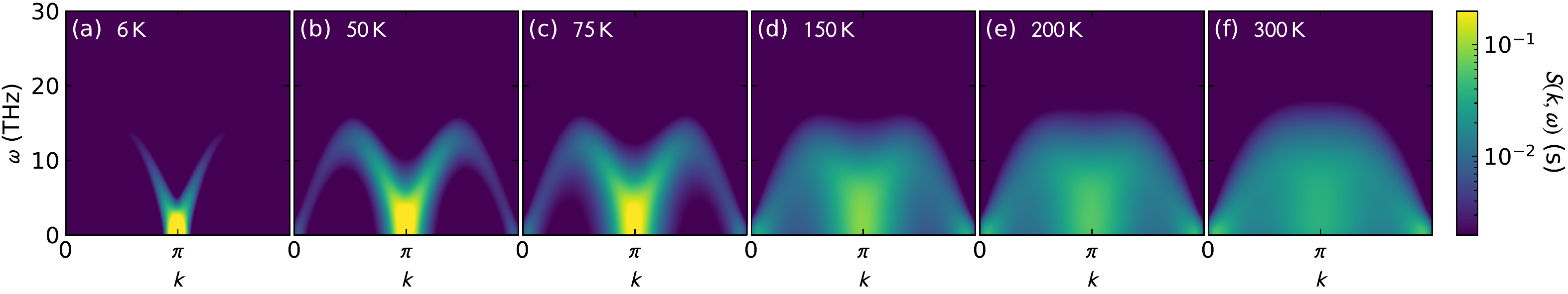}
	
	\caption{Landau-Lifshitz dynamics simulated spectrum for the classical $S\rightarrow \infty$ limit, with experimental resolution broadening applied. Because $\hbar \rightarrow 0$, the $y$ axis is in units of frequency. (For comparison to Fig. \ref{flo:QFIa} where $\hbar \neq 0$, 1~THz $\approx 4.1$~meV). Despite the finite spectral weight at non-zero frequency, nQFI vanishes at all temperatures when the classical limit is taken.}
	\label{flo:LLD_spectra}
\end{figure*}

We used Landau-Lifshitz dynamics to calculate spectra also for a fully classical ($S\rightarrow \infty$) spin system. In this method, spins are evolved by the classical equation of motion,
\begin{align}
    \frac{\mathrm{d}\mathbf{S}_i}{\mathrm{d}t} &= \mathbf{S}_i \times \mathbf{B}_i,
\end{align}
where $\mathbf{B}_i$ is the effective local magnetic field. Each $S(k,\omega)$ is calculated by Fourier-transforming real-space correlations into momentum space, and averaging over 192 independent simulation runs. In order to match experimental conditions, LLD spectra were convolved with a resolution function defined by the MAPS spectrometer. Fig.~\ref{flo:LLD_spectra} shows the resulting spectra. We stress that the spectra obtained with the classical simulation are in frequency-space --- \emph{i.e. not in energy-space}. To compare calculated $S(k,\omega)$ spectra with experiment it is common to introduce the semi-classical approximation $\epsilon^\mathrm{SCl} (k) = \hbar \omega^\mathrm{Cl} (k),$ with $\hbar$ finite, and where the superscripts $^\mathrm{Cl}$ and $^\mathrm{SCl}$ denote ``classical'' and ``semi-classical'', respectively. However, it is important to note that while this scaling by $\hbar$ will introduce apparent scattering at finite energy, it cannot by itself induce entanglement. Thus care needs to be taken to correctly take the classical limit when evaluating the QFI integral, Eq.~\eqref{eq:QFI}, from semiclassical simulations.

Taking the classical limit of a quantum spin system is, in general, a subtle problem. Here we will thus specialize to the HAF chain, for which we can make some precise statements. As $S\rightarrow \infty$, linear spin-wave theory (LSWT) becomes exact. At $T=0$ the system is in a classical N\'eel state, and the spectrum predicted by LSWT consists of a single sharp magnon mode dispersing as $\omega^\mathrm{Cl}(k)\propto \left| \sin (k)\right|$. This collapse of the continuous spectrum for $S=1/2$ to a discrete branch as $S\rightarrow \infty$ can be understood as a consequence of sum rules \cite{PhysRevB.26.1311}. It follows that the QFI density, \eqref{eq:QFI}, evaluated at $k=\pi$, vanishes in the classical limit at $T=0$. Resolution and thermal effects (at finite temperature) may broaden the sharp mode and induce scattering at finite frequency $\omega^\mathrm{Cl}$, as seen in Fig.~\ref{flo:LLD_spectra}. However, when we take $\hbar=0$ as is appropriate for a classical system (see below), the QFI density again vanishes. Hence, QFI correctly does not witness entanglement in the classical HAF chain.

%\textcolor{red}{
To formalize this statement at finite temperature, we take the classical limit of the HAF following the approach of Harris et al. \cite{PhysRevB.3.961}. We let $\hbar\rightarrow0$, $J\rightarrow0$, $S\rightarrow\infty$, while $\hbar S=\frac{1}{2}N_0$ and the characteristic temperature scale $k_B T_0 = 2JS^2$ remain finite. Here $N_0=1$ for spin-$1/2$. Using that $\tanh \left( x\right) \leq x$ in Eqs.~\eqref{eq:QFI}, \eqref{eq:QFIbound2} we obtain the inequality
\begin{align}
    \mathrm{nQFI}   &\leq \frac{\hbar}{6\pi S^2} \int_{0}^{\hbar \omega_\mathrm{max}} \mathrm{d} \left( \hbar \omega \right) \left( \frac{\hbar \omega}{k_B T}\right) \chi\prime\prime \left( \hbar\omega, T\right),    \label{eq:app:QFIbound}
\end{align}
where we have introduced an explicit cutoff frequency, $\omega_\mathrm{max}$, corresponding to the highest frequency in the spectrum of $\chi\prime\prime$. In the classical limit, the spectrum consists of a single magnon branch with dispersion relation $\omega^\mathrm{Cl}\left(Q\right)=2J^\mathrm{Cl}S^\mathrm{Cl}\left|\sin\left(k\right)\right|,$ and the highest frequency is the zone boundary frequency $\omega_\mathrm{ZB}^{Cl}=2 J^\mathrm{Cl} S^\mathrm{Cl},$ which may be large. However, the energy in any mode is $\epsilon\left( k\right) = \hbar \omega^\mathrm{Cl}\left( k\right) \leq \hbar \omega_\mathrm{ZB}^\mathrm{Cl}=2JS = k_B T_0/S$, which vanishes as $S\rightarrow \infty$. It is thus enough to note that the integral in \eqref{eq:app:QFIbound} must vanish since it is taken over an interval that vanishes in the classical limit. In Appendix~\ref{app:Semiclassical} we provide additional evidence that $\mathrm{nQFI}\rightarrow 0$ as the classical limit is approached.
%}

% In PRB, the bibliography goes after the appendices.


% KCuF3 quantum entanglement paper 

%%TC:ignore
%%%%%%%%%%%%%%%%%%%%%%%%%%%%%% Latex preamble

%%TC:endignore	

	%%TC:ignore
	
\onecolumngrid
\newpage

	\section*{\large Supplemental Material: Reproducing the DMRG results.}
	
\setcounter{page}{1}
\renewcommand{\thepage}{S\arabic{page}}  
	Here we provide detailed instructions on how to reproduce the DMRG results used in the main text. The results reported in this work were obtained with DMRG++ versions 5.67, 5.69, 5.70 and PsimagLite versions 2.67, 2.69, 2.70.
	
	The \textsc{DMRG}++ computer program \cite{Alvarez2009} can be obtained with:
	\begin{verbatim}
	git clone https://github.com/g1257/dmrgpp.git
	\end{verbatim}
	Dependencies include the BOOST and HDF5 libraries, and PsimagLite. The latter can be obtained with:
	\begin{verbatim}
	git clone https://github.com/g1257/PsimagLite.git
	\end{verbatim}
	To compile:
	\begin{verbatim}
	cd PsimagLite/lib; perl configure.pl; make
	cd ../../dmrgpp/src; perl configure.pl; make
	\end{verbatim}
	To simplify commands below we also run
	\begin{verbatim}
	export PATH="<PATH-TO-DMRG++>/src:$PATH"
	export SCRIPTS="<PATH-TO-DMRG++>/scripts"
	\end{verbatim}
	
	The documentation can be found at 
	\nolinkurl{https://g1257.github.io/dmrgPlusPlus/manual.html} or can be obtained
	by doing \verb!cd dmrgpp/doc; make manual.pdf!.
	
	\subsection{Obtaining zero-temperature spectra}
	The $T=0$ results can be reproduced as follows. First DMRG++ is run with an input file, \verb!dmrg -f inputGS.ain!, to obtain the ground state, where \texttt{inputGS.ain} has the form
	\begin{verbatim}
	##Ainur1.0
	TotalNumberOfSites=100;
	NumberOfTerms=2;
	
	### 1/2(S^+S^- + S^-S^+) part
	gt0:DegreesOfFreedom=1;
	gt0:GeometryKind="chain";
	gt0:GeometryOptions="ConstantValues";
	gt0:dir0:Connectors=[1.0];
	
	### S^zS^z part
	gt1:DegreesOfFreedom=1;
	gt1:GeometryKind="chain";
	gt1:GeometryOptions="ConstantValues";
	gt1:dir0:Connectors=[1.0];
	
	Model="Heisenberg";
	integer HeisenbergTwiceS=5;
	
	SolverOptions="twositedmrg,calcAndPrintEntropies";
	InfiniteLoopKeptStates=1000;
	FiniteLoops=[
	[ 49, 1000, 8],
	[-98, 1000, 8],
	[ 49, 1000, 8],
	[ 49, 1000, 2],
	[-98, 1000, 3]];
	
	# Keep a maximum of 1000 states, but allow
	# truncation with tolerance and minimum states
	string TruncationTolerance="1e-10,100";
	
	# Tolerance for Lanczos
	real LanczosEps=1e-10;
	int LanczosSteps=600;
	
	Threads=4;
	integer TargetSzPlusConst=250;
	\end{verbatim}
	Here we showed the input for $S=5/2$. The parameter \texttt{TargetSzPlusConst} should equal $S^z+S\cdot N$, where $S^z$ is the targeted $S^z$ sector and $N$ is the system size. The line may be left out for the $S=1/2$ case.
	
	The next step is to calculate dynamics, using the saved ground state as an input. It is convenient to do the dynamics run in a subdirectory \texttt{Szz}, so \verb!cp inputGS.ain Szz/inputSzz.ado! and add/modify the following lines in \texttt{inputSzz.ado},
	\begin{verbatim}
	SolverOptions="twositedmrg,restart,minimizeDisk,CorrectionVectorTargeting";
	
	### The finite loops pick up where gs run ended! I.e. the edge.
	FiniteLoops=[
	[98, 1000, 2],
	[-98, 1000, 2]];
	
	# RestartFilename is the name of the GS .hd5 file (extension is not needed)
	string RestartFilename="../inputGS";
	
	# The weight of the g.s. in the density matrix
	GsWeight=0.1;
	
	# Legacy thing, set to 0
	real CorrectionA=0;
	
	# Fermion spectra has sign changes in denominator. For boson operators (as in here) set it to 0
	integer DynamicDmrgType=0;
	
	# The site(s) where to apply the operator below. Here it is the center site.
	TSPSites=[50];
	
	# The delay in loop units before applying the operator. Set to 0 for all restarts to avoid delays.
	TSPLoops=[0];
	
	# If more than one operator is to be applied, how they should be combined.
	# Irrelevant if only one operator is applied, as is the case here.
	TSPProductOrSum="sum";
	
	# How the operator to be applied will be specified
	TSPOperator="expression";
	
	# The operator expression
	OperatorExpression="sz";
	
	# Apply operator to ground state
	string TSPApplyTo="|X0>";
	
	# How is the freq. given in the denominator (Matsubara is the other option)
	string CorrectionVectorFreqType="Real";
	
	# This is a dollarized input, so the omega will change from input to input.
	real CorrectionVectorOmega=$omega;
	
	# The broadening for the spectrum in omega + i*eta
	real CorrectionVectorEta=0.10;
	
	# The algorithm
	string CorrectionVectorAlgorithm="Krylov";
	
	#The labels below are ONLY read by manyOmegas.pl script
	
	# How many inputs files to create
	#OmegaTotal=601
	
	# Which one is the first omega value
	#OmegaBegin=0.0
	
	# Which is the "step" in omega
	#OmegaStep=0.025
	
	# Because the script will also be creating the batches, indicate what to measure in the batches
	#Observable=sz
	\end{verbatim}
	Then all individual inputs (one per $\omega$ in the correction vector approach) can be generated and submitted using the \texttt{manyOmegas.pl} script:
	\begin{verbatim}
	perl -I ${SCRIPTS} ${SCRIPTS}/manyOmegas.pl inputSzz.ado BatchTemplate <test/submit>.
	\end{verbatim}
	It is recommended to run with \texttt{test} first to verify correctness, before running with \texttt{submit}. Depending on the machine and scheduler, the BatchTemplate can be e.g. a PBS script.  The key is that it contains a line
	\begin{verbatim}
	dmrg -f $$input "<X0|$$obs|P1>,<X0|$$obs|P2>,<X0|$$obs|P3>" -p 10
	\end{verbatim}
	which allows \texttt{manyOmegas.pl} to fill in the appropriate input for each generated job batch. After all outputs have been generated,
	\begin{verbatim}
	perl -I ${SCRIPTS} ${SCRIPTS}/procOmegas.pl -f inputSzz.ado -p
	perl ${SCRIPTS}/pgfplot.pl
	\end{verbatim}
	can be used to process and plot the results.
	
	\subsection{Obtaining finite-temperature spectra}
	First note that the $T>0$ calculation discussed in the main text is very time consuming. It proceeds in three steps: First the $T=\infty$ state is found using a fictitious ``entangler'' Hamiltonian $H_\mathrm{E}$ acting in an enlarged Hilbert space \cite{PhysRevB.72.220401, PhysRevB.81.075108, PhysRevB.93.045137}. Second, the physical system is cooled through evolving in imaginary time with the physical Hamiltonian $H$ acting only on physical sites, i.e. we evolve with $H\otimes I$, where $I$ is the identity operator in the ancilla space. Third, dynamics is calculated using the operator $H\otimes I + I \otimes (-H)$.
	
	In the first step, we use a conventional (grand canonical) entangler, such that the enlarged system (physical and ancilla sites) can be described as a spin ladder, with physical sites on one leg (with even sites $0,2,4,\dots$) and ancilla sites on the other (with odd sites $1,3,5,\dots$). The entangler Hamiltonian is chosen such that its ground state corresponds to the $T=\infty$ state of the physical system. We find the ground state of $H_E$ by running \verb!dmrg -f Entangler.ain!, where \texttt{Entangler.ain} is given as
	\begin{verbatim}
	##Ainur1.0
	TotalNumberOfSites=100;
	NumberOfTerms=2;
	
	gt0:DegreesOfFreedom=1;
	gt0:GeometryKind="ladder";
	gt0:LadderLeg=2;
	gt0:GeometryOptions="ConstantValues";
	gt0:dir0:Connectors=[0.0];
	gt0:dir1:Connectors=[-10.0];
	integer gt0:IsPeriodicX=0;
	
	gt1:DegreesOfFreedom=1;
	gt1:GeometryKind="chain";
	gt1:GeometryOptions="ConstantValues";
	gt1:dir0:Connectors=[0];
	integer gt1:IsPeriodicX=0;
	
	Model="Heisenberg";
	integer HeisenbergTwiceS=1;
	
	SolverOptions="twositedmrg,MatrixVectorOnTheFly";
	InfiniteLoopKeptStates=1000;
	FiniteLoops=[
	[ 49, 1000, 0],
	[-98, 1000, 0],
	[ 98, 1000, 0],
	[-98, 1000, 0]];
	
	# Keep a maximum of 1000 states, but allow truncation with tolerance and minimum states as below
	string TruncationTolerance="1e-8,100";
	
	# Tolerance for Lanczos
	real LanczosEps=1e-8;
	int LanczosSteps=250;
	\end{verbatim}
	
	Next the imaginary time evolution is initiated with \verb!dmrg -f Evolution1.ain!, where \texttt{Evolution1.ain} adds and modifies some line compared to \texttt{Entangler.ain}. For brevity we only reproduce the modified lines below.
	\begin{verbatim}
	gt0:GeometryOptions="none";
	gt0:dir0:Connectors=[1.0, 0.0, 1.0, 0.0, 1.0, 0.0, 1.0, 0.0, ...];
	gt0:dir1:Connectors=[0.0, 0.0, 0.0, 0.0, ...];
	
	gt1:GeometryKind="ladder";
	gt1:LadderLeg=2;
	gt1:GeometryOptions="none";
	gt1:dir0:Connectors=[1.0, 0.0, 1.0, 0.0, 1.0, 0.0, 1.0, 0.0, ...];
	gt1:dir1:Connectors=[0.0, 0.0, 0.0, 0.0, ...];
	
	string PrintHamiltonianAverage="s==c";
	string RecoverySave="@M=100,@keep,1==1";
	SolverOptions="twositedmrg,restart,TargetingAncilla";
	FiniteLoops=[
	[ 98, 1000, 2],
	[-98, 1000, 2]];
	RepeatFiniteLoopsTimes=21;
	
	string RestartFilename="Entangler";
	
	TSPTau=0.1;
	TSPTimeSteps=5;
	TSPAdvanceEach=98;
	TSPAlgorithm="Krylov";
	TSPSites=[50];
	TSPLoops=[0];
	TSPProductOrSum="sum";
	GsWeight=0.1;
	
	TSPOperator="expression";
	OperatorExpression="identity";
	\end{verbatim}
	Above we have abbreviated the lines describing the connectors using $\dots$. The \texttt{gt?:dir1:Connectors} describe couplings across rungs, and should all be zero since we only act on physical sites. There are $N/2=50$ such rungs, and a value for each needs to be listed in the array in the input. The \texttt{gt?:dir0:Connectors} describe couplings along legs, starting at site $j$ corresponding to the $j$th position in the array (indexed from zero). To only couple physical sites every other leg coupling is set to zero. For open boundary conditions there are $N-2$ such bonds that need to be included in the array in the input.
	
	As further described in the input, we use Krylov imaginary time evolution with a time step $\tau=0.1$. The time evolution is done with an evolution operator $\exp \left[ -\frac{\beta H\tau}{2}\right]$, so $\tau$ is given in units of $\beta'=\beta/2$ for $J=1$. The imaginary time $\beta'$ can be obtained with \verb!h5dump -d /Def/FinalPsi/TimeSerializer/Time <hd5>!, where \texttt{<hd5>} should be replaced with the name of the \texttt{hd5} file of interest. To arrive at imaginary times corresponding to experimental temperatures we do additional restarts from appropriate \texttt{hd5} files output by the main temperature evolution loop, while tuning the $\tau$ value. Explicitly, the targeted $\beta'$ value can be found as $\beta' (J,T)=T/(2J)$, where $J$ and $T$ are given in Kelvin. Finally, the arguments to \texttt{RecoverySave} mean that we keep a maximum of $100$ \texttt{hd5} outputs, and output one in every loop (when the condition $1==1$ holds).
	
	Finally, the dynamics calculation proceeds similarly to the $T=0$ case, but with number of sites and precision as in the preceding $T>0$ step. We do, however, need to additionally add/modify the following lines 
	\begin{verbatim}
	string GeometrySubKind="GrandCanonical";
	
	gt0:dir0:Connectors=[1.0, -1.0, 1.0, -1.0, 1.0, -1.0, 1.0, -1.0, ...];
	gt0:dir1:Connectors=[0.0, 0.0, 0.0, 0.0, ...];
	
	gt1:dir0:Connectors=[1.0, -1.0, 1.0, -1.0, 1.0, -1.0, 1.0, -1.0, ...];
	gt1:dir1:Connectors=[0.0, 0.0, 0.0, 0.0, ...];
	
	string RestartFilename="../Evolution_150K";
	
	SolverOptions="CorrectionVectorTargeting,restart,twositedmrg,minimizeDisk,fixLegacyBugs";
	
	integer RestartSourceTvForPsi=0;
	vector RestartMappingTvs=[-1, -1, -1, -1];
	integer RestartMapStages=0;
	integer TridiagSteps=400;
	real TridiagEps=1e-9;
	\end{verbatim}
	The restart filename should be chosen to match the \texttt{hd5} file of interest. Note here that we calculate dynamics with $H\otimes I + I \otimes (-H)$, where $H\otimes I$ acts only on physical sites and $I\otimes (-H)$ acts only on ancilla sites. All rung couplings are zero. As before, we have abbreviated the arrays of coupling constants.


\begin{thebibliography}{103}%
	\makeatletter
	\providecommand \@ifxundefined [1]{%
		\@ifx{#1\undefined}
	}%
	\providecommand \@ifnum [1]{%
		\ifnum #1\expandafter \@firstoftwo
		\else \expandafter \@secondoftwo
		\fi
	}%
	\providecommand \@ifx [1]{%
		\ifx #1\expandafter \@firstoftwo
		\else \expandafter \@secondoftwo
		\fi
	}%
	\providecommand \natexlab [1]{#1}%
	\providecommand \enquote  [1]{``#1''}%
	\providecommand \bibnamefont  [1]{#1}%
	\providecommand \bibfnamefont [1]{#1}%
	\providecommand \citenamefont [1]{#1}%
	\providecommand \href@noop [0]{\@secondoftwo}%
	\providecommand \href [0]{\begingroup \@sanitize@url \@href}%
	\providecommand \@href[1]{\@@startlink{#1}\@@href}%
	\providecommand \@@href[1]{\endgroup#1\@@endlink}%
	\providecommand \@sanitize@url [0]{\catcode `\\12\catcode `\$12\catcode
		`\&12\catcode `\#12\catcode `\^12\catcode `\_12\catcode `\%12\relax}%
	\providecommand \@@startlink[1]{}%
	\providecommand \@@endlink[0]{}%
	\providecommand \url  [0]{\begingroup\@sanitize@url \@url }%
	\providecommand \@url [1]{\endgroup\@href {#1}{\urlprefix }}%
	\providecommand \urlprefix  [0]{URL }%
	\providecommand \Eprint [0]{\href }%
	\providecommand \doibase [0]{http://dx.doi.org/}%
	\providecommand \selectlanguage [0]{\@gobble}%
	\providecommand \bibinfo  [0]{\@secondoftwo}%
	\providecommand \bibfield  [0]{\@secondoftwo}%
	\providecommand \translation [1]{[#1]}%
	\providecommand \BibitemOpen [0]{}%
	\providecommand \bibitemStop [0]{}%
	\providecommand \bibitemNoStop [0]{.\EOS\space}%
	\providecommand \EOS [0]{\spacefactor3000\relax}%
	\providecommand \BibitemShut  [1]{\csname bibitem#1\endcsname}%
	\let\auto@bib@innerbib\@empty
	%</preamble>
	\bibitem [{\citenamefont {Bell}(1964)}]{PhysicsPhysiqueFizika.1.195}%
	\BibitemOpen
	\bibfield  {author} {\bibinfo {author} {\bibfnamefont {J.~S.}\ \bibnamefont
			{Bell}},\ }\bibfield  {title} {\enquote {\bibinfo {title} {On the {Einstein
					Podolsky Rosen} paradox},}\ }\href {\doibase
		10.1103/PhysicsPhysiqueFizika.1.195} {\bibfield  {journal} {\bibinfo
			{journal} {Physics Physique Fizika}\ }\textbf {\bibinfo {volume} {1}},\
		\bibinfo {pages} {195--200} (\bibinfo {year} {1964})}\BibitemShut {NoStop}%
	\bibitem [{\citenamefont {Clauser}\ \emph {et~al.}(1969)\citenamefont
		{Clauser}, \citenamefont {Horne}, \citenamefont {Shimony},\ and\
		\citenamefont {Holt}}]{PhysRevLett.23.880}%
	\BibitemOpen
	\bibfield  {author} {\bibinfo {author} {\bibfnamefont {John~F.}\ \bibnamefont
			{Clauser}}, \bibinfo {author} {\bibfnamefont {Michael~A.}\ \bibnamefont
			{Horne}}, \bibinfo {author} {\bibfnamefont {Abner}\ \bibnamefont {Shimony}},
		\ and\ \bibinfo {author} {\bibfnamefont {Richard~A.}\ \bibnamefont {Holt}},\
	}\bibfield  {title} {\enquote {\bibinfo {title} {Proposed experiment to test
				local hidden-variable theories},}\ }\href {\doibase
		10.1103/PhysRevLett.23.880} {\bibfield  {journal} {\bibinfo  {journal} {Phys.
				Rev. Lett.}\ }\textbf {\bibinfo {volume} {23}},\ \bibinfo {pages} {880--884}
		(\bibinfo {year} {1969})}\BibitemShut {NoStop}%
	\bibitem [{\citenamefont {Aspect}(1999)}]{Aspect_1999}%
	\BibitemOpen
	\bibfield  {author} {\bibinfo {author} {\bibfnamefont {Alain}\ \bibnamefont
			{Aspect}},\ }\bibfield  {title} {\enquote {\bibinfo {title} {Bell's
				inequality test: more ideal than ever},}\ }\href {\doibase 10.1038/18296}
	{\bibfield  {journal} {\bibinfo  {journal} {Nature (London)}\ }\textbf
		{\bibinfo {volume} {398}},\ \bibinfo {pages} {189--190} (\bibinfo {year}
		{1999})}\BibitemShut {NoStop}%
	\bibitem [{\citenamefont {Zeilinger}(1999)}]{RevModPhys.71.S288}%
	\BibitemOpen
	\bibfield  {author} {\bibinfo {author} {\bibfnamefont {Anton}\ \bibnamefont
			{Zeilinger}},\ }\bibfield  {title} {\enquote {\bibinfo {title} {Experiment
				and the foundations of quantum physics},}\ }\href {\doibase
		10.1103/RevModPhys.71.S288} {\bibfield  {journal} {\bibinfo  {journal} {Rev.
				Mod. Phys.}\ }\textbf {\bibinfo {volume} {71}},\ \bibinfo {pages}
		{S288--S297} (\bibinfo {year} {1999})}\BibitemShut {NoStop}%
	\bibitem [{\citenamefont {Horodecki}\ \emph {et~al.}(2009)\citenamefont
		{Horodecki}, \citenamefont {Horodecki}, \citenamefont {Horodecki},\ and\
		\citenamefont {Horodecki}}]{RevModPhys.81.865}%
	\BibitemOpen
	\bibfield  {author} {\bibinfo {author} {\bibfnamefont {Ryszard}\ \bibnamefont
			{Horodecki}}, \bibinfo {author} {\bibfnamefont {Pawe\l{}}\ \bibnamefont
			{Horodecki}}, \bibinfo {author} {\bibfnamefont {Micha\l{}}\ \bibnamefont
			{Horodecki}}, \ and\ \bibinfo {author} {\bibfnamefont {Karol}\ \bibnamefont
			{Horodecki}},\ }\bibfield  {title} {\enquote {\bibinfo {title} {Quantum
				entanglement},}\ }\href {\doibase 10.1103/RevModPhys.81.865} {\bibfield
		{journal} {\bibinfo  {journal} {Rev. Mod. Phys.}\ }\textbf {\bibinfo {volume}
			{81}},\ \bibinfo {pages} {865--942} (\bibinfo {year} {2009})}\BibitemShut
	{NoStop}%
	\bibitem [{\citenamefont {Schmied}\ \emph {et~al.}(2016)\citenamefont
		{Schmied}, \citenamefont {Bancal}, \citenamefont {Allard}, \citenamefont
		{Fadel}, \citenamefont {Scarani}, \citenamefont {Treutlein},\ and\
		\citenamefont {Sangouard}}]{Schmied16}%
	\BibitemOpen
	\bibfield  {author} {\bibinfo {author} {\bibfnamefont {Roman}\ \bibnamefont
			{Schmied}}, \bibinfo {author} {\bibfnamefont {Jean-Daniel}\ \bibnamefont
			{Bancal}}, \bibinfo {author} {\bibfnamefont {Baptiste}\ \bibnamefont
			{Allard}}, \bibinfo {author} {\bibfnamefont {Matteo}\ \bibnamefont {Fadel}},
		\bibinfo {author} {\bibfnamefont {Valerio}\ \bibnamefont {Scarani}}, \bibinfo
		{author} {\bibfnamefont {Philipp}\ \bibnamefont {Treutlein}}, \ and\ \bibinfo
		{author} {\bibfnamefont {Nicolas}\ \bibnamefont {Sangouard}},\ }\bibfield
	{title} {\enquote {\bibinfo {title} {Bell correlations in a {Bose-Einstein}
				condensate},}\ }\href {\doibase 10.1126/science.aad8665} {\bibfield
		{journal} {\bibinfo  {journal} {Science}\ }\textbf {\bibinfo {volume}
			{352}},\ \bibinfo {pages} {441--444} (\bibinfo {year} {2016})}\BibitemShut
	{NoStop}%
	\bibitem [{\citenamefont {Engelsen}\ \emph {et~al.}(2017)\citenamefont
		{Engelsen}, \citenamefont {Krishnakumar}, \citenamefont {Hosten},\ and\
		\citenamefont {Kasevich}}]{PhysRevLett.118.140401}%
	\BibitemOpen
	\bibfield  {author} {\bibinfo {author} {\bibfnamefont {Nils~J.}\ \bibnamefont
			{Engelsen}}, \bibinfo {author} {\bibfnamefont {Rajiv}\ \bibnamefont
			{Krishnakumar}}, \bibinfo {author} {\bibfnamefont {Onur}\ \bibnamefont
			{Hosten}}, \ and\ \bibinfo {author} {\bibfnamefont {Mark~A.}\ \bibnamefont
			{Kasevich}},\ }\bibfield  {title} {\enquote {\bibinfo {title} {Bell
				correlations in spin-squeezed states of 500 000 atoms},}\ }\href {\doibase
		10.1103/PhysRevLett.118.140401} {\bibfield  {journal} {\bibinfo  {journal}
			{Phys. Rev. Lett.}\ }\textbf {\bibinfo {volume} {118}},\ \bibinfo {pages}
		{140401} (\bibinfo {year} {2017})}\BibitemShut {NoStop}%
	\bibitem [{\citenamefont {Amico}\ \emph {et~al.}(2008)\citenamefont {Amico},
		\citenamefont {Fazio}, \citenamefont {Osterloh},\ and\ \citenamefont
		{Vedral}}]{RevModPhys.80.517}%
	\BibitemOpen
	\bibfield  {author} {\bibinfo {author} {\bibfnamefont {Luigi}\ \bibnamefont
			{Amico}}, \bibinfo {author} {\bibfnamefont {Rosario}\ \bibnamefont {Fazio}},
		\bibinfo {author} {\bibfnamefont {Andreas}\ \bibnamefont {Osterloh}}, \ and\
		\bibinfo {author} {\bibfnamefont {Vlatko}\ \bibnamefont {Vedral}},\
	}\bibfield  {title} {\enquote {\bibinfo {title} {Entanglement in many-body
				systems},}\ }\href {\doibase 10.1103/RevModPhys.80.517} {\bibfield  {journal}
		{\bibinfo  {journal} {Rev. Mod. Phys.}\ }\textbf {\bibinfo {volume} {80}},\
		\bibinfo {pages} {517--576} (\bibinfo {year} {2008})}\BibitemShut {NoStop}%
	\bibitem [{\citenamefont {Vedral}(2008)}]{Vedral2008}%
	\BibitemOpen
	\bibfield  {author} {\bibinfo {author} {\bibfnamefont {Vlatko}\ \bibnamefont
			{Vedral}},\ }\bibfield  {title} {\enquote {\bibinfo {title} {Quantifying
				entanglement in macroscopic systems},}\ }\href {\doibase 10.1038/nature07124}
	{\bibfield  {journal} {\bibinfo  {journal} {Nature (London)}\ }\textbf
		{\bibinfo {volume} {453}},\ \bibinfo {pages} {1004--1007} (\bibinfo {year}
		{2008})}\BibitemShut {NoStop}%
	\bibitem [{\citenamefont {Eisert}\ \emph {et~al.}(2010)\citenamefont {Eisert},
		\citenamefont {Cramer},\ and\ \citenamefont {Plenio}}]{RevModPhys.82.277}%
	\BibitemOpen
	\bibfield  {author} {\bibinfo {author} {\bibfnamefont {J.}~\bibnamefont
			{Eisert}}, \bibinfo {author} {\bibfnamefont {M.}~\bibnamefont {Cramer}}, \
		and\ \bibinfo {author} {\bibfnamefont {M.~B.}\ \bibnamefont {Plenio}},\
	}\bibfield  {title} {\enquote {\bibinfo {title} {Colloquium: Area laws for
				the entanglement entropy},}\ }\href {\doibase 10.1103/RevModPhys.82.277}
	{\bibfield  {journal} {\bibinfo  {journal} {Rev. Mod. Phys.}\ }\textbf
		{\bibinfo {volume} {82}},\ \bibinfo {pages} {277--306} (\bibinfo {year}
		{2010})}\BibitemShut {NoStop}%
	\bibitem [{\citenamefont {Laflorencie}(2016)}]{Laflorencie2016}%
	\BibitemOpen
	\bibfield  {author} {\bibinfo {author} {\bibfnamefont {Nicolas}\ \bibnamefont
			{Laflorencie}},\ }\bibfield  {title} {\enquote {\bibinfo {title} {Quantum
				entanglement in condensed matter systems},}\ }\href {\doibase
		https://doi.org/10.1016/j.physrep.2016.06.008} {\bibfield  {journal}
		{\bibinfo  {journal} {Phys. Rep.}\ }\textbf {\bibinfo {volume} {646}},\
		\bibinfo {pages} {1 -- 59} (\bibinfo {year} {2016})}\BibitemShut {NoStop}%
	\bibitem [{\citenamefont {Chiara}\ and\ \citenamefont
		{Sanpera}(2018)}]{Chiara2018}%
	\BibitemOpen
	\bibfield  {author} {\bibinfo {author} {\bibfnamefont {Gabriele~De}\
			\bibnamefont {Chiara}}\ and\ \bibinfo {author} {\bibfnamefont {Anna}\
			\bibnamefont {Sanpera}},\ }\bibfield  {title} {\enquote {\bibinfo {title}
			{Genuine quantum correlations in quantum many-body systems: a review of
				recent progress},}\ }\href {\doibase 10.1088/1361-6633/aabf61} {\bibfield
		{journal} {\bibinfo  {journal} {Rep. Prog. Phys.}\ }\textbf {\bibinfo
			{volume} {81}},\ \bibinfo {pages} {074002} (\bibinfo {year}
		{2018})}\BibitemShut {NoStop}%
	\bibitem [{\citenamefont {Friis}\ \emph {et~al.}(2019)\citenamefont {Friis},
		\citenamefont {Vitagliano}, \citenamefont {Malik},\ and\ \citenamefont
		{Huber}}]{Friis2019}%
	\BibitemOpen
	\bibfield  {author} {\bibinfo {author} {\bibfnamefont {Nicolai}\ \bibnamefont
			{Friis}}, \bibinfo {author} {\bibfnamefont {Giuseppe}\ \bibnamefont
			{Vitagliano}}, \bibinfo {author} {\bibfnamefont {Mehul}\ \bibnamefont
			{Malik}}, \ and\ \bibinfo {author} {\bibfnamefont {Marcus}\ \bibnamefont
			{Huber}},\ }\bibfield  {title} {\enquote {\bibinfo {title} {Entanglement
				certification from theory to experiment},}\ }\href {\doibase
		10.1038/s42254-018-0003-5} {\bibfield  {journal} {\bibinfo  {journal} {Nat.
				Rev. Phys.}\ }\textbf {\bibinfo {volume} {1}},\ \bibinfo {pages} {72--87}
		(\bibinfo {year} {2019})}\BibitemShut {NoStop}%
	\bibitem [{\citenamefont {Vidal}\ \emph {et~al.}(2003)\citenamefont {Vidal},
		\citenamefont {Latorre}, \citenamefont {Rico},\ and\ \citenamefont
		{Kitaev}}]{PhysRevLett.90.227902}%
	\BibitemOpen
	\bibfield  {author} {\bibinfo {author} {\bibfnamefont {G.}~\bibnamefont
			{Vidal}}, \bibinfo {author} {\bibfnamefont {J.~I.}\ \bibnamefont {Latorre}},
		\bibinfo {author} {\bibfnamefont {E.}~\bibnamefont {Rico}}, \ and\ \bibinfo
		{author} {\bibfnamefont {A.}~\bibnamefont {Kitaev}},\ }\bibfield  {title}
	{\enquote {\bibinfo {title} {Entanglement in quantum critical phenomena},}\
	}\href {\doibase 10.1103/PhysRevLett.90.227902} {\bibfield  {journal}
		{\bibinfo  {journal} {Phys. Rev. Lett.}\ }\textbf {\bibinfo {volume} {90}},\
		\bibinfo {pages} {227902} (\bibinfo {year} {2003})}\BibitemShut {NoStop}%
	\bibitem [{\citenamefont {Wen}(2017)}]{RevModPhys.89.041004}%
	\BibitemOpen
	\bibfield  {author} {\bibinfo {author} {\bibfnamefont {Xiao-Gang}\
			\bibnamefont {Wen}},\ }\bibfield  {title} {\enquote {\bibinfo {title}
			{Colloquium: Zoo of quantum-topological phases of matter},}\ }\href {\doibase
		10.1103/RevModPhys.89.041004} {\bibfield  {journal} {\bibinfo  {journal}
			{Rev. Mod. Phys.}\ }\textbf {\bibinfo {volume} {89}},\ \bibinfo {pages}
		{041004} (\bibinfo {year} {2017})}\BibitemShut {NoStop}%
	\bibitem [{\citenamefont {Abanin}\ \emph {et~al.}(2019)\citenamefont {Abanin},
		\citenamefont {Altman}, \citenamefont {Bloch},\ and\ \citenamefont
		{Serbyn}}]{RevModPhys.91.021001}%
	\BibitemOpen
	\bibfield  {author} {\bibinfo {author} {\bibfnamefont {Dmitry~A.}\
			\bibnamefont {Abanin}}, \bibinfo {author} {\bibfnamefont {Ehud}\ \bibnamefont
			{Altman}}, \bibinfo {author} {\bibfnamefont {Immanuel}\ \bibnamefont
			{Bloch}}, \ and\ \bibinfo {author} {\bibfnamefont {Maksym}\ \bibnamefont
			{Serbyn}},\ }\bibfield  {title} {\enquote {\bibinfo {title} {Colloquium:
				Many-body localization, thermalization, and entanglement},}\ }\href {\doibase
		10.1103/RevModPhys.91.021001} {\bibfield  {journal} {\bibinfo  {journal}
			{Rev. Mod. Phys.}\ }\textbf {\bibinfo {volume} {91}},\ \bibinfo {pages}
		{021001} (\bibinfo {year} {2019})}\BibitemShut {NoStop}%
	\bibitem [{\citenamefont {G\"uhne}\ and\ \citenamefont
		{T\'oth}(2009)}]{Guehne2009}%
	\BibitemOpen
	\bibfield  {author} {\bibinfo {author} {\bibfnamefont {Otfried}\ \bibnamefont
			{G\"uhne}}\ and\ \bibinfo {author} {\bibfnamefont {G\'eza}\ \bibnamefont
			{T\'oth}},\ }\bibfield  {title} {\enquote {\bibinfo {title} {Entanglement
				detection},}\ }\href {\doibase https://doi.org/10.1016/j.physrep.2009.02.004}
	{\bibfield  {journal} {\bibinfo  {journal} {Phys. Rep.}\ }\textbf {\bibinfo
			{volume} {474}},\ \bibinfo {pages} {1 -- 75} (\bibinfo {year}
		{2009})}\BibitemShut {NoStop}%
	\bibitem [{\citenamefont {Ghosh}\ \emph {et~al.}(2003)\citenamefont {Ghosh},
		\citenamefont {Rosenbaum}, \citenamefont {Aeppli},\ and\ \citenamefont
		{Coppersmith}}]{Ghosh2003}%
	\BibitemOpen
	\bibfield  {author} {\bibinfo {author} {\bibfnamefont {S.}~\bibnamefont
			{Ghosh}}, \bibinfo {author} {\bibfnamefont {T.~F.}\ \bibnamefont
			{Rosenbaum}}, \bibinfo {author} {\bibfnamefont {G.}~\bibnamefont {Aeppli}}, \
		and\ \bibinfo {author} {\bibfnamefont {S.~N.}\ \bibnamefont {Coppersmith}},\
	}\bibfield  {title} {\enquote {\bibinfo {title} {Entangled quantum state of
				magnetic dipoles},}\ }\href {\doibase 10.1038/nature01888} {\bibfield
		{journal} {\bibinfo  {journal} {Nature (London)}\ }\textbf {\bibinfo {volume}
			{425}},\ \bibinfo {pages} {48--51} (\bibinfo {year} {2003})}\BibitemShut
	{NoStop}%
	\bibitem [{\citenamefont {Islam}\ \emph {et~al.}(2015)\citenamefont {Islam},
		\citenamefont {Ma}, \citenamefont {Preiss}, \citenamefont {Tai},
		\citenamefont {Lukin}, \citenamefont {Rispoli},\ and\ \citenamefont
		{Greiner}}]{Islam2015}%
	\BibitemOpen
	\bibfield  {author} {\bibinfo {author} {\bibfnamefont {Rajibul}\ \bibnamefont
			{Islam}}, \bibinfo {author} {\bibfnamefont {Ruichao}\ \bibnamefont {Ma}},
		\bibinfo {author} {\bibfnamefont {Philipp~M.}\ \bibnamefont {Preiss}},
		\bibinfo {author} {\bibfnamefont {M.~Eric}\ \bibnamefont {Tai}}, \bibinfo
		{author} {\bibfnamefont {Alexander}\ \bibnamefont {Lukin}}, \bibinfo {author}
		{\bibfnamefont {Matthew}\ \bibnamefont {Rispoli}}, \ and\ \bibinfo {author}
		{\bibfnamefont {Markus}\ \bibnamefont {Greiner}},\ }\bibfield  {title}
	{\enquote {\bibinfo {title} {Measuring entanglement entropy in a quantum
				many-body system},}\ }\href {\doibase 10.1038/nature15750} {\bibfield
		{journal} {\bibinfo  {journal} {Nature (London)}\ }\textbf {\bibinfo {volume}
			{528}},\ \bibinfo {pages} {77--83} (\bibinfo {year} {2015})}\BibitemShut
	{NoStop}%
	\bibitem [{\citenamefont {Kaufman}\ \emph {et~al.}(2016)\citenamefont
		{Kaufman}, \citenamefont {Tai}, \citenamefont {Lukin}, \citenamefont
		{Rispoli}, \citenamefont {Schittko}, \citenamefont {Preiss},\ and\
		\citenamefont {Greiner}}]{Kaufman2016}%
	\BibitemOpen
	\bibfield  {author} {\bibinfo {author} {\bibfnamefont {Adam~M.}\ \bibnamefont
			{Kaufman}}, \bibinfo {author} {\bibfnamefont {M.~Eric}\ \bibnamefont {Tai}},
		\bibinfo {author} {\bibfnamefont {Alexander}\ \bibnamefont {Lukin}}, \bibinfo
		{author} {\bibfnamefont {Matthew}\ \bibnamefont {Rispoli}}, \bibinfo {author}
		{\bibfnamefont {Robert}\ \bibnamefont {Schittko}}, \bibinfo {author}
		{\bibfnamefont {Philipp~M.}\ \bibnamefont {Preiss}}, \ and\ \bibinfo {author}
		{\bibfnamefont {Markus}\ \bibnamefont {Greiner}},\ }\bibfield  {title}
	{\enquote {\bibinfo {title} {Quantum thermalization through entanglement in
				an isolated many-body system},}\ }\href {\doibase 10.1126/science.aaf6725}
	{\bibfield  {journal} {\bibinfo  {journal} {Science}\ }\textbf {\bibinfo
			{volume} {353}},\ \bibinfo {pages} {794--800} (\bibinfo {year}
		{2016})}\BibitemShut {NoStop}%
	\bibitem [{\citenamefont {Tennant}\ \emph {et~al.}(1993)\citenamefont
		{Tennant}, \citenamefont {Perring}, \citenamefont {Cowley},\ and\
		\citenamefont {Nagler}}]{Tennant1993}%
	\BibitemOpen
	\bibfield  {author} {\bibinfo {author} {\bibfnamefont {D.~A.}\ \bibnamefont
			{Tennant}}, \bibinfo {author} {\bibfnamefont {T.~G.}\ \bibnamefont
			{Perring}}, \bibinfo {author} {\bibfnamefont {R.~A.}\ \bibnamefont {Cowley}},
		\ and\ \bibinfo {author} {\bibfnamefont {S.~E.}\ \bibnamefont {Nagler}},\
	}\bibfield  {title} {\enquote {\bibinfo {title} {Unbound spinons in the
				{S}=1/2 antiferromagnetic chain {${\mathrm{KCuF}}_{3}$}},}\ }\href {\doibase
		10.1103/PhysRevLett.70.4003} {\bibfield  {journal} {\bibinfo  {journal}
			{Phys. Rev. Lett.}\ }\textbf {\bibinfo {volume} {70}},\ \bibinfo {pages}
		{4003--4006} (\bibinfo {year} {1993})}\BibitemShut {NoStop}%
	\bibitem [{\citenamefont {Brukner}\ \emph {et~al.}(2006)\citenamefont
		{Brukner}, \citenamefont {Vedral},\ and\ \citenamefont
		{Zeilinger}}]{PhysRevA.73.012110}%
	\BibitemOpen
	\bibfield  {author} {\bibinfo {author} {\bibfnamefont {{\v{C}}aslav}\
			\bibnamefont {Brukner}}, \bibinfo {author} {\bibfnamefont {Vlatko}\
			\bibnamefont {Vedral}}, \ and\ \bibinfo {author} {\bibfnamefont {Anton}\
			\bibnamefont {Zeilinger}},\ }\bibfield  {title} {\enquote {\bibinfo {title}
			{Crucial role of quantum entanglement in bulk properties of solids},}\ }\href
	{\doibase 10.1103/PhysRevA.73.012110} {\bibfield  {journal} {\bibinfo
			{journal} {Phys. Rev. A}\ }\textbf {\bibinfo {volume} {73}},\ \bibinfo
		{pages} {012110} (\bibinfo {year} {2006})}\BibitemShut {NoStop}%
	\bibitem [{\citenamefont {Christensen}\ \emph {et~al.}(2007)\citenamefont
		{Christensen}, \citenamefont {R{\o}nnow}, \citenamefont {McMorrow},
		\citenamefont {Harrison}, \citenamefont {Perring}, \citenamefont {Enderle},
		\citenamefont {Coldea}, \citenamefont {Regnault},\ and\ \citenamefont
		{Aeppli}}]{Christensen2007}%
	\BibitemOpen
	\bibfield  {author} {\bibinfo {author} {\bibfnamefont {N.~B.}\ \bibnamefont
			{Christensen}}, \bibinfo {author} {\bibfnamefont {H.~M.}\ \bibnamefont
			{R{\o}nnow}}, \bibinfo {author} {\bibfnamefont {D.~F.}\ \bibnamefont
			{McMorrow}}, \bibinfo {author} {\bibfnamefont {A.}~\bibnamefont {Harrison}},
		\bibinfo {author} {\bibfnamefont {T.~G.}\ \bibnamefont {Perring}}, \bibinfo
		{author} {\bibfnamefont {M.}~\bibnamefont {Enderle}}, \bibinfo {author}
		{\bibfnamefont {R.}~\bibnamefont {Coldea}}, \bibinfo {author} {\bibfnamefont
			{L.~P.}\ \bibnamefont {Regnault}}, \ and\ \bibinfo {author} {\bibfnamefont
			{G.}~\bibnamefont {Aeppli}},\ }\bibfield  {title} {\enquote {\bibinfo {title}
			{Quantum dynamics and entanglement of spins on a square lattice},}\ }\href
	{\doibase 10.1073/pnas.0703293104} {\bibfield  {journal} {\bibinfo  {journal}
			{Proc. Natl. Acad. Sci. USA}\ }\textbf {\bibinfo {volume} {104}},\ \bibinfo
		{pages} {15264--15269} (\bibinfo {year} {2007})}\BibitemShut {NoStop}%
	\bibitem [{\citenamefont {Mourigal}\ \emph {et~al.}(2013)\citenamefont
		{Mourigal}, \citenamefont {Enderle}, \citenamefont {Kl{\"o}pperpieper},
		\citenamefont {Caux}, \citenamefont {Stunault},\ and\ \citenamefont
		{R{\o}nnow}}]{mourigal2013fractional}%
	\BibitemOpen
	\bibfield  {author} {\bibinfo {author} {\bibfnamefont {Martin}\ \bibnamefont
			{Mourigal}}, \bibinfo {author} {\bibfnamefont {Mechthild}\ \bibnamefont
			{Enderle}}, \bibinfo {author} {\bibfnamefont {Axel}\ \bibnamefont
			{Kl{\"o}pperpieper}}, \bibinfo {author} {\bibfnamefont {Jean-S{\'e}bastien}\
			\bibnamefont {Caux}}, \bibinfo {author} {\bibfnamefont {Anne}\ \bibnamefont
			{Stunault}}, \ and\ \bibinfo {author} {\bibfnamefont {Henrik~M}\ \bibnamefont
			{R{\o}nnow}},\ }\bibfield  {title} {\enquote {\bibinfo {title} {Fractional
				spinon excitations in the quantum {Heisenberg} antiferromagnetic chain},}\
	}\href {https://doi.org/10.1038/nphys2652} {\bibfield  {journal} {\bibinfo
			{journal} {Nat. Phys.}\ }\textbf {\bibinfo {volume} {9}},\ \bibinfo {pages}
		{435--441} (\bibinfo {year} {2013})}\BibitemShut {NoStop}%
	\bibitem [{\citenamefont {Piazza}\ \emph {et~al.}(2015)\citenamefont {Piazza},
		\citenamefont {Mourigal}, \citenamefont {Christensen}, \citenamefont
		{Nilsen}, \citenamefont {Tregenna-Piggott}, \citenamefont {Perring},
		\citenamefont {Enderle}, \citenamefont {McMorrow}, \citenamefont {Ivanov},\
		and\ \citenamefont {R{\o}nnow}}]{Piazza2015}%
	\BibitemOpen
	\bibfield  {author} {\bibinfo {author} {\bibfnamefont {B.~Dalla}\
			\bibnamefont {Piazza}}, \bibinfo {author} {\bibfnamefont {M.}~\bibnamefont
			{Mourigal}}, \bibinfo {author} {\bibfnamefont {N.~B.}\ \bibnamefont
			{Christensen}}, \bibinfo {author} {\bibfnamefont {G.~J.}\ \bibnamefont
			{Nilsen}}, \bibinfo {author} {\bibfnamefont {P.}~\bibnamefont
			{Tregenna-Piggott}}, \bibinfo {author} {\bibfnamefont {T.~G.}\ \bibnamefont
			{Perring}}, \bibinfo {author} {\bibfnamefont {M.}~\bibnamefont {Enderle}},
		\bibinfo {author} {\bibfnamefont {D.~F.}\ \bibnamefont {McMorrow}}, \bibinfo
		{author} {\bibfnamefont {D.~A.}\ \bibnamefont {Ivanov}}, \ and\ \bibinfo
		{author} {\bibfnamefont {H.~M.}\ \bibnamefont {R{\o}nnow}},\ }\bibfield
	{title} {\enquote {\bibinfo {title} {Fractional excitations in the
				square-lattice quantum antiferromagnet},}\ }\href
	{https://doi.org/10.1038/nphys3172} {\bibfield  {journal} {\bibinfo
			{journal} {Nat. Phys.}\ }\textbf {\bibinfo {volume} {11}},\ \bibinfo {pages}
		{62--68} (\bibinfo {year} {2015})}\BibitemShut {NoStop}%
	\bibitem [{\citenamefont {Li}\ and\ \citenamefont
		{Haldane}(2008)}]{PhysRevLett.101.010504}%
	\BibitemOpen
	\bibfield  {author} {\bibinfo {author} {\bibfnamefont {Hui}\ \bibnamefont
			{Li}}\ and\ \bibinfo {author} {\bibfnamefont {F.~D.~M.}\ \bibnamefont
			{Haldane}},\ }\bibfield  {title} {\enquote {\bibinfo {title} {Entanglement
				spectrum as a generalization of entanglement entropy: Identification of
				topological order in non-{Abelian} fractional quantum {Hall} effect
				states},}\ }\href {\doibase 10.1103/PhysRevLett.101.010504} {\bibfield
		{journal} {\bibinfo  {journal} {Phys. Rev. Lett.}\ }\textbf {\bibinfo
			{volume} {101}},\ \bibinfo {pages} {010504} (\bibinfo {year}
		{2008})}\BibitemShut {NoStop}%
	\bibitem [{\citenamefont {Kitaev}\ and\ \citenamefont
		{Preskill}(2006)}]{PhysRevLett.96.110404}%
	\BibitemOpen
	\bibfield  {author} {\bibinfo {author} {\bibfnamefont {Alexei}\ \bibnamefont
			{Kitaev}}\ and\ \bibinfo {author} {\bibfnamefont {John}\ \bibnamefont
			{Preskill}},\ }\bibfield  {title} {\enquote {\bibinfo {title} {Topological
				entanglement entropy},}\ }\href {\doibase 10.1103/PhysRevLett.96.110404}
	{\bibfield  {journal} {\bibinfo  {journal} {Phys. Rev. Lett.}\ }\textbf
		{\bibinfo {volume} {96}},\ \bibinfo {pages} {110404} (\bibinfo {year}
		{2006})}\BibitemShut {NoStop}%
	\bibitem [{\citenamefont {Levin}\ and\ \citenamefont
		{Wen}(2006)}]{PhysRevLett.96.110405}%
	\BibitemOpen
	\bibfield  {author} {\bibinfo {author} {\bibfnamefont {Michael}\ \bibnamefont
			{Levin}}\ and\ \bibinfo {author} {\bibfnamefont {Xiao-Gang}\ \bibnamefont
			{Wen}},\ }\bibfield  {title} {\enquote {\bibinfo {title} {Detecting
				topological order in a ground state wave function},}\ }\href {\doibase
		10.1103/PhysRevLett.96.110405} {\bibfield  {journal} {\bibinfo  {journal}
			{Phys. Rev. Lett.}\ }\textbf {\bibinfo {volume} {96}},\ \bibinfo {pages}
		{110405} (\bibinfo {year} {2006})}\BibitemShut {NoStop}%
	\bibitem [{\citenamefont {Jiang}\ \emph {et~al.}(2012)\citenamefont {Jiang},
		\citenamefont {Wang},\ and\ \citenamefont {Balents}}]{Jiang2012}%
	\BibitemOpen
	\bibfield  {author} {\bibinfo {author} {\bibfnamefont {Hong-Chen}\
			\bibnamefont {Jiang}}, \bibinfo {author} {\bibfnamefont {Zhenghan}\
			\bibnamefont {Wang}}, \ and\ \bibinfo {author} {\bibfnamefont {Leon}\
			\bibnamefont {Balents}},\ }\bibfield  {title} {\enquote {\bibinfo {title}
			{Identifying topological order by entanglement entropy},}\ }\href {\doibase
		10.1038/nphys2465} {\bibfield  {journal} {\bibinfo  {journal} {Nat. Phys.}\
		}\textbf {\bibinfo {volume} {8}},\ \bibinfo {pages} {902--905} (\bibinfo
		{year} {2012})}\BibitemShut {NoStop}%
	\bibitem [{\citenamefont {Thomale}\ \emph {et~al.}(2010)\citenamefont
		{Thomale}, \citenamefont {Arovas},\ and\ \citenamefont
		{Bernevig}}]{PhysRevLett.105.116805}%
	\BibitemOpen
	\bibfield  {author} {\bibinfo {author} {\bibfnamefont {Ronny}\ \bibnamefont
			{Thomale}}, \bibinfo {author} {\bibfnamefont {D.~P.}\ \bibnamefont {Arovas}},
		\ and\ \bibinfo {author} {\bibfnamefont {B.~Andrei}\ \bibnamefont
			{Bernevig}},\ }\bibfield  {title} {\enquote {\bibinfo {title} {Nonlocal order
				in gapless systems: Entanglement spectrum in spin chains},}\ }\href {\doibase
		10.1103/PhysRevLett.105.116805} {\bibfield  {journal} {\bibinfo  {journal}
			{Phys. Rev. Lett.}\ }\textbf {\bibinfo {volume} {105}},\ \bibinfo {pages}
		{116805} (\bibinfo {year} {2010})}\BibitemShut {NoStop}%
	\bibitem [{\citenamefont {Chandran}\ \emph {et~al.}(2014)\citenamefont
		{Chandran}, \citenamefont {Khemani},\ and\ \citenamefont
		{Sondhi}}]{PhysRevLett.113.060501}%
	\BibitemOpen
	\bibfield  {author} {\bibinfo {author} {\bibfnamefont {Anushya}\ \bibnamefont
			{Chandran}}, \bibinfo {author} {\bibfnamefont {Vedika}\ \bibnamefont
			{Khemani}}, \ and\ \bibinfo {author} {\bibfnamefont {S.~L.}\ \bibnamefont
			{Sondhi}},\ }\bibfield  {title} {\enquote {\bibinfo {title} {How universal is
				the entanglement spectrum?}}\ }\href {\doibase
		10.1103/PhysRevLett.113.060501} {\bibfield  {journal} {\bibinfo  {journal}
			{Phys. Rev. Lett.}\ }\textbf {\bibinfo {volume} {113}},\ \bibinfo {pages}
		{060501} (\bibinfo {year} {2014})}\BibitemShut {NoStop}%
	\bibitem [{\citenamefont {Lundgren}\ \emph {et~al.}(2014)\citenamefont
		{Lundgren}, \citenamefont {Blair}, \citenamefont {Greiter}, \citenamefont
		{L\"auchli}, \citenamefont {Fiete},\ and\ \citenamefont
		{Thomale}}]{PhysRevLett.113.256404}%
	\BibitemOpen
	\bibfield  {author} {\bibinfo {author} {\bibfnamefont {Rex}\ \bibnamefont
			{Lundgren}}, \bibinfo {author} {\bibfnamefont {Jonathan}\ \bibnamefont
			{Blair}}, \bibinfo {author} {\bibfnamefont {Martin}\ \bibnamefont {Greiter}},
		\bibinfo {author} {\bibfnamefont {Andreas}\ \bibnamefont {L\"auchli}},
		\bibinfo {author} {\bibfnamefont {Gregory~A.}\ \bibnamefont {Fiete}}, \ and\
		\bibinfo {author} {\bibfnamefont {Ronny}\ \bibnamefont {Thomale}},\
	}\bibfield  {title} {\enquote {\bibinfo {title} {Momentum-space entanglement
				spectrum of bosons and fermions with interactions},}\ }\href {\doibase
		10.1103/PhysRevLett.113.256404} {\bibfield  {journal} {\bibinfo  {journal}
			{Phys. Rev. Lett.}\ }\textbf {\bibinfo {volume} {113}},\ \bibinfo {pages}
		{256404} (\bibinfo {year} {2014})}\BibitemShut {NoStop}%
	\bibitem [{\citenamefont {Lundgren}\ \emph {et~al.}(2016)\citenamefont
		{Lundgren}, \citenamefont {Blair}, \citenamefont {Laurell}, \citenamefont
		{Regnault}, \citenamefont {Fiete}, \citenamefont {Greiter},\ and\
		\citenamefont {Thomale}}]{PhysRevB.94.081112}%
	\BibitemOpen
	\bibfield  {author} {\bibinfo {author} {\bibfnamefont {Rex}\ \bibnamefont
			{Lundgren}}, \bibinfo {author} {\bibfnamefont {Jonathan}\ \bibnamefont
			{Blair}}, \bibinfo {author} {\bibfnamefont {Pontus}\ \bibnamefont {Laurell}},
		\bibinfo {author} {\bibfnamefont {Nicolas}\ \bibnamefont {Regnault}},
		\bibinfo {author} {\bibfnamefont {Gregory~A.}\ \bibnamefont {Fiete}},
		\bibinfo {author} {\bibfnamefont {Martin}\ \bibnamefont {Greiter}}, \ and\
		\bibinfo {author} {\bibfnamefont {Ronny}\ \bibnamefont {Thomale}},\
	}\bibfield  {title} {\enquote {\bibinfo {title} {Universal entanglement
				spectra in critical spin chains},}\ }\href {\doibase
		10.1103/PhysRevB.94.081112} {\bibfield  {journal} {\bibinfo  {journal} {Phys.
				Rev. B}\ }\textbf {\bibinfo {volume} {94}},\ \bibinfo {pages} {081112}
		(\bibinfo {year} {2016})}\BibitemShut {NoStop}%
	\bibitem [{\citenamefont {Pitsios}\ \emph {et~al.}(2017)\citenamefont
		{Pitsios}, \citenamefont {Banchi}, \citenamefont {Rab}, \citenamefont
		{Bentivegna}, \citenamefont {Caprara}, \citenamefont {Crespi}, \citenamefont
		{Spagnolo}, \citenamefont {Bose}, \citenamefont {Mataloni}, \citenamefont
		{Osellame},\ and\ \citenamefont {Sciarrino}}]{Pitsios2017}%
	\BibitemOpen
	\bibfield  {author} {\bibinfo {author} {\bibfnamefont {Ioannis}\ \bibnamefont
			{Pitsios}}, \bibinfo {author} {\bibfnamefont {Leonardo}\ \bibnamefont
			{Banchi}}, \bibinfo {author} {\bibfnamefont {Adil~S.}\ \bibnamefont {Rab}},
		\bibinfo {author} {\bibfnamefont {Marco}\ \bibnamefont {Bentivegna}},
		\bibinfo {author} {\bibfnamefont {Debora}\ \bibnamefont {Caprara}}, \bibinfo
		{author} {\bibfnamefont {Andrea}\ \bibnamefont {Crespi}}, \bibinfo {author}
		{\bibfnamefont {Nicol\`o}\ \bibnamefont {Spagnolo}}, \bibinfo {author}
		{\bibfnamefont {Sougato}\ \bibnamefont {Bose}}, \bibinfo {author}
		{\bibfnamefont {Paolo}\ \bibnamefont {Mataloni}}, \bibinfo {author}
		{\bibfnamefont {Roberto}\ \bibnamefont {Osellame}}, \ and\ \bibinfo {author}
		{\bibfnamefont {Fabio}\ \bibnamefont {Sciarrino}},\ }\bibfield  {title}
	{\enquote {\bibinfo {title} {Photonic simulation of entanglement growth and
				engineering after a spin chain quench},}\ }\href {\doibase
		10.1038/s41467-017-01589-y} {\bibfield  {journal} {\bibinfo  {journal} {Nat.
				Commun.}\ }\textbf {\bibinfo {volume} {8}},\ \bibinfo {pages} {1569}
		(\bibinfo {year} {2017})}\BibitemShut {NoStop}%
	\bibitem [{\citenamefont {Coffman}\ \emph {et~al.}(2000)\citenamefont
		{Coffman}, \citenamefont {Kundu},\ and\ \citenamefont
		{Wootters}}]{PhysRevA.61.052306}%
	\BibitemOpen
	\bibfield  {author} {\bibinfo {author} {\bibfnamefont {Valerie}\ \bibnamefont
			{Coffman}}, \bibinfo {author} {\bibfnamefont {Joydip}\ \bibnamefont {Kundu}},
		\ and\ \bibinfo {author} {\bibfnamefont {William~K.}\ \bibnamefont
			{Wootters}},\ }\bibfield  {title} {\enquote {\bibinfo {title} {Distributed
				entanglement},}\ }\href {\doibase 10.1103/PhysRevA.61.052306} {\bibfield
		{journal} {\bibinfo  {journal} {Phys. Rev. A}\ }\textbf {\bibinfo {volume}
			{61}},\ \bibinfo {pages} {052306} (\bibinfo {year} {2000})}\BibitemShut
	{NoStop}%
	\bibitem [{\citenamefont {Amico}\ \emph {et~al.}(2004)\citenamefont {Amico},
		\citenamefont {Osterloh}, \citenamefont {Plastina}, \citenamefont {Fazio},\
		and\ \citenamefont {Massimo~Palma}}]{PhysRevA.69.022304}%
	\BibitemOpen
	\bibfield  {author} {\bibinfo {author} {\bibfnamefont {Luigi}\ \bibnamefont
			{Amico}}, \bibinfo {author} {\bibfnamefont {Andreas}\ \bibnamefont
			{Osterloh}}, \bibinfo {author} {\bibfnamefont {Francesco}\ \bibnamefont
			{Plastina}}, \bibinfo {author} {\bibfnamefont {Rosario}\ \bibnamefont
			{Fazio}}, \ and\ \bibinfo {author} {\bibfnamefont {G.}~\bibnamefont
			{Massimo~Palma}},\ }\bibfield  {title} {\enquote {\bibinfo {title} {Dynamics
				of entanglement in one-dimensional spin systems},}\ }\href {\doibase
		10.1103/PhysRevA.69.022304} {\bibfield  {journal} {\bibinfo  {journal} {Phys.
				Rev. A}\ }\textbf {\bibinfo {volume} {69}},\ \bibinfo {pages} {022304}
		(\bibinfo {year} {2004})}\BibitemShut {NoStop}%
	\bibitem [{\citenamefont {Roscilde}\ \emph {et~al.}(2004)\citenamefont
		{Roscilde}, \citenamefont {Verrucchi}, \citenamefont {Fubini}, \citenamefont
		{Haas},\ and\ \citenamefont {Tognetti}}]{Roscilde_2004}%
	\BibitemOpen
	\bibfield  {author} {\bibinfo {author} {\bibfnamefont {Tommaso}\ \bibnamefont
			{Roscilde}}, \bibinfo {author} {\bibfnamefont {Paola}\ \bibnamefont
			{Verrucchi}}, \bibinfo {author} {\bibfnamefont {Andrea}\ \bibnamefont
			{Fubini}}, \bibinfo {author} {\bibfnamefont {Stephan}\ \bibnamefont {Haas}},
		\ and\ \bibinfo {author} {\bibfnamefont {Valerio}\ \bibnamefont {Tognetti}},\
	}\bibfield  {title} {\enquote {\bibinfo {title} {Studying quantum spin
				systems through entanglement estimators},}\ }\href {\doibase
		10.1103/PhysRevLett.93.167203} {\bibfield  {journal} {\bibinfo  {journal}
			{Phys. Rev. Lett.}\ }\textbf {\bibinfo {volume} {93}},\ \bibinfo {pages}
		{167203} (\bibinfo {year} {2004})}\BibitemShut {NoStop}%
	\bibitem [{\citenamefont {Baroni}\ \emph {et~al.}(2007)\citenamefont {Baroni},
		\citenamefont {Fubini}, \citenamefont {Tognetti},\ and\ \citenamefont
		{Verrucchi}}]{Baroni_2007}%
	\BibitemOpen
	\bibfield  {author} {\bibinfo {author} {\bibfnamefont {Fabrizio}\
			\bibnamefont {Baroni}}, \bibinfo {author} {\bibfnamefont {Andrea}\
			\bibnamefont {Fubini}}, \bibinfo {author} {\bibfnamefont {Valerio}\
			\bibnamefont {Tognetti}}, \ and\ \bibinfo {author} {\bibfnamefont {Paola}\
			\bibnamefont {Verrucchi}},\ }\bibfield  {title} {\enquote {\bibinfo {title}
			{Two-spin entanglement distribution near factorized states},}\ }\href
	{\doibase 10.1088/1751-8113/40/32/010} {\bibfield  {journal} {\bibinfo
			{journal} {J. Phys. A}\ }\textbf {\bibinfo {volume} {40}},\ \bibinfo {pages}
		{9845--9857} (\bibinfo {year} {2007})}\BibitemShut {NoStop}%
	\bibitem [{\citenamefont {Amico}\ \emph {et~al.}(2006)\citenamefont {Amico},
		\citenamefont {Baroni}, \citenamefont {Fubini}, \citenamefont {Patan\`e},
		\citenamefont {Tognetti},\ and\ \citenamefont {Verrucchi}}]{Amico_2006}%
	\BibitemOpen
	\bibfield  {author} {\bibinfo {author} {\bibfnamefont {L.}~\bibnamefont
			{Amico}}, \bibinfo {author} {\bibfnamefont {F.}~\bibnamefont {Baroni}},
		\bibinfo {author} {\bibfnamefont {A.}~\bibnamefont {Fubini}}, \bibinfo
		{author} {\bibfnamefont {D.}~\bibnamefont {Patan\`e}}, \bibinfo {author}
		{\bibfnamefont {V.}~\bibnamefont {Tognetti}}, \ and\ \bibinfo {author}
		{\bibfnamefont {Paola}\ \bibnamefont {Verrucchi}},\ }\bibfield  {title}
	{\enquote {\bibinfo {title} {Divergence of the entanglement range in
				low-dimensional quantum systems},}\ }\href {\doibase
		10.1103/PhysRevA.74.022322} {\bibfield  {journal} {\bibinfo  {journal} {Phys.
				Rev. A}\ }\textbf {\bibinfo {volume} {74}},\ \bibinfo {pages} {022322}
		(\bibinfo {year} {2006})}\BibitemShut {NoStop}%
	\bibitem [{\citenamefont {Hauke}\ \emph {et~al.}(2016)\citenamefont {Hauke},
		\citenamefont {Heyl}, \citenamefont {Tagliacozzo},\ and\ \citenamefont
		{Zoller}}]{Hauke2016}%
	\BibitemOpen
	\bibfield  {author} {\bibinfo {author} {\bibfnamefont {Philipp}\ \bibnamefont
			{Hauke}}, \bibinfo {author} {\bibfnamefont {Markus}\ \bibnamefont {Heyl}},
		\bibinfo {author} {\bibfnamefont {Luca}\ \bibnamefont {Tagliacozzo}}, \ and\
		\bibinfo {author} {\bibfnamefont {Peter}\ \bibnamefont {Zoller}},\ }\bibfield
	{title} {\enquote {\bibinfo {title} {Measuring multipartite entanglement
				through dynamic susceptibilities},}\ }\href {\doibase 10.1038/nphys3700}
	{\bibfield  {journal} {\bibinfo  {journal} {Nat. Phys.}\ }\textbf {\bibinfo
			{volume} {12}},\ \bibinfo {pages} {778--782} (\bibinfo {year}
		{2016})}\BibitemShut {NoStop}%
	\bibitem [{\citenamefont {Hyllus}\ \emph {et~al.}(2012)\citenamefont {Hyllus},
		\citenamefont {Laskowski}, \citenamefont {Krischek}, \citenamefont
		{Schwemmer}, \citenamefont {Wieczorek}, \citenamefont {Weinfurter},
		\citenamefont {Pezz\'e},\ and\ \citenamefont {Smerzi}}]{PhysRevA.85.022321}%
	\BibitemOpen
	\bibfield  {author} {\bibinfo {author} {\bibfnamefont {Philipp}\ \bibnamefont
			{Hyllus}}, \bibinfo {author} {\bibfnamefont {Wies\l{}aw}\ \bibnamefont
			{Laskowski}}, \bibinfo {author} {\bibfnamefont {Roland}\ \bibnamefont
			{Krischek}}, \bibinfo {author} {\bibfnamefont {Christian}\ \bibnamefont
			{Schwemmer}}, \bibinfo {author} {\bibfnamefont {Witlef}\ \bibnamefont
			{Wieczorek}}, \bibinfo {author} {\bibfnamefont {Harald}\ \bibnamefont
			{Weinfurter}}, \bibinfo {author} {\bibfnamefont {Luca}\ \bibnamefont
			{Pezz\'e}}, \ and\ \bibinfo {author} {\bibfnamefont {Augusto}\ \bibnamefont
			{Smerzi}},\ }\bibfield  {title} {\enquote {\bibinfo {title} {Fisher
				information and multiparticle entanglement},}\ }\href {\doibase
		10.1103/PhysRevA.85.022321} {\bibfield  {journal} {\bibinfo  {journal} {Phys.
				Rev. A}\ }\textbf {\bibinfo {volume} {85}},\ \bibinfo {pages} {022321}
		(\bibinfo {year} {2012})}\BibitemShut {NoStop}%
	\bibitem [{\citenamefont {Laurell}\ \emph {et~al.}(2020)\citenamefont
		{Laurell}, \citenamefont {Scheie}, \citenamefont {Mukherjee}, \citenamefont
		{Koza}, \citenamefont {Enderle}, \citenamefont {Tylczynski}, \citenamefont
		{Okamoto}, \citenamefont {Coldea}, \citenamefont {Tennant},\ and\
		\citenamefont {Alvarez}}]{Laurell2020}%
	\BibitemOpen
	\bibfield  {author} {\bibinfo {author} {\bibfnamefont {Pontus}\ \bibnamefont
			{Laurell}}, \bibinfo {author} {\bibfnamefont {Allen}\ \bibnamefont {Scheie}},
		\bibinfo {author} {\bibfnamefont {Chiron~J.}\ \bibnamefont {Mukherjee}},
		\bibinfo {author} {\bibfnamefont {Michael~M.}\ \bibnamefont {Koza}}, \bibinfo
		{author} {\bibfnamefont {Mechtild}\ \bibnamefont {Enderle}}, \bibinfo
		{author} {\bibfnamefont {Zbigniew}\ \bibnamefont {Tylczynski}}, \bibinfo
		{author} {\bibfnamefont {Satoshi}\ \bibnamefont {Okamoto}}, \bibinfo {author}
		{\bibfnamefont {Radu}\ \bibnamefont {Coldea}}, \bibinfo {author}
		{\bibfnamefont {D.~Alan}\ \bibnamefont {Tennant}}, \ and\ \bibinfo {author}
		{\bibfnamefont {Gonzalo}\ \bibnamefont {Alvarez}},\ }\href
	{https://arxiv.org/abs/2010.11164} {\enquote {\bibinfo {title} {Dynamics,
				entanglement, and the classical point in the transverse-field {XXZ} chain
				{Cs}$_2${CoCl}$_4$},}\ } (\bibinfo {year} {2020}),\ \bibinfo {note}
	{arXiv:2010.11164}\BibitemShut {NoStop}%
	\bibitem [{\citenamefont {Stone}\ \emph {et~al.}(2007)\citenamefont {Stone},
		\citenamefont {Tian}, \citenamefont {Lumsden}, \citenamefont {Granroth},
		\citenamefont {Mandrus}, \citenamefont {Chung}, \citenamefont {Harrison},\
		and\ \citenamefont {Nagler}}]{StonePhysRevLett.99.087204}%
	\BibitemOpen
	\bibfield  {author} {\bibinfo {author} {\bibfnamefont {M.~B.}\ \bibnamefont
			{Stone}}, \bibinfo {author} {\bibfnamefont {W.}~\bibnamefont {Tian}},
		\bibinfo {author} {\bibfnamefont {M.~D.}\ \bibnamefont {Lumsden}}, \bibinfo
		{author} {\bibfnamefont {G.~E.}\ \bibnamefont {Granroth}}, \bibinfo {author}
		{\bibfnamefont {D.}~\bibnamefont {Mandrus}}, \bibinfo {author} {\bibfnamefont
			{J.-H.}\ \bibnamefont {Chung}}, \bibinfo {author} {\bibfnamefont
			{N.}~\bibnamefont {Harrison}}, \ and\ \bibinfo {author} {\bibfnamefont
			{S.~E.}\ \bibnamefont {Nagler}},\ }\bibfield  {title} {\enquote {\bibinfo
			{title} {Quantum spin correlations in an organometallic alternating-sign
				chain},}\ }\href {\doibase 10.1103/PhysRevLett.99.087204} {\bibfield
		{journal} {\bibinfo  {journal} {Phys. Rev. Lett.}\ }\textbf {\bibinfo
			{volume} {99}},\ \bibinfo {pages} {087204} (\bibinfo {year}
		{2007})}\BibitemShut {NoStop}%
	\bibitem [{\citenamefont {Garlatti}\ \emph {et~al.}(2017)\citenamefont
		{Garlatti}, \citenamefont {Guidi}, \citenamefont {Ansbro}, \citenamefont
		{Santini}, \citenamefont {Amoretti}, \citenamefont {Ollivier}, \citenamefont
		{Mutka}, \citenamefont {Timco}, \citenamefont {Vitorica-Yrezabal},
		\citenamefont {Whitehead}, \citenamefont {Winpenny},\ and\ \citenamefont
		{Carretta}}]{Garlatti2017}%
	\BibitemOpen
	\bibfield  {author} {\bibinfo {author} {\bibfnamefont {E.}~\bibnamefont
			{Garlatti}}, \bibinfo {author} {\bibfnamefont {T.}~\bibnamefont {Guidi}},
		\bibinfo {author} {\bibfnamefont {S.}~\bibnamefont {Ansbro}}, \bibinfo
		{author} {\bibfnamefont {P.}~\bibnamefont {Santini}}, \bibinfo {author}
		{\bibfnamefont {G.}~\bibnamefont {Amoretti}}, \bibinfo {author}
		{\bibfnamefont {J.}~\bibnamefont {Ollivier}}, \bibinfo {author}
		{\bibfnamefont {H.}~\bibnamefont {Mutka}}, \bibinfo {author} {\bibfnamefont
			{G.}~\bibnamefont {Timco}}, \bibinfo {author} {\bibfnamefont {I.~J.}\
			\bibnamefont {Vitorica-Yrezabal}}, \bibinfo {author} {\bibfnamefont
			{G.~F.~S.}\ \bibnamefont {Whitehead}}, \bibinfo {author} {\bibfnamefont
			{R.~E.~P.}\ \bibnamefont {Winpenny}}, \ and\ \bibinfo {author} {\bibfnamefont
			{S.}~\bibnamefont {Carretta}},\ }\bibfield  {title} {\enquote {\bibinfo
			{title} {Portraying entanglement between molecular qubits with
				four-dimensional inelastic neutron scattering},}\ }\href
	{https://doi.org/10.1038/ncomms14543} {\bibfield  {journal} {\bibinfo
			{journal} {Nat. Commun.}\ }\textbf {\bibinfo {volume} {8}},\ \bibinfo {pages}
		{14543} (\bibinfo {year} {2017})}\BibitemShut {NoStop}%
	\bibitem [{\citenamefont {Mathew}\ \emph {et~al.}(2020)\citenamefont {Mathew},
		\citenamefont {Silva}, \citenamefont {Jain}, \citenamefont {Mohan},
		\citenamefont {Adroja}, \citenamefont {Sakai}, \citenamefont {Tomy},
		\citenamefont {Banerjee}, \citenamefont {Goreti}, \citenamefont {N.},
		\citenamefont {Singh},\ and\ \citenamefont
		{Jaiswal-Nagar}}]{PhysRevResearch.2.043329}%
	\BibitemOpen
	\bibfield  {author} {\bibinfo {author} {\bibfnamefont {George}\ \bibnamefont
			{Mathew}}, \bibinfo {author} {\bibfnamefont {Saulo L.~L.}\ \bibnamefont
			{Silva}}, \bibinfo {author} {\bibfnamefont {Anil}\ \bibnamefont {Jain}},
		\bibinfo {author} {\bibfnamefont {Arya}\ \bibnamefont {Mohan}}, \bibinfo
		{author} {\bibfnamefont {D.~T.}\ \bibnamefont {Adroja}}, \bibinfo {author}
		{\bibfnamefont {V.~G.}\ \bibnamefont {Sakai}}, \bibinfo {author}
		{\bibfnamefont {C.~V.}\ \bibnamefont {Tomy}}, \bibinfo {author}
		{\bibfnamefont {Alok}\ \bibnamefont {Banerjee}}, \bibinfo {author}
		{\bibfnamefont {Rajendar}\ \bibnamefont {Goreti}}, \bibinfo {author}
		{\bibfnamefont {Aswathi~V.}\ \bibnamefont {N.}}, \bibinfo {author}
		{\bibfnamefont {Ranjit}\ \bibnamefont {Singh}}, \ and\ \bibinfo {author}
		{\bibfnamefont {D.}~\bibnamefont {Jaiswal-Nagar}},\ }\bibfield  {title}
	{\enquote {\bibinfo {title} {Experimental realization of multipartite
				entanglement via quantum {Fisher} information in a uniform antiferromagnetic
				quantum spin chain},}\ }\href {\doibase 10.1103/PhysRevResearch.2.043329}
	{\bibfield  {journal} {\bibinfo  {journal} {Phys. Rev. Research}\ }\textbf
		{\bibinfo {volume} {2}},\ \bibinfo {pages} {043329} (\bibinfo {year}
		{2020})}\BibitemShut {NoStop}%
	\bibitem [{\citenamefont {Wie{\'{s}}niak}\ \emph {et~al.}(2005)\citenamefont
		{Wie{\'{s}}niak}, \citenamefont {Vedral},\ and\ \citenamefont
		{Brukner}}]{Wiesniak_2005}%
	\BibitemOpen
	\bibfield  {author} {\bibinfo {author} {\bibfnamefont {Marcin}\ \bibnamefont
			{Wie{\'{s}}niak}}, \bibinfo {author} {\bibfnamefont {Vlatko}\ \bibnamefont
			{Vedral}}, \ and\ \bibinfo {author} {\bibfnamefont {{\v{C}}aslav}\
			\bibnamefont {Brukner}},\ }\bibfield  {title} {\enquote {\bibinfo {title}
			{Magnetic susceptibility as a macroscopic entanglement witness},}\ }\href
	{\doibase 10.1088/1367-2630/7/1/258} {\bibfield  {journal} {\bibinfo
			{journal} {New J. Phys.}\ }\textbf {\bibinfo {volume} {7}},\ \bibinfo {pages}
		{258--258} (\bibinfo {year} {2005})}\BibitemShut {NoStop}%
	\bibitem [{\citenamefont {Wie\ifmmode~\acute{s}\else \'{s}\fi{}niak}\ \emph
		{et~al.}(2008)\citenamefont {Wie\ifmmode~\acute{s}\else \'{s}\fi{}niak},
		\citenamefont {Vedral},\ and\ \citenamefont {Brukner}}]{PhysRevB.78.064108}%
	\BibitemOpen
	\bibfield  {author} {\bibinfo {author} {\bibfnamefont {Marcin}\ \bibnamefont
			{Wie\ifmmode~\acute{s}\else \'{s}\fi{}niak}}, \bibinfo {author}
		{\bibfnamefont {Vlatko}\ \bibnamefont {Vedral}}, \ and\ \bibinfo {author}
		{\bibfnamefont {{\v{C}}aslav}\ \bibnamefont {Brukner}},\ }\bibfield  {title}
	{\enquote {\bibinfo {title} {Heat capacity as an indicator of
				entanglement},}\ }\href {\doibase 10.1103/PhysRevB.78.064108} {\bibfield
		{journal} {\bibinfo  {journal} {Phys. Rev. B}\ }\textbf {\bibinfo {volume}
			{78}},\ \bibinfo {pages} {064108} (\bibinfo {year} {2008})}\BibitemShut
	{NoStop}%
	\bibitem [{\citenamefont {Rappoport}\ \emph {et~al.}(2007)\citenamefont
		{Rappoport}, \citenamefont {Ghivelder}, \citenamefont {Fernandes},
		\citenamefont {Guimar\~aes},\ and\ \citenamefont
		{Continentino}}]{PhysRevB.75.054422}%
	\BibitemOpen
	\bibfield  {author} {\bibinfo {author} {\bibfnamefont {T.~G.}\ \bibnamefont
			{Rappoport}}, \bibinfo {author} {\bibfnamefont {L.}~\bibnamefont
			{Ghivelder}}, \bibinfo {author} {\bibfnamefont {J.~C.}\ \bibnamefont
			{Fernandes}}, \bibinfo {author} {\bibfnamefont {R.~B.}\ \bibnamefont
			{Guimar\~aes}}, \ and\ \bibinfo {author} {\bibfnamefont {M.~A.}\ \bibnamefont
			{Continentino}},\ }\bibfield  {title} {\enquote {\bibinfo {title}
			{Experimental observation of quantum entanglement in low-dimensional spin
				systems},}\ }\href {\doibase 10.1103/PhysRevB.75.054422} {\bibfield
		{journal} {\bibinfo  {journal} {Phys. Rev. B}\ }\textbf {\bibinfo {volume}
			{75}},\ \bibinfo {pages} {054422} (\bibinfo {year} {2007})}\BibitemShut
	{NoStop}%
	\bibitem [{\citenamefont {Das}\ \emph {et~al.}(2013)\citenamefont {Das},
		\citenamefont {Singh}, \citenamefont {Chakraborty}, \citenamefont {Gopal},\
		and\ \citenamefont {Mitra}}]{Das_2013}%
	\BibitemOpen
	\bibfield  {author} {\bibinfo {author} {\bibfnamefont {Diptaranjan}\
			\bibnamefont {Das}}, \bibinfo {author} {\bibfnamefont {Harkirat}\
			\bibnamefont {Singh}}, \bibinfo {author} {\bibfnamefont {Tanmoy}\
			\bibnamefont {Chakraborty}}, \bibinfo {author} {\bibfnamefont
			{Radha~Krishna}\ \bibnamefont {Gopal}}, \ and\ \bibinfo {author}
		{\bibfnamefont {Chiranjib}\ \bibnamefont {Mitra}},\ }\bibfield  {title}
	{\enquote {\bibinfo {title} {Experimental detection of quantum information
				sharing and its quantification in quantum spin systems},}\ }\href {\doibase
		10.1088/1367-2630/15/1/013047} {\bibfield  {journal} {\bibinfo  {journal}
			{New J. Phys.}\ }\textbf {\bibinfo {volume} {15}},\ \bibinfo {pages} {013047}
		(\bibinfo {year} {2013})}\BibitemShut {NoStop}%
	\bibitem [{\citenamefont {Singh}\ \emph {et~al.}(2013)\citenamefont {Singh},
		\citenamefont {Chakraborty}, \citenamefont {Das}, \citenamefont {Jeevan},
		\citenamefont {Tokiwa}, \citenamefont {Gegenwart},\ and\ \citenamefont
		{Mitra}}]{Singh2013}%
	\BibitemOpen
	\bibfield  {author} {\bibinfo {author} {\bibfnamefont {H}~\bibnamefont
			{Singh}}, \bibinfo {author} {\bibfnamefont {T}~\bibnamefont {Chakraborty}},
		\bibinfo {author} {\bibfnamefont {D}~\bibnamefont {Das}}, \bibinfo {author}
		{\bibfnamefont {H~S}\ \bibnamefont {Jeevan}}, \bibinfo {author}
		{\bibfnamefont {Y}~\bibnamefont {Tokiwa}}, \bibinfo {author} {\bibfnamefont
			{P}~\bibnamefont {Gegenwart}}, \ and\ \bibinfo {author} {\bibfnamefont
			{C}~\bibnamefont {Mitra}},\ }\bibfield  {title} {\enquote {\bibinfo {title}
			{Experimental quantification of entanglement through heat capacity},}\ }\href
	{\doibase 10.1088/1367-2630/15/11/113001} {\bibfield  {journal} {\bibinfo
			{journal} {New J. Phys.}\ }\textbf {\bibinfo {volume} {15}},\ \bibinfo
		{pages} {113001} (\bibinfo {year} {2013})}\BibitemShut {NoStop}%
	\bibitem [{\citenamefont {Sahling}\ \emph {et~al.}(2015)\citenamefont
		{Sahling}, \citenamefont {Remenyi}, \citenamefont {Paulsen}, \citenamefont
		{Monceau}, \citenamefont {Saligrama}, \citenamefont {Marin}, \citenamefont
		{Revcolevschi}, \citenamefont {Regnault}, \citenamefont {Raymond},\ and\
		\citenamefont {Lorenzo}}]{Sahling2015}%
	\BibitemOpen
	\bibfield  {author} {\bibinfo {author} {\bibfnamefont {S.}~\bibnamefont
			{Sahling}}, \bibinfo {author} {\bibfnamefont {G.}~\bibnamefont {Remenyi}},
		\bibinfo {author} {\bibfnamefont {C.}~\bibnamefont {Paulsen}}, \bibinfo
		{author} {\bibfnamefont {P.}~\bibnamefont {Monceau}}, \bibinfo {author}
		{\bibfnamefont {V.}~\bibnamefont {Saligrama}}, \bibinfo {author}
		{\bibfnamefont {C.}~\bibnamefont {Marin}}, \bibinfo {author} {\bibfnamefont
			{A.}~\bibnamefont {Revcolevschi}}, \bibinfo {author} {\bibfnamefont {L.~P.}\
			\bibnamefont {Regnault}}, \bibinfo {author} {\bibfnamefont {S.}~\bibnamefont
			{Raymond}}, \ and\ \bibinfo {author} {\bibfnamefont {J.~E.}\ \bibnamefont
			{Lorenzo}},\ }\bibfield  {title} {\enquote {\bibinfo {title} {Experimental
				realization of long-distance entanglement between spins in antiferromagnetic
				quantum spin chains},}\ }\href {\doibase 10.1038/nphys3186} {\bibfield
		{journal} {\bibinfo  {journal} {Nat. Phys.}\ }\textbf {\bibinfo {volume}
			{11}},\ \bibinfo {pages} {255--260} (\bibinfo {year} {2015})}\BibitemShut
	{NoStop}%
	\bibitem [{\citenamefont {Broholm}\ \emph {et~al.}(2020)\citenamefont
		{Broholm}, \citenamefont {Cava}, \citenamefont {Kivelson}, \citenamefont
		{Nocera}, \citenamefont {Norman},\ and\ \citenamefont
		{Senthil}}]{broholm2020quantum}%
	\BibitemOpen
	\bibfield  {author} {\bibinfo {author} {\bibfnamefont {C}~\bibnamefont
			{Broholm}}, \bibinfo {author} {\bibfnamefont {RJ}~\bibnamefont {Cava}},
		\bibinfo {author} {\bibfnamefont {SA}~\bibnamefont {Kivelson}}, \bibinfo
		{author} {\bibfnamefont {DG}~\bibnamefont {Nocera}}, \bibinfo {author}
		{\bibfnamefont {MR}~\bibnamefont {Norman}}, \ and\ \bibinfo {author}
		{\bibfnamefont {T}~\bibnamefont {Senthil}},\ }\bibfield  {title} {\enquote
		{\bibinfo {title} {Quantum spin liquids},}\ }\href {\doibase
		https://doi.org/10.1126/science.aay0668} {\bibfield  {journal} {\bibinfo
			{journal} {Science}\ }\textbf {\bibinfo {volume} {367}} (\bibinfo {year}
		{2020}),\ https://doi.org/10.1126/science.aay0668}\BibitemShut {NoStop}%
	\bibitem [{\citenamefont {Knolle}\ and\ \citenamefont
		{Moessner}(2019)}]{knolle2019field}%
	\BibitemOpen
	\bibfield  {author} {\bibinfo {author} {\bibfnamefont {Johannes}\
			\bibnamefont {Knolle}}\ and\ \bibinfo {author} {\bibfnamefont {Roderich}\
			\bibnamefont {Moessner}},\ }\bibfield  {title} {\enquote {\bibinfo {title} {A
				field guide to spin liquids},}\ }\href {\doibase
		https://doi.org/10.1146/annurev-conmatphys-031218-013401} {\bibfield
		{journal} {\bibinfo  {journal} {Annu. Rev. Condens. Matter Phys.}\ }\textbf
		{\bibinfo {volume} {10}},\ \bibinfo {pages} {451--472} (\bibinfo {year}
		{2019})}\BibitemShut {NoStop}%
	\bibitem [{\citenamefont {Savary}\ and\ \citenamefont
		{Balents}(2016)}]{Savary_2016}%
	\BibitemOpen
	\bibfield  {author} {\bibinfo {author} {\bibfnamefont {Lucile}\ \bibnamefont
			{Savary}}\ and\ \bibinfo {author} {\bibfnamefont {Leon}\ \bibnamefont
			{Balents}},\ }\bibfield  {title} {\enquote {\bibinfo {title} {Quantum spin
				liquids: a review},}\ }\href {\doibase 10.1088/0034-4885/80/1/016502}
	{\bibfield  {journal} {\bibinfo  {journal} {Rep. Prog. Phys.}\ }\textbf
		{\bibinfo {volume} {80}},\ \bibinfo {pages} {016502} (\bibinfo {year}
		{2016})}\BibitemShut {NoStop}%
	\bibitem [{\citenamefont {Lake}\ \emph {et~al.}(2013)\citenamefont {Lake},
		\citenamefont {Tennant}, \citenamefont {Caux}, \citenamefont {Barthel},
		\citenamefont {Schollw\"ock}, \citenamefont {Nagler},\ and\ \citenamefont
		{Frost}}]{Lake2013}%
	\BibitemOpen
	\bibfield  {author} {\bibinfo {author} {\bibfnamefont {B.}~\bibnamefont
			{Lake}}, \bibinfo {author} {\bibfnamefont {D.~A.}\ \bibnamefont {Tennant}},
		\bibinfo {author} {\bibfnamefont {J.-S.}\ \bibnamefont {Caux}}, \bibinfo
		{author} {\bibfnamefont {T.}~\bibnamefont {Barthel}}, \bibinfo {author}
		{\bibfnamefont {U.}~\bibnamefont {Schollw\"ock}}, \bibinfo {author}
		{\bibfnamefont {S.~E.}\ \bibnamefont {Nagler}}, \ and\ \bibinfo {author}
		{\bibfnamefont {C.~D.}\ \bibnamefont {Frost}},\ }\bibfield  {title} {\enquote
		{\bibinfo {title} {Multispinon continua at zero and finite temperature in a
				near-ideal {Heisenberg} chain},}\ }\href {\doibase
		10.1103/PhysRevLett.111.137205} {\bibfield  {journal} {\bibinfo  {journal}
			{Phys. Rev. Lett.}\ }\textbf {\bibinfo {volume} {111}},\ \bibinfo {pages}
		{137205} (\bibinfo {year} {2013})}\BibitemShut {NoStop}%
	\bibitem [{\citenamefont {Lake}\ \emph
		{et~al.}(2005{\natexlab{a}})\citenamefont {Lake}, \citenamefont {Tennant},\
		and\ \citenamefont {Nagler}}]{Lake2005_PRB}%
	\BibitemOpen
	\bibfield  {author} {\bibinfo {author} {\bibfnamefont {B.}~\bibnamefont
			{Lake}}, \bibinfo {author} {\bibfnamefont {D.~A.}\ \bibnamefont {Tennant}}, \
		and\ \bibinfo {author} {\bibfnamefont {S.~E.}\ \bibnamefont {Nagler}},\
	}\bibfield  {title} {\enquote {\bibinfo {title} {Longitudinal magnetic
				dynamics and dimensional crossover in the quasi-one-dimensional
				spin-{$\frac{1}{2}$ Heisenberg} antiferromagnet {${\mathrm{KCuF}}_{3}$}},}\
	}\href {\doibase 10.1103/PhysRevB.71.134412} {\bibfield  {journal} {\bibinfo
			{journal} {Phys. Rev. B}\ }\textbf {\bibinfo {volume} {71}},\ \bibinfo
		{pages} {134412} (\bibinfo {year} {2005}{\natexlab{a}})}\BibitemShut
	{NoStop}%
	\bibitem [{\citenamefont {Ollivier}\ and\ \citenamefont
		{Zurek}(2001)}]{PhysRevLett.88.017901}%
	\BibitemOpen
	\bibfield  {author} {\bibinfo {author} {\bibfnamefont {Harold}\ \bibnamefont
			{Ollivier}}\ and\ \bibinfo {author} {\bibfnamefont {Wojciech~H.}\
			\bibnamefont {Zurek}},\ }\bibfield  {title} {\enquote {\bibinfo {title}
			{Quantum discord: A measure of the quantumness of correlations},}\ }\href
	{\doibase 10.1103/PhysRevLett.88.017901} {\bibfield  {journal} {\bibinfo
			{journal} {Phys. Rev. Lett.}\ }\textbf {\bibinfo {volume} {88}},\ \bibinfo
		{pages} {017901} (\bibinfo {year} {2001})}\BibitemShut {NoStop}%
	\bibitem [{\citenamefont {Ferraro}\ \emph {et~al.}(2010)\citenamefont
		{Ferraro}, \citenamefont {Aolita}, \citenamefont {Cavalcanti}, \citenamefont
		{Cucchietti},\ and\ \citenamefont {Ac\'{\i}n}}]{PhysRevA.81.052318}%
	\BibitemOpen
	\bibfield  {author} {\bibinfo {author} {\bibfnamefont {A.}~\bibnamefont
			{Ferraro}}, \bibinfo {author} {\bibfnamefont {L.}~\bibnamefont {Aolita}},
		\bibinfo {author} {\bibfnamefont {D.}~\bibnamefont {Cavalcanti}}, \bibinfo
		{author} {\bibfnamefont {F.~M.}\ \bibnamefont {Cucchietti}}, \ and\ \bibinfo
		{author} {\bibfnamefont {A.}~\bibnamefont {Ac\'{\i}n}},\ }\bibfield  {title}
	{\enquote {\bibinfo {title} {Almost all quantum states have nonclassical
				correlations},}\ }\href {\doibase 10.1103/PhysRevA.81.052318} {\bibfield
		{journal} {\bibinfo  {journal} {Phys. Rev. A}\ }\textbf {\bibinfo {volume}
			{81}},\ \bibinfo {pages} {052318} (\bibinfo {year} {2010})}\BibitemShut
	{NoStop}%
	\bibitem [{\citenamefont {Lanyon}\ \emph {et~al.}(2008)\citenamefont {Lanyon},
		\citenamefont {Barbieri}, \citenamefont {Almeida},\ and\ \citenamefont
		{White}}]{PhysRevLett.101.200501}%
	\BibitemOpen
	\bibfield  {author} {\bibinfo {author} {\bibfnamefont {B.~P.}\ \bibnamefont
			{Lanyon}}, \bibinfo {author} {\bibfnamefont {M.}~\bibnamefont {Barbieri}},
		\bibinfo {author} {\bibfnamefont {M.~P.}\ \bibnamefont {Almeida}}, \ and\
		\bibinfo {author} {\bibfnamefont {A.~G.}\ \bibnamefont {White}},\ }\bibfield
	{title} {\enquote {\bibinfo {title} {Experimental quantum computing without
				entanglement},}\ }\href {\doibase 10.1103/PhysRevLett.101.200501} {\bibfield
		{journal} {\bibinfo  {journal} {Phys. Rev. Lett.}\ }\textbf {\bibinfo
			{volume} {101}},\ \bibinfo {pages} {200501} (\bibinfo {year}
		{2008})}\BibitemShut {NoStop}%
	\bibitem [{\citenamefont {Wootters}(1998)}]{Wootters_1998}%
	\BibitemOpen
	\bibfield  {author} {\bibinfo {author} {\bibfnamefont {William~K.}\
			\bibnamefont {Wootters}},\ }\bibfield  {title} {\enquote {\bibinfo {title}
			{Entanglement of formation of an arbitrary state of two qubits},}\ }\href
	{\doibase 10.1103/PhysRevLett.80.2245} {\bibfield  {journal} {\bibinfo
			{journal} {Phys. Rev. Lett.}\ }\textbf {\bibinfo {volume} {80}},\ \bibinfo
		{pages} {2245--2248} (\bibinfo {year} {1998})}\BibitemShut {NoStop}%
	\bibitem [{\citenamefont {Osborne}\ and\ \citenamefont
		{Verstraete}(2006)}]{PhysRevLett.96.220503}%
	\BibitemOpen
	\bibfield  {author} {\bibinfo {author} {\bibfnamefont {Tobias~J.}\
			\bibnamefont {Osborne}}\ and\ \bibinfo {author} {\bibfnamefont {Frank}\
			\bibnamefont {Verstraete}},\ }\bibfield  {title} {\enquote {\bibinfo {title}
			{General monogamy inequality for bipartite qubit entanglement},}\ }\href
	{\doibase 10.1103/PhysRevLett.96.220503} {\bibfield  {journal} {\bibinfo
			{journal} {Phys. Rev. Lett.}\ }\textbf {\bibinfo {volume} {96}},\ \bibinfo
		{pages} {220503} (\bibinfo {year} {2006})}\BibitemShut {NoStop}%
	\bibitem [{\citenamefont {Baskaran}\ \emph {et~al.}(2007)\citenamefont
		{Baskaran}, \citenamefont {Mandal},\ and\ \citenamefont
		{Shankar}}]{PhysRevLett.98.247201}%
	\BibitemOpen
	\bibfield  {author} {\bibinfo {author} {\bibfnamefont {G.}~\bibnamefont
			{Baskaran}}, \bibinfo {author} {\bibfnamefont {Saptarshi}\ \bibnamefont
			{Mandal}}, \ and\ \bibinfo {author} {\bibfnamefont {R.}~\bibnamefont
			{Shankar}},\ }\bibfield  {title} {\enquote {\bibinfo {title} {Exact results
				for spin dynamics and fractionalization in the {Kitaev} model},}\ }\href
	{\doibase 10.1103/PhysRevLett.98.247201} {\bibfield  {journal} {\bibinfo
			{journal} {Phys. Rev. Lett.}\ }\textbf {\bibinfo {volume} {98}},\ \bibinfo
		{pages} {247201} (\bibinfo {year} {2007})}\BibitemShut {NoStop}%
	\bibitem [{\citenamefont {Li}\ and\ \citenamefont {Zhu}(2008)}]{Li2008}%
	\BibitemOpen
	\bibfield  {author} {\bibinfo {author} {\bibfnamefont {You-Quan}\
			\bibnamefont {Li}}\ and\ \bibinfo {author} {\bibfnamefont {Guo-Qiang}\
			\bibnamefont {Zhu}},\ }\bibfield  {title} {\enquote {\bibinfo {title}
			{Concurrence vectors for entanglement of high-dimensional systems},}\ }\href
	{\doibase 10.1007/s11467-008-0022-2} {\bibfield  {journal} {\bibinfo
			{journal} {Front. Phys. China}\ }\textbf {\bibinfo {volume} {3}},\ \bibinfo
		{pages} {250--257} (\bibinfo {year} {2008})}\BibitemShut {NoStop}%
	\bibitem [{\citenamefont {Osterloh}(2015)}]{Osterloh2015}%
	\BibitemOpen
	\bibfield  {author} {\bibinfo {author} {\bibfnamefont {Andreas}\ \bibnamefont
			{Osterloh}},\ }\bibfield  {title} {\enquote {\bibinfo {title} {{SL}-invariant
				entanglement measures in higher dimensions: the case of spin 1 and 3/2},}\
	}\href {\doibase 10.1088/1751-8113/48/6/065303} {\bibfield  {journal}
		{\bibinfo  {journal} {J. Phys. A}\ }\textbf {\bibinfo {volume} {48}},\
		\bibinfo {pages} {065303} (\bibinfo {year} {2015})}\BibitemShut {NoStop}%
	\bibitem [{\citenamefont {Bahmani}\ \emph {et~al.}(2020)\citenamefont
		{Bahmani}, \citenamefont {Najarbashi},\ and\ \citenamefont
		{Tavana}}]{Bahmani2020}%
	\BibitemOpen
	\bibfield  {author} {\bibinfo {author} {\bibfnamefont {H.}~\bibnamefont
			{Bahmani}}, \bibinfo {author} {\bibfnamefont {G.}~\bibnamefont {Najarbashi}},
		\ and\ \bibinfo {author} {\bibfnamefont {A.}~\bibnamefont {Tavana}},\
	}\bibfield  {title} {\enquote {\bibinfo {title} {Generalized concurrence and
				quantum phase transition in spin-1 {Heisenberg} model},}\ }\href
	{https://doi.org/10.1088/1402-4896/ab606e} {\bibfield  {journal} {\bibinfo
			{journal} {Phys. Scr.}\ }\textbf {\bibinfo {volume} {95}},\ \bibinfo {pages}
		{055701} (\bibinfo {year} {2020})}\BibitemShut {NoStop}%
	\bibitem [{\citenamefont {Braunstein}\ and\ \citenamefont
		{Caves}(1994)}]{PhysRevLett.72.3439}%
	\BibitemOpen
	\bibfield  {author} {\bibinfo {author} {\bibfnamefont {Samuel~L.}\
			\bibnamefont {Braunstein}}\ and\ \bibinfo {author} {\bibfnamefont
			{Carlton~M.}\ \bibnamefont {Caves}},\ }\bibfield  {title} {\enquote {\bibinfo
			{title} {Statistical distance and the geometry of quantum states},}\ }\href
	{\doibase 10.1103/PhysRevLett.72.3439} {\bibfield  {journal} {\bibinfo
			{journal} {Phys. Rev. Lett.}\ }\textbf {\bibinfo {volume} {72}},\ \bibinfo
		{pages} {3439--3443} (\bibinfo {year} {1994})}\BibitemShut {NoStop}%
	\bibitem [{\citenamefont {T{\'{o}}th}\ and\ \citenamefont
		{Apellaniz}(2014)}]{Toth_2014}%
	\BibitemOpen
	\bibfield  {author} {\bibinfo {author} {\bibfnamefont {G{\'{e}}za}\
			\bibnamefont {T{\'{o}}th}}\ and\ \bibinfo {author} {\bibfnamefont {Iagoba}\
			\bibnamefont {Apellaniz}},\ }\bibfield  {title} {\enquote {\bibinfo {title}
			{Quantum metrology from a quantum information science perspective},}\ }\href
	{\doibase 10.1088/1751-8113/47/42/424006} {\bibfield  {journal} {\bibinfo
			{journal} {J. Phys. A}\ }\textbf {\bibinfo {volume} {47}},\ \bibinfo {pages}
		{424006} (\bibinfo {year} {2014})}\BibitemShut {NoStop}%
	\bibitem [{\citenamefont {T\'oth}(2012)}]{PhysRevA.85.022322}%
	\BibitemOpen
	\bibfield  {author} {\bibinfo {author} {\bibfnamefont {G\'eza}\ \bibnamefont
			{T\'oth}},\ }\bibfield  {title} {\enquote {\bibinfo {title} {Multipartite
				entanglement and high-precision metrology},}\ }\href {\doibase
		10.1103/PhysRevA.85.022322} {\bibfield  {journal} {\bibinfo  {journal} {Phys.
				Rev. A}\ }\textbf {\bibinfo {volume} {85}},\ \bibinfo {pages} {022322}
		(\bibinfo {year} {2012})}\BibitemShut {NoStop}%
	\bibitem [{\citenamefont {Brenes}\ \emph {et~al.}(2020)\citenamefont {Brenes},
		\citenamefont {Pappalardi}, \citenamefont {Goold},\ and\ \citenamefont
		{Silva}}]{PhysRevLett.124.040605}%
	\BibitemOpen
	\bibfield  {author} {\bibinfo {author} {\bibfnamefont {Marlon}\ \bibnamefont
			{Brenes}}, \bibinfo {author} {\bibfnamefont {Silvia}\ \bibnamefont
			{Pappalardi}}, \bibinfo {author} {\bibfnamefont {John}\ \bibnamefont
			{Goold}}, \ and\ \bibinfo {author} {\bibfnamefont {Alessandro}\ \bibnamefont
			{Silva}},\ }\bibfield  {title} {\enquote {\bibinfo {title} {Multipartite
				entanglement structure in the eigenstate thermalization hypothesis},}\ }\href
	{\doibase 10.1103/PhysRevLett.124.040605} {\bibfield  {journal} {\bibinfo
			{journal} {Phys. Rev. Lett.}\ }\textbf {\bibinfo {volume} {124}},\ \bibinfo
		{pages} {040605} (\bibinfo {year} {2020})}\BibitemShut {NoStop}%
	\bibitem [{\citenamefont {Lake}\ \emph
		{et~al.}(2005{\natexlab{b}})\citenamefont {Lake}, \citenamefont {Tennant},
		\citenamefont {Frost},\ and\ \citenamefont {Nagler}}]{Lake2005}%
	\BibitemOpen
	\bibfield  {author} {\bibinfo {author} {\bibfnamefont {Bella}\ \bibnamefont
			{Lake}}, \bibinfo {author} {\bibfnamefont {D.~Alan}\ \bibnamefont {Tennant}},
		\bibinfo {author} {\bibfnamefont {Chris~D.}\ \bibnamefont {Frost}}, \ and\
		\bibinfo {author} {\bibfnamefont {Stephen~E.}\ \bibnamefont {Nagler}},\
	}\bibfield  {title} {\enquote {\bibinfo {title} {Quantum criticality and
				universal scaling of a quantum antiferromagnet},}\ }\href {\doibase
		10.1038/nmat1327} {\bibfield  {journal} {\bibinfo  {journal} {Nat. Mater.}\
		}\textbf {\bibinfo {volume} {4}},\ \bibinfo {pages} {329--334} (\bibinfo
		{year} {2005}{\natexlab{b}})}\BibitemShut {NoStop}%
	\bibitem [{\citenamefont {White}(1992)}]{PhysRevLett.69.2863}%
	\BibitemOpen
	\bibfield  {author} {\bibinfo {author} {\bibfnamefont {Steven~R.}\
			\bibnamefont {White}},\ }\bibfield  {title} {\enquote {\bibinfo {title}
			{Density matrix formulation for quantum renormalization groups},}\ }\href
	{\doibase 10.1103/PhysRevLett.69.2863} {\bibfield  {journal} {\bibinfo
			{journal} {Phys. Rev. Lett.}\ }\textbf {\bibinfo {volume} {69}},\ \bibinfo
		{pages} {2863--2866} (\bibinfo {year} {1992})}\BibitemShut {NoStop}%
	\bibitem [{\citenamefont {White}(1993)}]{PhysRevB.48.10345}%
	\BibitemOpen
	\bibfield  {author} {\bibinfo {author} {\bibfnamefont {Steven~R.}\
			\bibnamefont {White}},\ }\bibfield  {title} {\enquote {\bibinfo {title}
			{Density-matrix algorithms for quantum renormalization groups},}\ }\href
	{\doibase 10.1103/PhysRevB.48.10345} {\bibfield  {journal} {\bibinfo
			{journal} {Phys. Rev. B}\ }\textbf {\bibinfo {volume} {48}},\ \bibinfo
		{pages} {10345--10356} (\bibinfo {year} {1993})}\BibitemShut {NoStop}%
	\bibitem [{\citenamefont {Alvarez}(2009)}]{Alvarez2009}%
	\BibitemOpen
	\bibfield  {author} {\bibinfo {author} {\bibfnamefont {G.}~\bibnamefont
			{Alvarez}},\ }\bibfield  {title} {\enquote {\bibinfo {title} {The density
				matrix renormalization group for strongly correlated electron systems: A
				generic implementation},}\ }\href {\doibase 10.1016/j.cpc.2009.02.016}
	{\bibfield  {journal} {\bibinfo  {journal} {Comp. Phys. Comms.}\ }\textbf
		{\bibinfo {volume} {180}},\ \bibinfo {pages} {1572--1578} (\bibinfo {year}
		{2009})}\BibitemShut {NoStop}%
	\bibitem [{\citenamefont {K\"uhner}\ and\ \citenamefont
		{White}(1999)}]{PhysRevB.60.335}%
	\BibitemOpen
	\bibfield  {author} {\bibinfo {author} {\bibfnamefont {Till~D.}\ \bibnamefont
			{K\"uhner}}\ and\ \bibinfo {author} {\bibfnamefont {Steven~R.}\ \bibnamefont
			{White}},\ }\bibfield  {title} {\enquote {\bibinfo {title} {Dynamical
				correlation functions using the density matrix renormalization group},}\
	}\href {\doibase 10.1103/PhysRevB.60.335} {\bibfield  {journal} {\bibinfo
			{journal} {Phys. Rev. B}\ }\textbf {\bibinfo {volume} {60}},\ \bibinfo
		{pages} {335--343} (\bibinfo {year} {1999})}\BibitemShut {NoStop}%
	\bibitem [{\citenamefont {Nocera}\ and\ \citenamefont
		{Alvarez}(2016{\natexlab{a}})}]{PhysRevE.94.053308}%
	\BibitemOpen
	\bibfield  {author} {\bibinfo {author} {\bibfnamefont {A.}~\bibnamefont
			{Nocera}}\ and\ \bibinfo {author} {\bibfnamefont {G.}~\bibnamefont
			{Alvarez}},\ }\bibfield  {title} {\enquote {\bibinfo {title} {Spectral
				functions with the density matrix renormalization group: Krylov-space
				approach for correction vectors},}\ }\href {\doibase
		10.1103/PhysRevE.94.053308} {\bibfield  {journal} {\bibinfo  {journal} {Phys.
				Rev. E}\ }\textbf {\bibinfo {volume} {94}},\ \bibinfo {pages} {053308}
		(\bibinfo {year} {2016}{\natexlab{a}})}\BibitemShut {NoStop}%
	\bibitem [{\citenamefont {Feiguin}\ and\ \citenamefont
		{White}(2005)}]{PhysRevB.72.220401}%
	\BibitemOpen
	\bibfield  {author} {\bibinfo {author} {\bibfnamefont {Adrian~E.}\
			\bibnamefont {Feiguin}}\ and\ \bibinfo {author} {\bibfnamefont {Steven~R.}\
			\bibnamefont {White}},\ }\bibfield  {title} {\enquote {\bibinfo {title}
			{Finite-temperature density matrix renormalization using an enlarged
				{Hilbert} space},}\ }\href {\doibase 10.1103/PhysRevB.72.220401} {\bibfield
		{journal} {\bibinfo  {journal} {Phys. Rev. B}\ }\textbf {\bibinfo {volume}
			{72}},\ \bibinfo {pages} {220401} (\bibinfo {year} {2005})}\BibitemShut
	{NoStop}%
	\bibitem [{\citenamefont {Feiguin}\ and\ \citenamefont
		{Fiete}(2010)}]{PhysRevB.81.075108}%
	\BibitemOpen
	\bibfield  {author} {\bibinfo {author} {\bibfnamefont {Adrian~E.}\
			\bibnamefont {Feiguin}}\ and\ \bibinfo {author} {\bibfnamefont {Gregory~A.}\
			\bibnamefont {Fiete}},\ }\bibfield  {title} {\enquote {\bibinfo {title}
			{Spectral properties of a spin-incoherent {Luttinger} liquid},}\ }\href
	{\doibase 10.1103/PhysRevB.81.075108} {\bibfield  {journal} {\bibinfo
			{journal} {Phys. Rev. B}\ }\textbf {\bibinfo {volume} {81}},\ \bibinfo
		{pages} {075108} (\bibinfo {year} {2010})}\BibitemShut {NoStop}%
	\bibitem [{\citenamefont {Nocera}\ and\ \citenamefont
		{Alvarez}(2016{\natexlab{b}})}]{PhysRevB.93.045137}%
	\BibitemOpen
	\bibfield  {author} {\bibinfo {author} {\bibfnamefont {A.}~\bibnamefont
			{Nocera}}\ and\ \bibinfo {author} {\bibfnamefont {G.}~\bibnamefont
			{Alvarez}},\ }\bibfield  {title} {\enquote {\bibinfo {title}
			{Symmetry-conserving purification of quantum states within the density matrix
				renormalization group},}\ }\href {\doibase 10.1103/PhysRevB.93.045137}
	{\bibfield  {journal} {\bibinfo  {journal} {Phys. Rev. B}\ }\textbf {\bibinfo
			{volume} {93}},\ \bibinfo {pages} {045137} (\bibinfo {year}
		{2016}{\natexlab{b}})}\BibitemShut {NoStop}%
	\bibitem [{Sup()}]{SuppMat}%
	\BibitemOpen
	\href@noop {} {}\bibinfo {note} {See Supplemental Material at [URL will be
		inserted by publisher] for more details on reproducing the
		calculations.}\BibitemShut {Stop}%
	\bibitem [{\citenamefont {Samarakoon}\ \emph {et~al.}(2017)\citenamefont
		{Samarakoon}, \citenamefont {Banerjee}, \citenamefont {Zhang}, \citenamefont
		{Kamiya}, \citenamefont {Nagler}, \citenamefont {Tennant}, \citenamefont
		{Lee},\ and\ \citenamefont {Batista}}]{samarakoon2017comprehensive}%
	\BibitemOpen
	\bibfield  {author} {\bibinfo {author} {\bibfnamefont {A.~M.}\ \bibnamefont
			{Samarakoon}}, \bibinfo {author} {\bibfnamefont {A.}~\bibnamefont
			{Banerjee}}, \bibinfo {author} {\bibfnamefont {S.-S.}\ \bibnamefont {Zhang}},
		\bibinfo {author} {\bibfnamefont {Y.}~\bibnamefont {Kamiya}}, \bibinfo
		{author} {\bibfnamefont {S.~E.}\ \bibnamefont {Nagler}}, \bibinfo {author}
		{\bibfnamefont {D.~A.}\ \bibnamefont {Tennant}}, \bibinfo {author}
		{\bibfnamefont {S.-H.}\ \bibnamefont {Lee}}, \ and\ \bibinfo {author}
		{\bibfnamefont {C.~D.}\ \bibnamefont {Batista}},\ }\bibfield  {title}
	{\enquote {\bibinfo {title} {Comprehensive study of the dynamics of a
				classical {Kitaev} spin liquid},}\ }\href {\doibase
		10.1103/PhysRevB.96.134408} {\bibfield  {journal} {\bibinfo  {journal} {Phys.
				Rev. B}\ }\textbf {\bibinfo {volume} {96}},\ \bibinfo {pages} {134408}
		(\bibinfo {year} {2017})}\BibitemShut {NoStop}%
	\bibitem [{\citenamefont {Hutchings}\ \emph {et~al.}(1969)\citenamefont
		{Hutchings}, \citenamefont {Samuelsen}, \citenamefont {Shirane},\ and\
		\citenamefont {Hirakawa}}]{Hutchings_1969}%
	\BibitemOpen
	\bibfield  {author} {\bibinfo {author} {\bibfnamefont {M.~T.}\ \bibnamefont
			{Hutchings}}, \bibinfo {author} {\bibfnamefont {E.~J.}\ \bibnamefont
			{Samuelsen}}, \bibinfo {author} {\bibfnamefont {G.}~\bibnamefont {Shirane}},
		\ and\ \bibinfo {author} {\bibfnamefont {K.}~\bibnamefont {Hirakawa}},\
	}\bibfield  {title} {\enquote {\bibinfo {title} {Neutron-diffraction
				determination of the antiferromagnetic structure of
				{KCu${\mathrm{F}}_{3}$}},}\ }\href {\doibase 10.1103/PhysRev.188.919}
	{\bibfield  {journal} {\bibinfo  {journal} {Phys. Rev.}\ }\textbf {\bibinfo
			{volume} {188}},\ \bibinfo {pages} {919--923} (\bibinfo {year}
		{1969})}\BibitemShut {NoStop}%
	\bibitem [{\citenamefont {Lambert}\ and\ \citenamefont
		{S\o{}rensen}(2019)}]{Lambert_2019}%
	\BibitemOpen
	\bibfield  {author} {\bibinfo {author} {\bibfnamefont {J.}~\bibnamefont
			{Lambert}}\ and\ \bibinfo {author} {\bibfnamefont {E.~S.}\ \bibnamefont
			{S\o{}rensen}},\ }\bibfield  {title} {\enquote {\bibinfo {title} {Estimates
				of the quantum {Fisher} information in the {$S=1$} antiferromagnetic
				{Heisenberg} spin chain with uniaxial anisotropy},}\ }\href {\doibase
		10.1103/PhysRevB.99.045117} {\bibfield  {journal} {\bibinfo  {journal} {Phys.
				Rev. B}\ }\textbf {\bibinfo {volume} {99}},\ \bibinfo {pages} {045117}
		(\bibinfo {year} {2019})}\BibitemShut {NoStop}%
	\bibitem [{\citenamefont {Gozel}\ \emph {et~al.}(2019)\citenamefont {Gozel},
		\citenamefont {Mila},\ and\ \citenamefont {Affleck}}]{Gozel_2019}%
	\BibitemOpen
	\bibfield  {author} {\bibinfo {author} {\bibfnamefont {Samuel}\ \bibnamefont
			{Gozel}}, \bibinfo {author} {\bibfnamefont {Fr\'ed\'eric}\ \bibnamefont
			{Mila}}, \ and\ \bibinfo {author} {\bibfnamefont {Ian}\ \bibnamefont
			{Affleck}},\ }\bibfield  {title} {\enquote {\bibinfo {title} {Asymptotic
				freedom and large spin antiferromagnetic chains},}\ }\href {\doibase
		10.1103/PhysRevLett.123.037202} {\bibfield  {journal} {\bibinfo  {journal}
			{Phys. Rev. Lett.}\ }\textbf {\bibinfo {volume} {123}},\ \bibinfo {pages}
		{037202} (\bibinfo {year} {2019})}\BibitemShut {NoStop}%
	\bibitem [{\citenamefont {de~Almeida}\ and\ \citenamefont
		{Hauke}(2020)}]{Almeida2020}%
	\BibitemOpen
	\bibfield  {author} {\bibinfo {author} {\bibfnamefont {Ricardo~Costa}\
			\bibnamefont {de~Almeida}}\ and\ \bibinfo {author} {\bibfnamefont {Philipp}\
			\bibnamefont {Hauke}},\ }\href {https://arxiv.org/abs/2005.03049} {\enquote
		{\bibinfo {title} {From entanglement certification with quench dynamics to
				multipartite entanglement of interacting fermions},}\ } (\bibinfo {year}
	{2020}),\ \bibinfo {note} {arXiv:2005.03049}\BibitemShut {NoStop}%
	\bibitem [{\citenamefont {Fr\'erot}\ and\ \citenamefont
		{Roscilde}(2016)}]{PhysRevB.94.075121}%
	\BibitemOpen
	\bibfield  {author} {\bibinfo {author} {\bibfnamefont {Ir\'en\'ee}\
			\bibnamefont {Fr\'erot}}\ and\ \bibinfo {author} {\bibfnamefont {Tommaso}\
			\bibnamefont {Roscilde}},\ }\bibfield  {title} {\enquote {\bibinfo {title}
			{Quantum variance: A measure of quantum coherence and quantum correlations
				for many-body systems},}\ }\href {\doibase 10.1103/PhysRevB.94.075121}
	{\bibfield  {journal} {\bibinfo  {journal} {Phys. Rev. B}\ }\textbf {\bibinfo
			{volume} {94}},\ \bibinfo {pages} {075121} (\bibinfo {year}
		{2016})}\BibitemShut {NoStop}%
	\bibitem [{\citenamefont {Fr\'erot}\ and\ \citenamefont
		{Roscilde}(2019)}]{Frerot2019}%
	\BibitemOpen
	\bibfield  {author} {\bibinfo {author} {\bibfnamefont {Ir\'en\'ee}\
			\bibnamefont {Fr\'erot}}\ and\ \bibinfo {author} {\bibfnamefont {Tommaso}\
			\bibnamefont {Roscilde}},\ }\bibfield  {title} {\enquote {\bibinfo {title}
			{Reconstructing the quantum critical fan of strongly correlated systems using
				quantum correlations},}\ }\href {\doibase 10.1038/s41467-019-08324-9}
	{\bibfield  {journal} {\bibinfo  {journal} {Nat. Commun.}\ }\textbf {\bibinfo
			{volume} {10}},\ \bibinfo {pages} {577} (\bibinfo {year} {2019})}\BibitemShut
	{NoStop}%
	\bibitem [{\citenamefont {Gabbrielli}\ \emph {et~al.}(2018)\citenamefont
		{Gabbrielli}, \citenamefont {Smerzi},\ and\ \citenamefont
		{Pezz\`e}}]{Gabbrielli2018}%
	\BibitemOpen
	\bibfield  {author} {\bibinfo {author} {\bibfnamefont {Marco}\ \bibnamefont
			{Gabbrielli}}, \bibinfo {author} {\bibfnamefont {Augusto}\ \bibnamefont
			{Smerzi}}, \ and\ \bibinfo {author} {\bibfnamefont {Luca}\ \bibnamefont
			{Pezz\`e}},\ }\bibfield  {title} {\enquote {\bibinfo {title} {Multipartite
				entanglement at finite temperature},}\ }\href {\doibase
		10.1038/s41598-018-31761-3} {\bibfield  {journal} {\bibinfo  {journal} {Sci.
				Rep.}\ }\textbf {\bibinfo {volume} {8}},\ \bibinfo {pages} {15663} (\bibinfo
		{year} {2018})}\BibitemShut {NoStop}%
	\bibitem [{\citenamefont {Rajabpour}(2017)}]{PhysRevD.96.126007}%
	\BibitemOpen
	\bibfield  {author} {\bibinfo {author} {\bibfnamefont {M.~A.}\ \bibnamefont
			{Rajabpour}},\ }\bibfield  {title} {\enquote {\bibinfo {title} {Multipartite
				entanglement and quantum {Fisher} information in conformal field theories},}\
	}\href {\doibase 10.1103/PhysRevD.96.126007} {\bibfield  {journal} {\bibinfo
			{journal} {Phys. Rev. D}\ }\textbf {\bibinfo {volume} {96}},\ \bibinfo
		{pages} {126007} (\bibinfo {year} {2017})}\BibitemShut {NoStop}%
	\bibitem [{\citenamefont {Tennant}\ \emph {et~al.}(2003)\citenamefont
		{Tennant}, \citenamefont {Broholm}, \citenamefont {Reich}, \citenamefont
		{Nagler}, \citenamefont {Granroth}, \citenamefont {Barnes}, \citenamefont
		{Damle}, \citenamefont {Xu}, \citenamefont {Chen},\ and\ \citenamefont
		{Sales}}]{Tennant2003}%
	\BibitemOpen
	\bibfield  {author} {\bibinfo {author} {\bibfnamefont {D.~A.}\ \bibnamefont
			{Tennant}}, \bibinfo {author} {\bibfnamefont {C.}~\bibnamefont {Broholm}},
		\bibinfo {author} {\bibfnamefont {D.~H.}\ \bibnamefont {Reich}}, \bibinfo
		{author} {\bibfnamefont {S.~E.}\ \bibnamefont {Nagler}}, \bibinfo {author}
		{\bibfnamefont {G.~E.}\ \bibnamefont {Granroth}}, \bibinfo {author}
		{\bibfnamefont {T.}~\bibnamefont {Barnes}}, \bibinfo {author} {\bibfnamefont
			{K.}~\bibnamefont {Damle}}, \bibinfo {author} {\bibfnamefont
			{G.}~\bibnamefont {Xu}}, \bibinfo {author} {\bibfnamefont {Y.}~\bibnamefont
			{Chen}}, \ and\ \bibinfo {author} {\bibfnamefont {B.~C.}\ \bibnamefont
			{Sales}},\ }\bibfield  {title} {\enquote {\bibinfo {title} {Neutron
				scattering study of two-magnon states in the quantum magnet copper
				nitrate},}\ }\href {\doibase 10.1103/PhysRevB.67.054414} {\bibfield
		{journal} {\bibinfo  {journal} {Phys. Rev. B}\ }\textbf {\bibinfo {volume}
			{67}},\ \bibinfo {pages} {054414} (\bibinfo {year} {2003})}\BibitemShut
	{NoStop}%
	\bibitem [{\citenamefont {Wernsdorfer}\ \emph {et~al.}(2002)\citenamefont
		{Wernsdorfer}, \citenamefont {Aliaga-Alcalde}, \citenamefont {Hendrickson},\
		and\ \citenamefont {Christou}}]{Wernsdorfer2002}%
	\BibitemOpen
	\bibfield  {author} {\bibinfo {author} {\bibfnamefont {Wolfgang}\
			\bibnamefont {Wernsdorfer}}, \bibinfo {author} {\bibfnamefont {N{\'u}ria}\
			\bibnamefont {Aliaga-Alcalde}}, \bibinfo {author} {\bibfnamefont {David~N.}\
			\bibnamefont {Hendrickson}}, \ and\ \bibinfo {author} {\bibfnamefont
			{George}\ \bibnamefont {Christou}},\ }\bibfield  {title} {\enquote {\bibinfo
			{title} {Exchange-biased quantum tunnelling in a supramolecular dimer of
				single-molecule magnets},}\ }\href {\doibase 10.1038/416406a} {\bibfield
		{journal} {\bibinfo  {journal} {Nature (London)}\ }\textbf {\bibinfo {volume}
			{416}},\ \bibinfo {pages} {406--409} (\bibinfo {year} {2002})}\BibitemShut
	{NoStop}%
	\bibitem [{\citenamefont {Uematsu}\ and\ \citenamefont
		{Kawamura}(2018)}]{Uematsu_2018}%
	\BibitemOpen
	\bibfield  {author} {\bibinfo {author} {\bibfnamefont {Kazuki}\ \bibnamefont
			{Uematsu}}\ and\ \bibinfo {author} {\bibfnamefont {Hikaru}\ \bibnamefont
			{Kawamura}},\ }\bibfield  {title} {\enquote {\bibinfo {title}
			{Randomness-induced quantum spin liquid behavior in the $s=\frac{1}{2}$
				random ${J}_{1}\text{\ensuremath{-}}{J}_{2}$ {Heisenberg} antiferromagnet on
				the square lattice},}\ }\href {\doibase 10.1103/PhysRevB.98.134427}
	{\bibfield  {journal} {\bibinfo  {journal} {Phys. Rev. B}\ }\textbf {\bibinfo
			{volume} {98}},\ \bibinfo {pages} {134427} (\bibinfo {year}
		{2018})}\BibitemShut {NoStop}%
	\bibitem [{\citenamefont {Uematsu}\ and\ \citenamefont
		{Kawamura}(2019)}]{Uematsu_2019}%
	\BibitemOpen
	\bibfield  {author} {\bibinfo {author} {\bibfnamefont {Kazuki}\ \bibnamefont
			{Uematsu}}\ and\ \bibinfo {author} {\bibfnamefont {Hikaru}\ \bibnamefont
			{Kawamura}},\ }\bibfield  {title} {\enquote {\bibinfo {title}
			{Randomness-induced quantum spin liquid behavior in the $s=1/2$ random-bond
				{Heisenberg} antiferromagnet on the pyrochlore lattice},}\ }\href {\doibase
		10.1103/PhysRevLett.123.087201} {\bibfield  {journal} {\bibinfo  {journal}
			{Phys. Rev. Lett.}\ }\textbf {\bibinfo {volume} {123}},\ \bibinfo {pages}
		{087201} (\bibinfo {year} {2019})}\BibitemShut {NoStop}%
	\bibitem [{\citenamefont {Liu}\ \emph {et~al.}(2018)\citenamefont {Liu},
		\citenamefont {Shao}, \citenamefont {Lin}, \citenamefont {Guo},\ and\
		\citenamefont {Sandvik}}]{Liu_2018}%
	\BibitemOpen
	\bibfield  {author} {\bibinfo {author} {\bibfnamefont {Lu}~\bibnamefont
			{Liu}}, \bibinfo {author} {\bibfnamefont {Hui}\ \bibnamefont {Shao}},
		\bibinfo {author} {\bibfnamefont {Yu-Cheng}\ \bibnamefont {Lin}}, \bibinfo
		{author} {\bibfnamefont {Wenan}\ \bibnamefont {Guo}}, \ and\ \bibinfo
		{author} {\bibfnamefont {Anders~W.}\ \bibnamefont {Sandvik}},\ }\bibfield
	{title} {\enquote {\bibinfo {title} {Random-singlet phase in disordered
				two-dimensional quantum magnets},}\ }\href {\doibase
		10.1103/PhysRevX.8.041040} {\bibfield  {journal} {\bibinfo  {journal} {Phys.
				Rev. X}\ }\textbf {\bibinfo {volume} {8}},\ \bibinfo {pages} {041040}
		(\bibinfo {year} {2018})}\BibitemShut {NoStop}%
	\bibitem [{\citenamefont {Shen}\ \emph {et~al.}(2020)\citenamefont {Shen},
		\citenamefont {Kuhn}, \citenamefont {Dalgliesh}, \citenamefont {de~Haan},
		\citenamefont {Geerits}, \citenamefont {Irfan}, \citenamefont {Li},
		\citenamefont {Lu}, \citenamefont {Parnell}, \citenamefont {Plomp},
		\citenamefont {van Well}, \citenamefont {Washington}, \citenamefont {Baxter},
		\citenamefont {Ortiz}, \citenamefont {Snow},\ and\ \citenamefont
		{Pynn}}]{Shen20}%
	\BibitemOpen
	\bibfield  {author} {\bibinfo {author} {\bibfnamefont {J.}~\bibnamefont
			{Shen}}, \bibinfo {author} {\bibfnamefont {S.~J.}\ \bibnamefont {Kuhn}},
		\bibinfo {author} {\bibfnamefont {R.~M.}\ \bibnamefont {Dalgliesh}}, \bibinfo
		{author} {\bibfnamefont {V.~O.}\ \bibnamefont {de~Haan}}, \bibinfo {author}
		{\bibfnamefont {N.}~\bibnamefont {Geerits}}, \bibinfo {author} {\bibfnamefont
			{A.~A.~M.}\ \bibnamefont {Irfan}}, \bibinfo {author} {\bibfnamefont
			{F.}~\bibnamefont {Li}}, \bibinfo {author} {\bibfnamefont {S.}~\bibnamefont
			{Lu}}, \bibinfo {author} {\bibfnamefont {S.~R.}\ \bibnamefont {Parnell}},
		\bibinfo {author} {\bibfnamefont {J.}~\bibnamefont {Plomp}}, \bibinfo
		{author} {\bibfnamefont {A.~A.}\ \bibnamefont {van Well}}, \bibinfo {author}
		{\bibfnamefont {A.}~\bibnamefont {Washington}}, \bibinfo {author}
		{\bibfnamefont {D.~V.}\ \bibnamefont {Baxter}}, \bibinfo {author}
		{\bibfnamefont {G.}~\bibnamefont {Ortiz}}, \bibinfo {author} {\bibfnamefont
			{W.~M.}\ \bibnamefont {Snow}}, \ and\ \bibinfo {author} {\bibfnamefont
			{R.}~\bibnamefont {Pynn}},\ }\bibfield  {title} {\enquote {\bibinfo {title}
			{Unveiling contextual realities by microscopically entangling a neutron},}\
	}\href {\doibase 10.1038/s41467-020-14741-y} {\bibfield  {journal} {\bibinfo
			{journal} {Nat. Commun.}\ }\textbf {\bibinfo {volume} {11}},\ \bibinfo
		{pages} {930} (\bibinfo {year} {2020})}\BibitemShut {NoStop}%
	\bibitem [{\citenamefont {Boothroyd}(2020)}]{boothroyd2020principles}%
	\BibitemOpen
	\bibfield  {author} {\bibinfo {author} {\bibfnamefont {Andrew}\ \bibnamefont
			{Boothroyd}},\ }\href@noop {} {\emph {\bibinfo {title} {Principles of Neutron
				Scattering from Condensed Matter}}}\ (\bibinfo  {publisher} {Oxford
		University Press, USA},\ \bibinfo {year} {2020})\BibitemShut {NoStop}%
	\bibitem [{\citenamefont {Brown}(1998)}]{BrownFF}%
	\BibitemOpen
	\bibfield  {author} {\bibinfo {author} {\bibfnamefont {P.~J.}\ \bibnamefont
			{Brown}},\ }\href {https://www.ill.eu/sites/ccsl/ffacts/} {\enquote {\bibinfo
			{title} {Magnetic form factors},}\ }\bibinfo {howpublished} {The Cambridge
		Crystallographic Subroutine Library} (\bibinfo {year} {1998})\BibitemShut
	{NoStop}%
	\bibitem [{\citenamefont {Abernathy}\ \emph {et~al.}(2012)\citenamefont
		{Abernathy}, \citenamefont {Stone}, \citenamefont {Loguillo}, \citenamefont
		{Lucas}, \citenamefont {Delaire}, \citenamefont {Tang}, \citenamefont {Lin},\
		and\ \citenamefont {Fultz}}]{abernathy2012design}%
	\BibitemOpen
	\bibfield  {author} {\bibinfo {author} {\bibfnamefont {Douglas~L}\
			\bibnamefont {Abernathy}}, \bibinfo {author} {\bibfnamefont {Matthew~B}\
			\bibnamefont {Stone}}, \bibinfo {author} {\bibfnamefont {MJ}~\bibnamefont
			{Loguillo}}, \bibinfo {author} {\bibfnamefont {MS}~\bibnamefont {Lucas}},
		\bibinfo {author} {\bibfnamefont {O}~\bibnamefont {Delaire}}, \bibinfo
		{author} {\bibfnamefont {Xiaoli}\ \bibnamefont {Tang}}, \bibinfo {author}
		{\bibfnamefont {JYY}\ \bibnamefont {Lin}}, \ and\ \bibinfo {author}
		{\bibfnamefont {B}~\bibnamefont {Fultz}},\ }\bibfield  {title} {\enquote
		{\bibinfo {title} {Design and operation of the wide angular-range chopper
				spectrometer arcs at the spallation neutron source},}\ }\href {\doibase
		10.1063/1.3680104} {\bibfield  {journal} {\bibinfo  {journal} {Rev. Sci.
				Instrum.}\ }\textbf {\bibinfo {volume} {83}},\ \bibinfo {pages} {015114}
		(\bibinfo {year} {2012})}\BibitemShut {NoStop}%
	\bibitem [{\citenamefont {Arnold}\ \emph {et~al.}(2014)\citenamefont {Arnold},
		\citenamefont {Bilheux}, \citenamefont {Borreguero}, \citenamefont {Buts},
		\citenamefont {Campbell}, \citenamefont {Chapon}, \citenamefont {Doucet},
		\citenamefont {Draper}, \citenamefont {Leal}, \citenamefont {Gigg} \emph
		{et~al.}}]{arnold2014mantid}%
	\BibitemOpen
	\bibfield  {author} {\bibinfo {author} {\bibfnamefont {Owen}\ \bibnamefont
			{Arnold}}, \bibinfo {author} {\bibfnamefont {Jean-Christophe}\ \bibnamefont
			{Bilheux}}, \bibinfo {author} {\bibfnamefont {JM}~\bibnamefont {Borreguero}},
		\bibinfo {author} {\bibfnamefont {Alex}\ \bibnamefont {Buts}}, \bibinfo
		{author} {\bibfnamefont {Stuart~I}\ \bibnamefont {Campbell}}, \bibinfo
		{author} {\bibfnamefont {L}~\bibnamefont {Chapon}}, \bibinfo {author}
		{\bibfnamefont {Mathieu}\ \bibnamefont {Doucet}}, \bibinfo {author}
		{\bibfnamefont {N}~\bibnamefont {Draper}}, \bibinfo {author} {\bibfnamefont
			{R~Ferraz}\ \bibnamefont {Leal}}, \bibinfo {author} {\bibfnamefont
			{MA}~\bibnamefont {Gigg}},  \emph {et~al.},\ }\bibfield  {title} {\enquote
		{\bibinfo {title} {Mantid---{Data} analysis and visualization package for
				neutron scattering and $\mu$ sr experiments},}\ }\href {\doibase
		10.1016/j.nima.2014.07.029} {\bibfield  {journal} {\bibinfo  {journal}
			{Nuclear Instruments and Methods in Physics Research Section A: Accelerators,
				Spectrometers, Detectors and Associated Equipment}\ }\textbf {\bibinfo
			{volume} {764}},\ \bibinfo {pages} {156--166} (\bibinfo {year}
		{2014})}\BibitemShut {NoStop}%
	\bibitem [{\citenamefont {Jeckelmann}(2002)}]{PhysRevB.66.045114}%
	\BibitemOpen
	\bibfield  {author} {\bibinfo {author} {\bibfnamefont {Eric}\ \bibnamefont
			{Jeckelmann}},\ }\bibfield  {title} {\enquote {\bibinfo {title} {Dynamical
				density-matrix renormalization-group method},}\ }\href {\doibase
		10.1103/PhysRevB.66.045114} {\bibfield  {journal} {\bibinfo  {journal} {Phys.
				Rev. B}\ }\textbf {\bibinfo {volume} {66}},\ \bibinfo {pages} {045114}
		(\bibinfo {year} {2002})}\BibitemShut {NoStop}%
	\bibitem [{\citenamefont {Itoh}\ \emph {et~al.}(1995)\citenamefont {Itoh},
		\citenamefont {Kakurai}, \citenamefont {Endoh},\ and\ \citenamefont
		{Tanaka}}]{ITOH1995}%
	\BibitemOpen
	\bibfield  {author} {\bibinfo {author} {\bibfnamefont {S.}~\bibnamefont
			{Itoh}}, \bibinfo {author} {\bibfnamefont {K.}~\bibnamefont {Kakurai}},
		\bibinfo {author} {\bibfnamefont {Y.}~\bibnamefont {Endoh}}, \ and\ \bibinfo
		{author} {\bibfnamefont {H.}~\bibnamefont {Tanaka}},\ }\bibfield  {title}
	{\enquote {\bibinfo {title} {Inelastic pulsed-neutron scattering from
				{CsVCl}$_3$},}\ }\href {\doibase
		https://doi.org/10.1016/0921-4526(95)00091-M} {\bibfield  {journal} {\bibinfo
			{journal} {Physica B: Condens. Matter}\ }\textbf {\bibinfo {volume}
			{213-214}},\ \bibinfo {pages} {161--163} (\bibinfo {year}
		{1995})}\BibitemShut {NoStop}%
	\bibitem [{\citenamefont {Hutchings}\ \emph {et~al.}(1972)\citenamefont
		{Hutchings}, \citenamefont {Shirane}, \citenamefont {Birgeneau},\ and\
		\citenamefont {Holt}}]{Hutchings_1972}%
	\BibitemOpen
	\bibfield  {author} {\bibinfo {author} {\bibfnamefont {M.~T.}\ \bibnamefont
			{Hutchings}}, \bibinfo {author} {\bibfnamefont {G.}~\bibnamefont {Shirane}},
		\bibinfo {author} {\bibfnamefont {R.~J.}\ \bibnamefont {Birgeneau}}, \ and\
		\bibinfo {author} {\bibfnamefont {S.~L.}\ \bibnamefont {Holt}},\ }\bibfield
	{title} {\enquote {\bibinfo {title} {Spin dynamics in the one-dimensional
				antiferromagnet {${(\mathrm{C}{\mathrm{D}}_{3})}_{4}$
					NMn${\mathrm{Cl}}_{3}$}},}\ }\href {\doibase 10.1103/PhysRevB.5.1999}
	{\bibfield  {journal} {\bibinfo  {journal} {Phys. Rev. B}\ }\textbf {\bibinfo
			{volume} {5}},\ \bibinfo {pages} {1999--2014} (\bibinfo {year}
		{1972})}\BibitemShut {NoStop}%
	\bibitem [{\citenamefont {M\"uller}(1982)}]{PhysRevB.26.1311}%
	\BibitemOpen
	\bibfield  {author} {\bibinfo {author} {\bibfnamefont {Gerhard}\ \bibnamefont
			{M\"uller}},\ }\bibfield  {title} {\enquote {\bibinfo {title} {Sum rules in
				the dynamics of quantum spin chains},}\ }\href {\doibase
		10.1103/PhysRevB.26.1311} {\bibfield  {journal} {\bibinfo  {journal} {Phys.
				Rev. B}\ }\textbf {\bibinfo {volume} {26}},\ \bibinfo {pages} {1311--1320}
		(\bibinfo {year} {1982})}\BibitemShut {NoStop}%
	\bibitem [{\citenamefont {Harris}\ \emph {et~al.}(1971)\citenamefont {Harris},
		\citenamefont {Kumar}, \citenamefont {Halperin},\ and\ \citenamefont
		{Hohenberg}}]{PhysRevB.3.961}%
	\BibitemOpen
	\bibfield  {author} {\bibinfo {author} {\bibfnamefont {A.~B.}\ \bibnamefont
			{Harris}}, \bibinfo {author} {\bibfnamefont {D.}~\bibnamefont {Kumar}},
		\bibinfo {author} {\bibfnamefont {B.~I.}\ \bibnamefont {Halperin}}, \ and\
		\bibinfo {author} {\bibfnamefont {P.~C.}\ \bibnamefont {Hohenberg}},\
	}\bibfield  {title} {\enquote {\bibinfo {title} {Dynamics of an
				antiferromagnet at low temperatures: Spin-wave damping and hydrodynamics},}\
	}\href {\doibase 10.1103/PhysRevB.3.961} {\bibfield  {journal} {\bibinfo
			{journal} {Phys. Rev. B}\ }\textbf {\bibinfo {volume} {3}},\ \bibinfo {pages}
		{961--1024} (\bibinfo {year} {1971})}\BibitemShut {NoStop}%
\end{thebibliography}

\begin{thebibliography}{4}%
	\makeatletter
	\providecommand \@ifxundefined [1]{%
		\@ifx{#1\undefined}
	}%
	\providecommand \@ifnum [1]{%
		\ifnum #1\expandafter \@firstoftwo
		\else \expandafter \@secondoftwo
		\fi
	}%
	\providecommand \@ifx [1]{%
		\ifx #1\expandafter \@firstoftwo
		\else \expandafter \@secondoftwo
		\fi
	}%
	\providecommand \natexlab [1]{#1}%
	\providecommand \enquote  [1]{``#1''}%
	\providecommand \bibnamefont  [1]{#1}%
	\providecommand \bibfnamefont [1]{#1}%
	\providecommand \citenamefont [1]{#1}%
	\providecommand \href@noop [0]{\@secondoftwo}%
	\providecommand \href [0]{\begingroup \@sanitize@url \@href}%
	\providecommand \@href[1]{\@@startlink{#1}\@@href}%
	\providecommand \@@href[1]{\endgroup#1\@@endlink}%
	\providecommand \@sanitize@url [0]{\catcode `\\12\catcode `\$12\catcode
		`\&12\catcode `\#12\catcode `\^12\catcode `\_12\catcode `\%12\relax}%
	\providecommand \@@startlink[1]{}%
	\providecommand \@@endlink[0]{}%
	\providecommand \url  [0]{\begingroup\@sanitize@url \@url }%
	\providecommand \@url [1]{\endgroup\@href {#1}{\urlprefix }}%
	\providecommand \urlprefix  [0]{URL }%
	\providecommand \Eprint [0]{\href }%
	\providecommand \doibase [0]{http://dx.doi.org/}%
	\providecommand \selectlanguage [0]{\@gobble}%
	\providecommand \bibinfo  [0]{\@secondoftwo}%
	\providecommand \bibfield  [0]{\@secondoftwo}%
	\providecommand \translation [1]{[#1]}%
	\providecommand \BibitemOpen [0]{}%
	\providecommand \bibitemStop [0]{}%
	\providecommand \bibitemNoStop [0]{.\EOS\space}%
	\providecommand \EOS [0]{\spacefactor3000\relax}%
	\providecommand \BibitemShut  [1]{\csname bibitem#1\endcsname}%
	\let\auto@bib@innerbib\@empty
	%</preamble>
	\bibitem [{\citenamefont {Alvarez}(2009)}]{Alvarez2009}%
	\BibitemOpen
	\bibfield  {author} {\bibinfo {author} {\bibfnamefont {G.}~\bibnamefont
			{Alvarez}},\ }\bibfield  {title} {\enquote {\bibinfo {title} {The density
				matrix renormalization group for strongly correlated electron systems: A
				generic implementation},}\ }\href {\doibase 10.1016/j.cpc.2009.02.016}
	{\bibfield  {journal} {\bibinfo  {journal} {Comp. Phys. Comms.}\ }\textbf
		{\bibinfo {volume} {180}},\ \bibinfo {pages} {1572--1578} (\bibinfo {year}
		{2009})}\BibitemShut {NoStop}%
	\bibitem [{\citenamefont {Feiguin}\ and\ \citenamefont
		{White}(2005)}]{PhysRevB.72.220401}%
	\BibitemOpen
	\bibfield  {author} {\bibinfo {author} {\bibfnamefont {Adrian~E.}\
			\bibnamefont {Feiguin}}\ and\ \bibinfo {author} {\bibfnamefont {Steven~R.}\
			\bibnamefont {White}},\ }\bibfield  {title} {\enquote {\bibinfo {title}
			{Finite-temperature density matrix renormalization using an enlarged
				{Hilbert} space},}\ }\href {\doibase 10.1103/PhysRevB.72.220401} {\bibfield
		{journal} {\bibinfo  {journal} {Phys. Rev. B}\ }\textbf {\bibinfo {volume}
			{72}},\ \bibinfo {pages} {220401} (\bibinfo {year} {2005})}\BibitemShut
	{NoStop}%
	\bibitem [{\citenamefont {Feiguin}\ and\ \citenamefont
		{Fiete}(2010)}]{PhysRevB.81.075108}%
	\BibitemOpen
	\bibfield  {author} {\bibinfo {author} {\bibfnamefont {Adrian~E.}\
			\bibnamefont {Feiguin}}\ and\ \bibinfo {author} {\bibfnamefont {Gregory~A.}\
			\bibnamefont {Fiete}},\ }\bibfield  {title} {\enquote {\bibinfo {title}
			{Spectral properties of a spin-incoherent {Luttinger} liquid},}\ }\href
	{\doibase 10.1103/PhysRevB.81.075108} {\bibfield  {journal} {\bibinfo
			{journal} {Phys. Rev. B}\ }\textbf {\bibinfo {volume} {81}},\ \bibinfo
		{pages} {075108} (\bibinfo {year} {2010})}\BibitemShut {NoStop}%
	\bibitem [{\citenamefont {Nocera}\ and\ \citenamefont
		{Alvarez}(2016)}]{PhysRevB.93.045137}%
	\BibitemOpen
	\bibfield  {author} {\bibinfo {author} {\bibfnamefont {A.}~\bibnamefont
			{Nocera}}\ and\ \bibinfo {author} {\bibfnamefont {G.}~\bibnamefont
			{Alvarez}},\ }\bibfield  {title} {\enquote {\bibinfo {title}
			{Symmetry-conserving purification of quantum states within the density matrix
				renormalization group},}\ }\href {\doibase 10.1103/PhysRevB.93.045137}
	{\bibfield  {journal} {\bibinfo  {journal} {Phys. Rev. B}\ }\textbf {\bibinfo
			{volume} {93}},\ \bibinfo {pages} {045137} (\bibinfo {year}
		{2016})}\BibitemShut {NoStop}%
\end{thebibliography}
\end{document}